%% file: GRRT.tex
\documentclass[11pt,a4paper]{article}

\usepackage{graphicx}
\usepackage{float}
\usepackage{afterpage}
\usepackage{epsfig,cite}
\usepackage{amssymb}
\usepackage{amsmath}
\usepackage{dsfont}
\usepackage{multirow}
\usepackage{url}

\newcommand{\be}{\begin{equation}}
\newcommand{\ee}{\end{equation}}
\newcommand{\bea}{\begin{eqnarray}}
\newcommand{\eea}{\end{eqnarray}}
\newcommand{\bi}{\begin{itemize}}
\newcommand{\ei}{\end{itemize}}
\newcommand{\ben}{\begin{enumerate}}
\newcommand{\een}{\end{enumerate}}
\newcommand{\la}{\left\langle}
\newcommand{\ra}{\right\rangle}
\newcommand{\lc}{\left[}
\newcommand{\rc}{\right]}
\newcommand{\lp}{\left(}
\newcommand{\rp}{\right)}

\def\frac#1#2{{{#1}\over {#2}}}
\def\gsim{\mathrel{\rlap{\lower4pt\hbox{\hskip1pt$\sim$}}
    \raise1pt\hbox{$>$}}}         
\def\lsim{\mathrel{\rlap{\lower4pt\hbox{\hskip1pt$\sim$}}
    \raise1pt\hbox{$<$}}}         

\newcommand{\draft}[1]{}

\def\beq{\begin{equation}}  
\def\eeq{\end{equation}}

\def \n0{N_j^{(0)}}

\def\lapprox{\lower .7ex\hbox{$\;\stackrel{\textstyle <}{\sim}\;$}}
\def\gapprox{\lower .7ex\hbox{$\;\stackrel{\textstyle >}{\sim}\;$}}

\begin{document}
\vspace{-2.0cm}
\begin{flushright}
OUTP-15-06P \\
DCPT/15/32 \\
IPPP/15/16 \\
\end{flushright}

\begin{center}
  {\Large \bf Charm production in the forward
    region: constraints on \\[0.1cm] the small-$x$ gluon and backgrounds for neutrino
    astronomy}
\vspace{.7cm}

Rhorry~Gauld$^{1}$, Juan~Rojo$^{2}$, Luca~Rottoli$^{2}$ and Jim~Talbert$^{2}$

\vspace{.3cm}
       {\it ~$^1$ Institute for Particle Physics Phenomenology, \\
         Durham University, 
           Durham DH1 3LE, UK\\
  ~$^2$ Rudolf Peierls Centre for Theoretical Physics, 1 Keble Road,\\
        University of Oxford, OX1 3NP Oxford, UK}
\end{center}   

\vspace{0.1cm}

\begin{center}
{\bf \large Abstract}
\end{center}

The recent observation by the IceCube experiment of cosmic neutrinos
at energies up to a few PeV heralds the beginning of neutrino
astronomy.
At such high energies, the `conventional' neutrino flux is
suppressed and the `prompt' component from charm meson decays is
expected to become the dominant background to astrophysical neutrinos.
Charm production at high energies is however theoretically uncertain,
both since the scale uncertainties of the NLO calculation are large,
and also because it is directly sensitive to the poorly-known gluon PDF at small-$x$.
In this work we provide detailed perturbative QCD predictions for charm
and bottom production in
the forward region, and validate them by comparing with recent data
from the LHCb experiment at 7 TeV.
Finding good agreement between data and theory, we use the LHCb measurements
to constrain the small-$x$ gluon PDF, achieving a substantial reduction
in its uncertainties.
Using these improved PDFs, we
provide predictions for charm and bottom production at LHCb at
13 TeV, as well as for the
ratio of cross-sections between 13 and 7 TeV.
The same calculations are used to compute
the energy distribution of neutrinos from charm decays in $pA$
collisions, a key ingredient towards achieving a theoretically
robust estimate of charm-induced
backgrounds at neutrino telescopes.

\clearpage

\tableofcontents

\clearpage

\input{sec-introduction}

\input{sec-pqcd}

\input{sec-rw}

\input{sec-13tev}

\input{sec-results}

\input{sec-delivery}

\appendix
\input{sec-appendix}

\input{GRRT.bbl}

\end{document}

%% file: sec-introduction.tex
\section{Introduction}
\label{sec:introduction}

The recent observation of very high-energy cosmic neutrinos by the
IceCube experiment at the South Pole marks the beginning of neutrino
astronomy~\cite{Aartsen:2013bka,Aartsen:2013jdh}.
The most recent (2010-12) dataset~\cite{Aartsen:2014gkd} contains 37
neutrino candidates with energies between 30 and 2000 TeV, and arrival
directions consistent with isotropy.
At these high energies, the `conventional' atmospheric neutrino flux,
arising from the decays of pions and kaons produced by the collisions
of cosmic rays with nuclei in the
atmosphere~\cite{Barr:2004br,GonzalezGarcia:2006ay,Honda:2006qj} is
highly suppressed due to energy loss before the decays occur.
However charmed mesons decay almost instantaneously so at high
energies, despite their smaller production cross-section, the
so-called `prompt' neutrino flux from charm
decays~\cite{Lipari:1993hd,Pasquali:1998ji,Enberg:2008te,Gondolo:1995fq,Martin:2003us,Gelmini:1999ve,Bhattacharya:2015jpa,Garzelli:2015psa,Engel:2015dxa,Arguelles:2015wba}
becomes the dominant background to astrophysical neutrinos.
The prompt flux has a harder spectrum than the conventional flux and
is thus difficult to distinguish from the expected astrophysical
neutrinos on this basis.

It is therefore essential to have a reliable
estimate of this prompt neutrino background.
Unfortunately, charm production at high energies is affected by
substantial theoretical uncertainties when computed in perturbative
QCD (pQCD).
First of all, the small value of the charm quark mass ($m_c$), 
close to $\Lambda_{\rm QCD}$, leads to a large value for
$\alpha_s(m_c)$, which translates into substantial scale
uncertainties in the NLO calculation.
In addition, this process probes the gluon PDF at very small values of
$x$, around $x\simeq 10^{-5}$, where there are no direct
experimental constraints and consequently large
uncertainties~\cite{Rojo:2015acz,Watt:2011kp,Ball:2012wy,Gao:2013xoa,Ball:2010de,Harland-Lang:2014zoa}.
Another source of theoretical uncertainty is the
choice of the value of $m_c$ itself.

For these reasons, alternative calculations based on saturation 
models or non-linear evolution dynamics have been proposed.
However, these calculations are model dependent, seldom
validated with collider data, and often based on
outdated PDF sets.
A possible alternative would be to use high-energy resummation for
heavy quark production~\cite{Ball:2001pq}, but for consistency this approach
requires a small-$x$ resummed PDF fit~\cite{Altarelli:2008aj,Ciafaloni:2007gf} which is 
currently not available.
While there are some hints for deviations with respect to fixed-order
DGLAP evolution in inclusive HERA
data~\cite{Aaron:2009aa,Caola:2009iy,H1:2015mha}, there is so far no conclusive
evidence that fixed-order pQCD cannot be reliably applied to the
region relevant for calculations of atmospheric charm production.
Therefore, our predictions will be based on next-to-leading order (NLO) QCD,
where charm fragmentation is accounted for either analytically or
by the matching to parton showers.

With the above
motivation, in this work
we provide state-of-the-art pQCD predictions
for charm and bottom production in the forward region.
Our calculations are based both on the semi-analytical {\sc\small
  FONLL} approach~\cite{Cacciari:1993mq},
as well as the fully exclusive
description of the final state provided by the {\sc\small
  MadGraph5\_aMC@NLO}~\cite{Alwall:2014hca} and {\sc\small POWHEG} 
Monte Carlo programs,
where the NLO result is matched to the {\sc\small
  Pythia8}~\cite{Sjostrand:2007gs,Sjostrand:2014zea} parton shower.
As input in the calculation, we use the recent NNPDF3.0 NLO PDF
set~\cite{Ball:2014uwa} and verify the stability
of the results when other modern PDF sets are used, 
in particular MMHT14~\cite{Harland-Lang:2014zoa} and CT10~\cite{Lai:2010vv}.

One central ingredient of our approach
is the validation of our pQCD calculations with the 
data from the LHCb experiment on charm and bottom production in the
forward region at 7 TeV~\cite{Aaij:2013noa,Aaij:2013mga}.
The LHCb measurements cover a similar kinematical range as that of charm
production relevant to the prompt neutrino background for IceCube.
For instance, an incoming cosmic ray with energy $E=100$~PeV 
corresponds to a centre-of-mass energy of $\sqrt{s} =\sqrt{2m_NE}\simeq 14$~TeV.
Measurements of forwardly produced
heavy flavour hadrons therefore provide a perfect environment for testing
the validity of pQCD prompt neutrino flux predictions.
As we will show, both the analytical {\sc\small FONLL} calculation and
the exclusive {\sc\small MadGraph5\_aMC@NLO} and {\sc\small POWHEG}
results are consistent with the LHCb charm and bottom data
within theoretical uncertainties.
Therefore, we can
be confident that these calculations
can be reliably applied to predictions
of the atmospheric prompt neutrino flux.

The compatibility between the NLO QCD predictions and the 7 TeV
charm production data from LHCb indicates that it is possible
to use this process to constrain the small-$x$ gluon
PDF~\cite{Zenaiev:2015rfa}.
To partially cancel the large scale uncertainties of the NLO
calculation, we construct normalised differential cross-sections using
a fixed bin as reference.
We then include the LHCb charm data into NNPDF3.0 fit
using the Bayesian reweighting method~\cite{Ball:2010gb,Ball:2011gg},
finding a substantial reduction of the small-$x$ gluon PDF
uncertainties.
The resulting PDF set, NNPDF3.0+LHCb, is particularly suitable
for providing predictions for both heavy quark production within the LHCb
acceptance at 13 TeV, as well as to provide a reliable estimate of the rate of high
energy neutrino production in $pA$ collisions relevant for estimations of
the prompt neutrino flux at IceCube.

In this work we also provide detailed predictions for charm and bottom production
at LHCb Run II, using the improved NNPDF3.0+LHCb PDFs, including the 
evaluation of theoretical uncertainties arising from missing higher-orders, PDFs, and 
the value of the heavy quark mass.
Our results are tabulated using the binning scheme adopted for the 7~TeV
measurements, and predictions for other binning choices are available upon request.
In addition, we also provide predictions for the ratio of differential
distributions of charm and bottom production between 13 and 7 TeV,
$R_{13/7}$, which provides complementary information on PDF
discrimination~\cite{Mangano:2012mh}.
After computing this observable and its corresponding theoretical 
uncertainty for $B$ and $D$ mesons, we apply our calculations
to the LHCb 7~TeV data to provide robust predictions for the fiducial
cross-sections within the LHCb acceptance for Run II. These predictions 
are useful for estimating $B$ and $D$ yields at 13~TeV, which can in
turn be used to assess the statistical precision of future measurements ---
such as rare $B$ decays for example.

Using the same theoretical set-up as outlined for the LHC calculations, 
we provide predictions for the neutrino energy spectrum arising from 
the decays of charmed mesons in high-energy proton-air
collisions.
These results are an important ingredient for the computation of
the expected number of prompt neutrino events at
IceCube.
While it is beyond the scope of this paper to
compare with the IceCube measurements,
our $pA \to \nu X$ cross-sections are available
in the form of an interpolation code for the relevant
range of incoming cosmic ray energies. These results can be 
used as an input for well-established frameworks such as the
$Z$-moment approach~\cite{Gondolo:1995fq,Bhattacharya:2015jpa}
to construct predictions for IceCube.

The outline of this paper is as follows.
In Sect.~\ref{sec:pqcd} we review the framework for pQCD
computations of heavy quark production and perform
an extensive comparison with the LHCb 7 TeV data on charm
and bottom cross-sections.
In Sect.~\ref{sec:rw} we include the normalised LHCb charm data
into NNPDF3.0 using the Bayesian 
reweighting method, obtaining an improved PDF set with reduced
uncertainties in the small-$x$ region which will be the central
ingredient of our subsequent calculations.
In Sect.~\ref{sec:results13tev} we provide
predictions for heavy quark production within the LHCb acceptance
at 13~TeV, as well as ratio of 13 over 7 TeV cross-sections.
In Sect.~\ref{sec:results} we present our 
predictions for the energy distributions of neutrinos from charm
decays in $pA$ collisions for a range of incoming cosmic ray
energies relevant for neutrino telescopes.
In Sect.~\ref{sec:delivery}
we summarise our findings and discuss possible next steps.
Appendix~\ref{sec:appendix} contains a tabulation our
theory predictions for
charm and bottom production at LHCb
at 13~TeV, as well as the ratio of cross-sections
between 13 and 7~TeV.

%% file: sec-pqcd.tex
\section{Heavy quark production in the forward region and LHCb data}
\label{sec:pqcd}

In pQCD, the NLO calculation of heavy quark pair production
in hadronic collisions
has been available for a long time, both at the level of total
inclusive cross-sections~\cite{Nason:1987xz}, and of differential
distributions~\cite{Nason:1989zy,Mangano:1991jk,Beenakker:1990maa,Beenakker:1988bq}.
Subsequently, the fixed-order calculation has been improved with the
resummation of soft gluons at NLL~\cite{Bonciani:1998vc,Kidonakis:1997gm} and
NNLL~\cite{Czakon:2009zw,Ahrens:2010zv} accuracy.
Another way of refining the fixed-order result is by matching it to
the massless calculation, valid in the limit where the heavy quark
transverse momentum ($p_T^h$) greatly exceeds the heavy quark mass ($m_h$),
thus obtaining a result which is valid both at small and at large
values of $p_T^h$~\cite{Cacciari:1993mq,Cacciari:1998it,Kniehl:2008zza}, 
and has the benefit of reduced of scale uncertainties as compared to the NLO calculation.
More recently, the next-to-NLO (NNLO) calculation for inclusive heavy quark pair production 
has become available~\cite{Czakon:2012pz,Czakon:2013goa,Baernreuther:2012ws},
and results for the differential distributions for the case of top quark production have also been
presented~\cite{Czakon:2014xsa,Czakon:2015pga}.
These calculations will eventually be applied to charm
and bottom production as well.

In this section, we begin by discussing our set-up for providing
pQCD calculations of charm and beauty production, and their
subsequent fragmentation and decay.
We then demonstrate that the kinematic coverage of charm
production at LHCb data overlaps with that
relevant for the calculation of prompt neutrino fluxes at IceCube.
With this in mind, we present a detailed comparison of the pQCD
calculations for charm and bottom production in the forward region with
the 7~TeV LHCb data, and examine relevant sources of theoretical uncertainty.
Throughout this work, the NNPDF3.0 NLO PDF set will be used as a 
baseline for our predictions, and we also
study the dependence of our predictions on the 
choice of input PDF set.

\subsection{Heavy quark production in the forward region}
\label{sec:thsettings}
In this work we will provide pQCD predictions of heavy quark pair 
production using three different approaches: {\sc\small FONLL}, {\sc\small
  POWHEG} and {\sc\small MadGraph5\_{\rm a}MC@NLO}. 
We discuss briefly each of these approaches in turn.
A similar comparison between different calculations for
heavy quark production at the LHC and other hadron colliders, focussed
on data in the central rapidity region, was presented in~\cite{Cacciari:2012ny}.

\begin{itemize}
\item {\sc\small
  FONLL}~\cite{Cacciari:1993mq,Cacciari:1998it,Cacciari:2001td} is a
  semi-analytical calculation based on the matching of the NLO fixed-order
  calculation~\cite{Nason:1989zy}, including full dependence on the heavy quark mass $m_h$,
  with the resummed NLL calculation where the heavy quark is treated
  as a massless parton.
  This matching allows a consistent description of the $p_T^h$
  spectrum, from low to high transverse momenta.\footnote{ The FONLL
  approach can also be applied
  to other processes, such as DIS structure functions~\cite{Forte:2010ta}.}
  The fragmentation of heavy quarks into heavy flavored hadrons is
  then described analytically~\cite{Cacciari:2005uk}, with parameters
  extracted from LEP data.
  It is also possible to include the decays of the $D$ mesons using this approach.

  In the region relevant for the LHCb data, where $p_T^h$ does not greatly exceed $m_h$, 
  the {\sc\small FONLL} result corresponds to the fixed-order NLO massive
  calculation, and thus for simplicity in this work
  by ``{\sc\small
    FONLL} calculation''
  we denote the fixed-order NLO obtained from the {\sc\small
  FONLL} code.
  
\item The {\sc\small POWHEG}~\cite{Nason:2004rx,Frixione:2007vw,Alioli:2010xd} method
	allows NLO calculations to be matched to a Monte Carlo parton shower. In the case of 
	heavy quark production~\cite{Frixione:2007nw}, the massive NLO calculation performed
	in a fixed-flavour scheme is matched achieving NLO+LL accuracy ---
	thanks to the resummation achieved by the parton shower.
  The fragmentation and hadronisation
  of heavy quarks into heavy hadrons and their
  subsequent decay into leptons is then modeled by the specific parton
  shower which has been matched too, with modelling parameters tuned to data.
  In this work we use {\sc\small POWHEG} matched to the {\sc\small
    Pythia8} shower~\cite{Sjostrand:2007gs,Sjostrand:2014zea}, using the
  {\sc\small
    Monash 2013} tune for the modelling of the soft and semi-hard
  physics~\cite{Skands:2014pea}. We will refer to this set-up as the 
  {\sc\small POWHEG} calculation.

\item {\sc\small MadGraph5\_{\rm a}MC@NLO}~\cite{Alwall:2014hca} provides
  automated calculations of arbitrary processes at LO and NLO,
  both at fixed-order and matched to a variety of
  parton showers using the {\sc\small MC@NLO} method~\cite{Frixione:2002ik}.
  For consistency with the {\sc\small POWHEG} calculation, the
  {\sc\small Pythia8} parton shower and {\sc\small Monash 2013} tune
  are also used for this prediction,
  therefore treating charm hadronisation and decay with universal
  settings.
  Note that the {\sc\small MC@NLO} and {\sc\small POWHEG} methods
  to match fixed-order calculations with parton showers are different,
  and thus the spread between the two 
 calculations provide an estimate of the underlying theoretical
 uncertainties introduced by the various matching processes. This set-up 
 is referred to as the {\sc\small  {\rm a}MC@NLO} calculation.

 In the kinematic region relevant for charm and bottom production
 at LHCb, the effects of parton shower resummation in {\sc\small POWHEG}
 and {\sc\small {\rm a}MC@NLO} are expected to be moderate,
 and thus the comparison of the three generators allows a meaningful
  validation of the pQCD calculations for the heavy
  quark production and fragmentation using three
  independent approaches.
  
\end{itemize}

  The following common set of theory input
  parameters are adopted for all three calculations:
\begin{itemize}
\item As the input set of parton distributions, we use the
  NNPDF3.0 NLO PDF set~\cite{Ball:2014uwa} with five
  active flavours ($n_f=5$).
  The dependence of the results with respect to the choice of
  input PDF set will be discussed in Sect.~\ref{sec:deppdfset}, and
  comparisons with recent PDF fits will be made in Sect.~\ref{sec:rw}.
   At the LHC, charm and bottom pairs are predominantly produced 
   through the gluon-gluon initial state, and therefore
   our calculation will be sensitive to the details of
   the gluon PDF and the associated uncertainties at small-$x$.
   
   Charm production in the presented {\sc\small FONLL} 
   predictions only includes the matching between the $n_f=3$
   to the $n_f=4$ schemes, so in principle one should use
   a $n_f=4$ PDF set for consistency.
   However, it 
   has been verified that the results are
   unchanged in the latter scenario:
   differences between using {\sc\small FONLL}
  with $n_f=4$ and $n_f=5$ PDFs for charm production are at most
    1.5\% at the highest values of $p_T^D$ covered by the LHCb data, much
    smaller than any other theoretical
    or experimental uncertainty.

   In the case of both {\sc\small POWHEG}
   and {\sc\small {\rm a}MC@NLO} calculations, the matching between 
   schemes is not included.
      We have verified however, by explicitly including these
   terms in the {\sc\small POWHEG}~\cite{Gauld:2015lxa} calculation, that
   such effects are also in this case unimportant.
   In particular,
   the effect of including the 
    $n_f=3$ to $n_f=5$ compensation terms in the {\sc\small POWHEG} calcaulation
    with a $n_f=5$ PDF set leads to an increase in scale variation of (2-3)\% above $m_b$,
    while the central value is essentially unaltered ($< 1\%$) due to a compensation of $n_f$ dependent 
    $\alpha_s$ modifications and a depletion of the gluon PDF due to $g\rightarrow Q\bar{Q}$ splittings.

    For completeness, we provide here the explicit expressions
    of the compensation terms
    that must be dynamically applied to the partonic heavy-quark
     production cross-section to transform from the $n_f$ to the $n_f+1$ scheme:
\beq \label{comp}
\begin{aligned}
-\hat{\sigma}^{(0)}_{q\bar{q}} 	& \frac{2 T_F \alpha_s(\mu_R^2)}{3\pi} \rm{Log}\left[ \frac{\mu_R^2}{m_Q^2}\right]  \,,\\
-\hat{\sigma}^{(0)}_{gg} 		& \frac{2 T_F \alpha_s(\mu_R^2)}{3\pi} \left( \rm{Log}\left[ \frac{\mu_R^2}{m_Q^2}\right] -  \rm{Log}\left[ \frac{\mu_F^2}{m_Q^2}\right] \right)\, .
\end{aligned}
\eeq

  These expressions are valid for the choice $\mu_F$ and 
  $\mu_R > m_Q$. If only the value of $\mu_F$ exceeds 
  $m_Q$, then only the $\mu_F$-dependent correction to 
  the gluon-gluon induced process should be applied 
  (and similarly for the $\mu_R$-dependent corrections). In the 
  case of charm production, if $\mu_F$ and $\mu_R$ exceed $m_b$, 
  then the corrections~(\ref{comp}) should be applied at both 
  charm and bottom thresholds.
  
  In addition, let us recall that differences between  $n_f=4$  and
  $n_f=5$ PDFs are only sizeable far above the bottom
  threshold~\cite{Ball:2011mu}, thus not relevant for the analysis
  of the LHCb production data.
   
 \item The value of the strong coupling constant is taken to
   be $\alpha_s(m_Z)=0.118$, consistent with the latest
   PDG average~\cite{Beringer:1900zz}.
   The uncertainties due to the uncertainty of the value of $\alpha_s(m_Z)$ are negligible
   as compared to other sources of theory uncertainty and are
   thus not considered here.

   \item Concerning the treatment of $\alpha_s(Q)$, in this
  work we always use
  consistently the same heavy flavour scheme as the
  corresponding input PDF set.
  Since we use $n_f=5$ PDF sets, then $\alpha_s(Q)$ runs with up to
  $n_f=5$ active flavours depending on the value of $Q$.
  Close to the charm threshold, $\alpha_s^{(n_f=3)}(Q)$
  and $\alpha_s^{(n_f=5)}(Q)$ are extremely similar by construction.

  Note also that the VFN running of $\alpha_s(Q)$ is essential to
  obtain agreement with the PDG global average of $\alpha_s(m_Z)$:
  using the $n_f=3$ scheme all the way up to $Q=m_Z$ will lead to a value of
  $\alpha_s(m_Z)$ much smaller than the PDG average.
  
 \item The central renormalisation and factorisation scales are varied event-by-event,
   and taken to be
\be
\mu_F=\mu_R=\sqrt{m_h^2 + p_{T,h}^2} \, .
\ee
	To estimate the size of missing higher-order corrections, 
	$\mu_F$ and $\mu_R$ are varied by a factor of two around the
	central scale, with the restriction $1/2 \le \mu_F/\mu_R \le 2$ to
	avoid introducing artificially large logarithms. Uncertainties computed 
	in this way are referred to as scale uncertainties.

\item The charm quark pole mass is taken to be $m_c=1.5 \pm 0.2$ GeV,
  while for the bottom quark pole mass we use $m_b=4.75\pm 0.25$ GeV.
  The uncertainty of $m_c$ and $m_b$ will be included in the theory uncertainty
  of our calculation.
  While it should be possible to reduce the theory uncertainty due to
  the choice of heavy quark masses by using calculations in the $\overline{\rm MS}$
  scheme~\cite{Dowling:2013baa},
  where the latest PDG values are $m_c(m_c)=1.275\pm 0.025$ GeV
  and $m_b(m_b)=4.18\pm 0.03$ GeV~\cite{Beringer:1900zz},
  this would not affect our results since the
  uncertainties due to $\delta m_c$ (and even more due to $\delta m_b$) are subleading
  as compared to other theory uncertainties.
  
\item The fragmentation probabilities $f(c\to D)$ for the
  different types of charmed mesons are taken to be the same as those
  of the LHCb measurement~\cite{Aaij:2013mga}, viz. $f(c\to D^0)=0.565$, $f(c\to D^{\pm})=0.246$,
   $f(c\to D_s^{\pm})=0.080$, and $f(c\to \Lambda_c)=0.094$. When uncertainties
   are considered, the sum of these fragmentation probabilities is consistent with unity.
   In comparison to the other sources of theoretical uncertainty, the
   impact of the uncertainty of these values for the considered observables is negligible.
   
\item When semi-leptonic decays of $D$ hadrons are considered, 
 the following branching fractions are enforced:
 $\mathcal{B}( D^0 \to \nu_l X ) = 0.101$, $\mathcal{B}( D^{\pm} \to \nu_l X ) = 0.153$,
 $\mathcal{B}( D_s^{\pm} \to \nu_l X ) = 0.06$, and $\mathcal{B}( \Lambda_c \to \nu_l X ) = 0.02$.
 Combined with the fragmentation probabilities, this corresponds to
 a partial decay width
 $\Gamma(c\to \nu_lX)/\Gamma(c\to \rm{anything}) = 0.102$ for prompt $D$ hadron decays.
 
\item The fragmentation probabilities $f(b\to B)$ for bottom mesons
  are taken to be $f(b\to B_u)=f(b\to B_d)=0.337$, as determined
  by the LHCb analysis of Ref.~\cite{Aaij:2011jp}. 
\end{itemize}

\subsection{Sensitivity to the small-$x$ gluon PDF}

\begin{figure}[t]
\centering 
\includegraphics[scale=0.39]{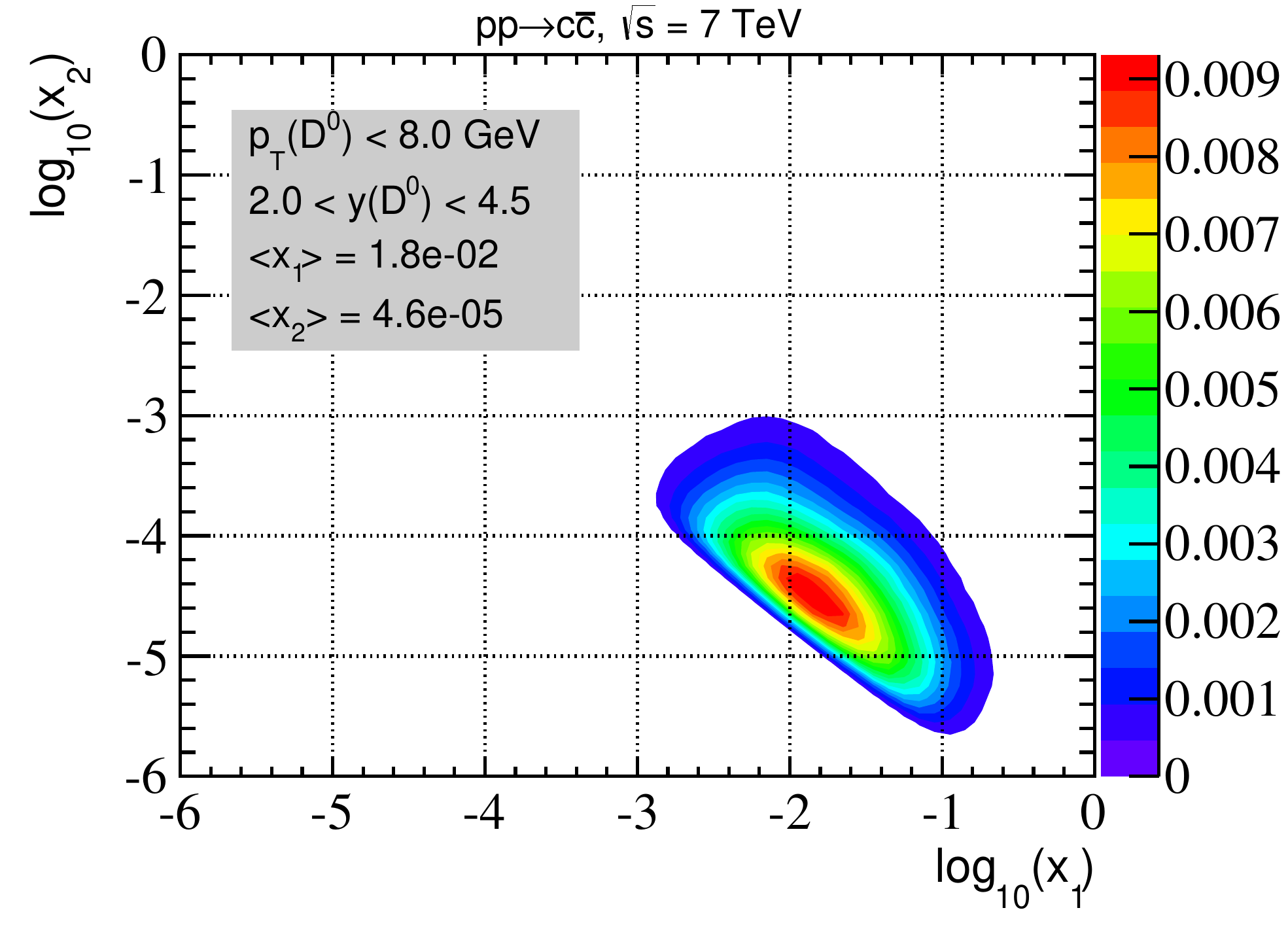}
\includegraphics[scale=0.39]{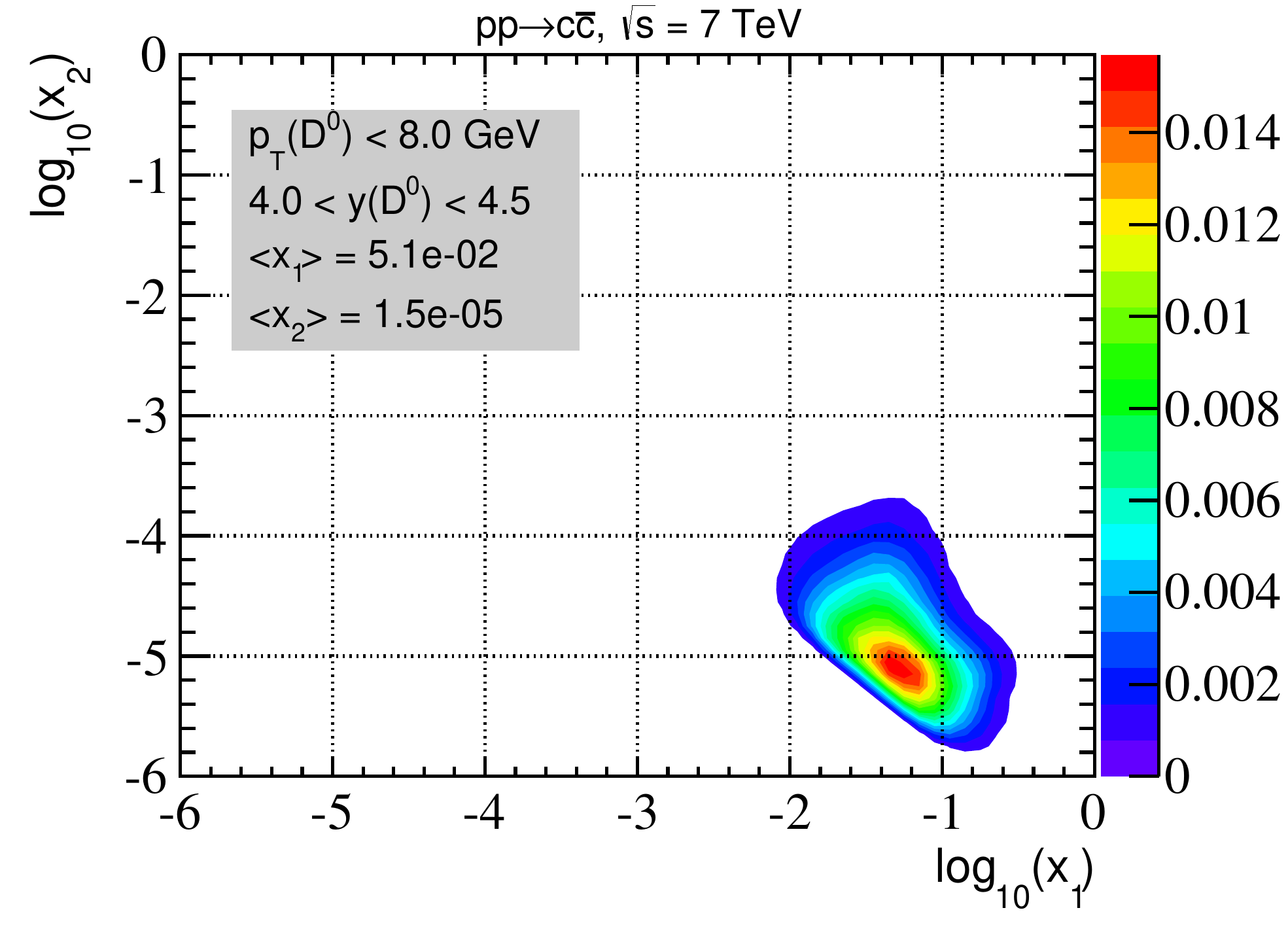}
\includegraphics[scale=0.39]{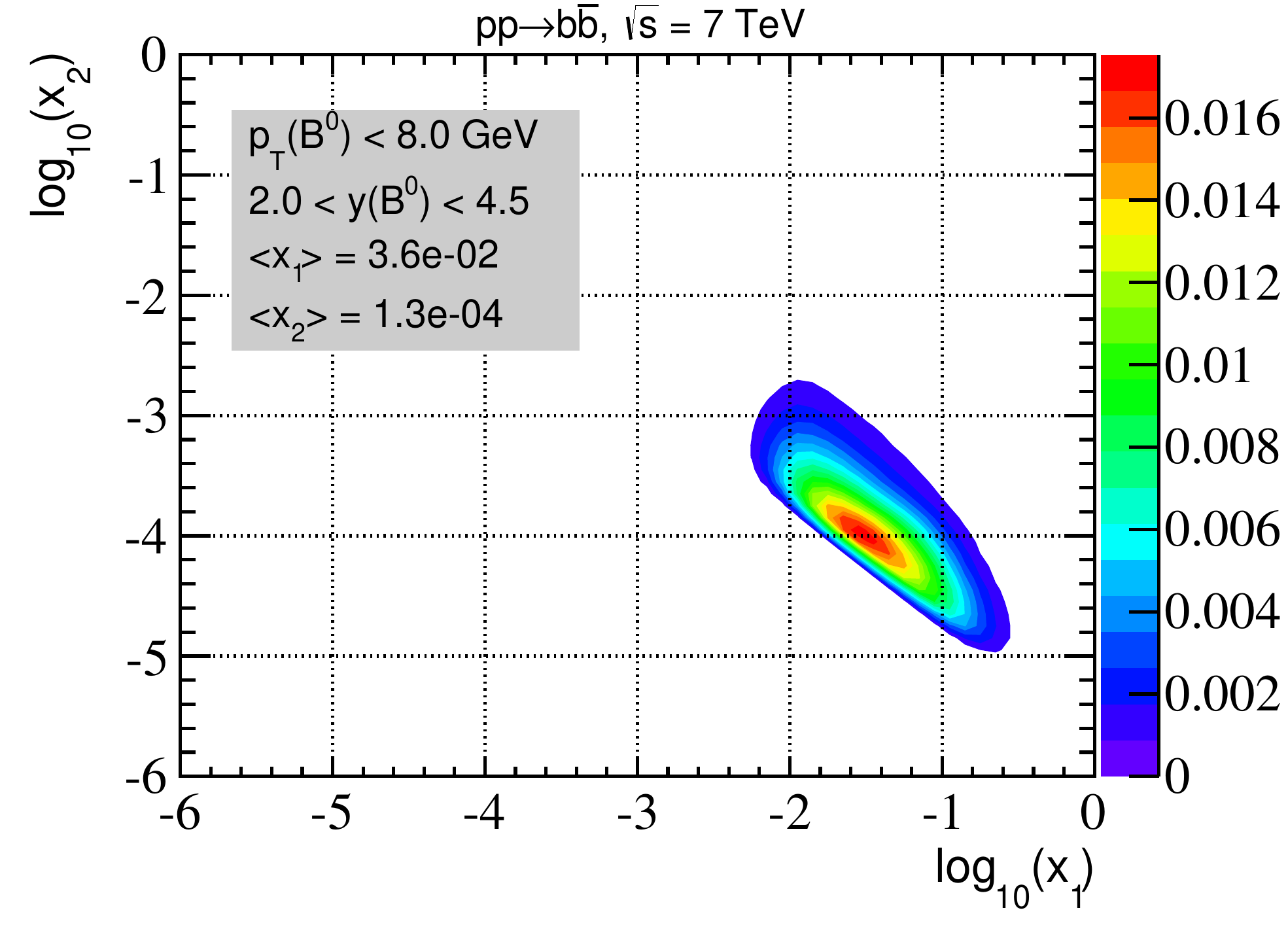}
\includegraphics[scale=0.39]{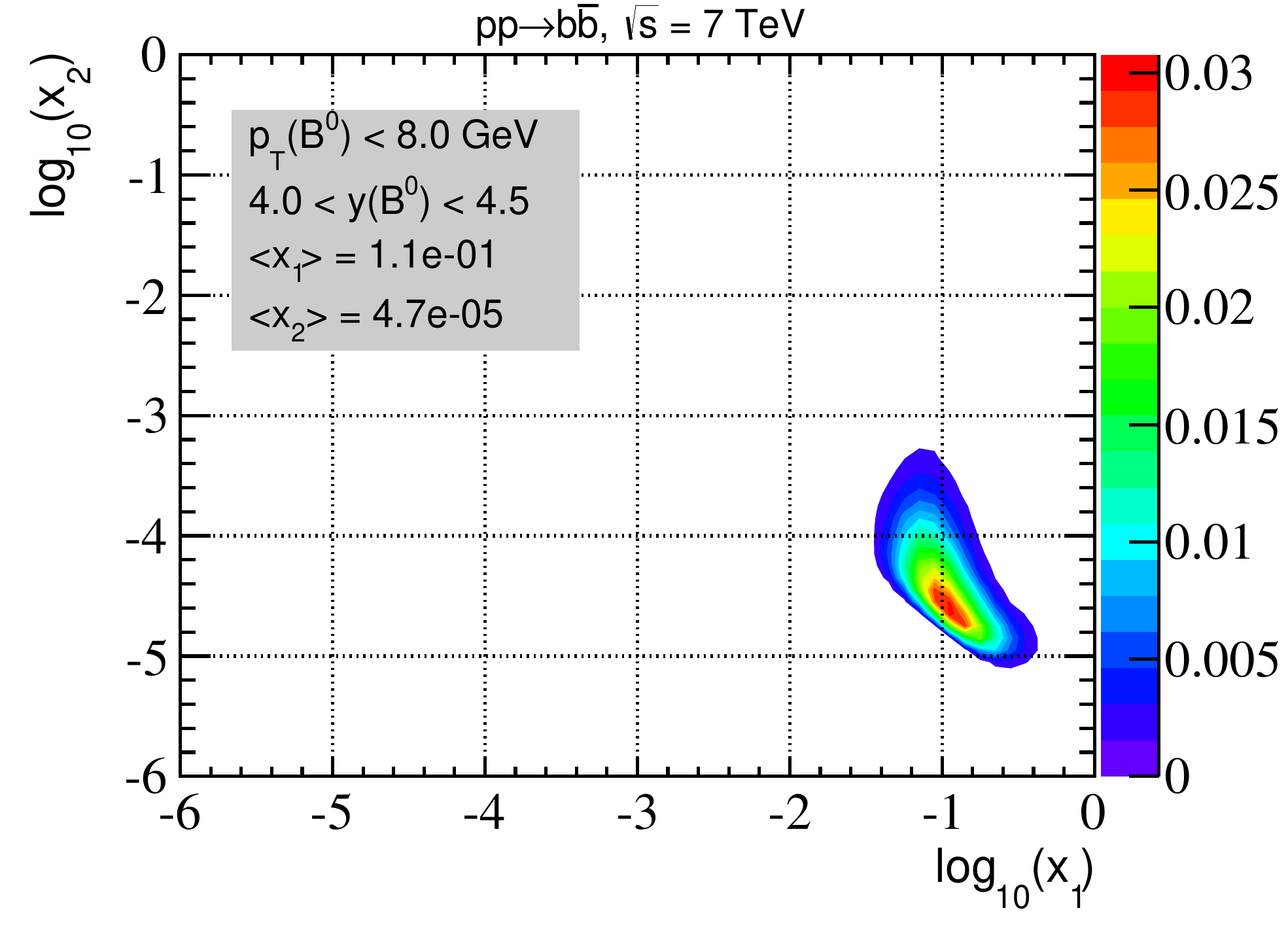}
\caption{\small Contour plot for the values of $(x_1,x_2)$
  sampled in the LO calculation of charm (upper plots)
  and bottom (lower plots) production at 7 TeV, within the
  LHCb fiducial acceptance.
  The calculation has been performed with {\sc\small POWHEG}
  using the NNPDF3.0 LO set.
  The regions in red indicate where the PDFs are sampled
  more frequently, while those in blue indicate less frequent
  sampling.
  The left plots have been computed in the full fiducial region,
  while the right plots are restricted to the
  forward region $4.0 \le y \le 4.5$.
}
\label{fig:kincovLO}
\end{figure}

%

In order to better understand the relation between heavy quark production kinematics
and the gluon PDF, it is useful to 
determine the coverage in the $(x_1,x_2)$ plane of the LHCb charm and
bottom measurements, where $x_1$ and $x_2$ are the values of
Bjorken-$x$ corresponding to the PDFs in each of the two incoming protons.
This coverage is illustrated by
the various contour plots shown in Fig.~\ref{fig:kincovLO}.
These plots contain the values of $(x_1,x_2)$
  sampled by the LO calculation of charm (upper)
  and bottom (lower) production at 7 TeV, within the
  LHCb acceptance.
  In the left plots, $D^0$ and $B^0$ hadrons are required to be within
  the LHCb rapidity acceptance ($2.0 \le y \le 4.5$) and have been 
  restricted to a low $p_T$ region ($p_T < 8$~GeV). In the right plots,
  the hadrons are further restricted in rapidity to the most forward region
  with $4.0 \le y \le 4.5$.
  The calculation has been performed with {\sc\small POWHEG}
  using NNPDF3.0 LO.
  In all plots, the contours have been normalised to the corresponding 
  fiducial region, and therefore the regions in red indicate where the PDFs 
  are sampled more frequently, while those in blue indicate less frequent
  sampling.
  Note that due to the asymmetric acceptance of LHCb, events with $x_1 \ge x_2$,
  where the first parton is a constituent of the proton travelling in the direction of the 
  LHCb detector (positive rapidity), will be typically selected.
 
  As shown in Fig.~\ref{fig:kincovLO}, measurements of charm
  production probe average values of Bjorken-$x$
  as low as $\la x_2\ra\simeq 4.6\cdot10^{-5}$, and even knowledge of the gluon
  PDF for values
  below $x\le 10^{-5}$ is required for particular bins.
  This is demonstrated by the plot restricted to the forward region,
  where $\la x_2\ra \simeq 1.5\cdot10^{-5}$.
  In this region, there is very limited
  direct experimental information, since HERA inclusive structure
  function data~\cite{Aaron:2009aa} is only
  available down to $x_{\rm min}\sim 6\cdot10^{-5}$.
  For this reason, it is of paramount importance to validate
  our pQCD calculation with the LHCb data itself, since
  we are using as input PDFs in a region where uncertainties
  are extremely large.
  In contrast, the situation for bottom production is under
  better control since $\la x_2\ra\simeq 1.3\cdot10^{-4}$,
  a region well covered by the HERA data.
  This said, for bottom production in the most forward bin,
  $4.0 \le y \le 4.5$, we find that $\la x_2\ra\simeq 4.7\cdot10^{-5}$,
  just below the limit of HERA data, demonstrating that PDF
  uncertainties also have a sizable impact in this region.

\begin{figure}[t]
\centering 
\includegraphics[scale=0.39]{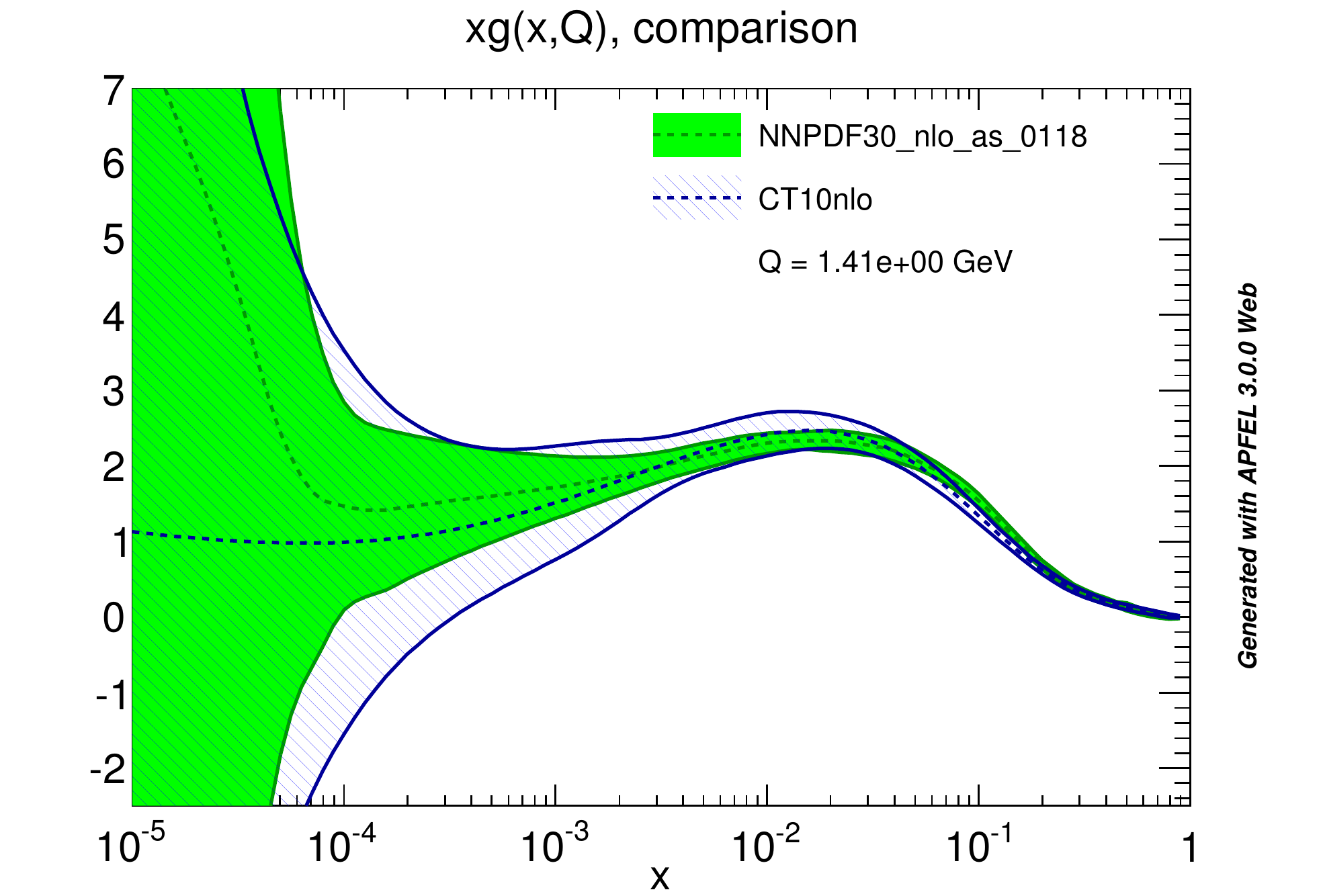}
\includegraphics[scale=0.39]{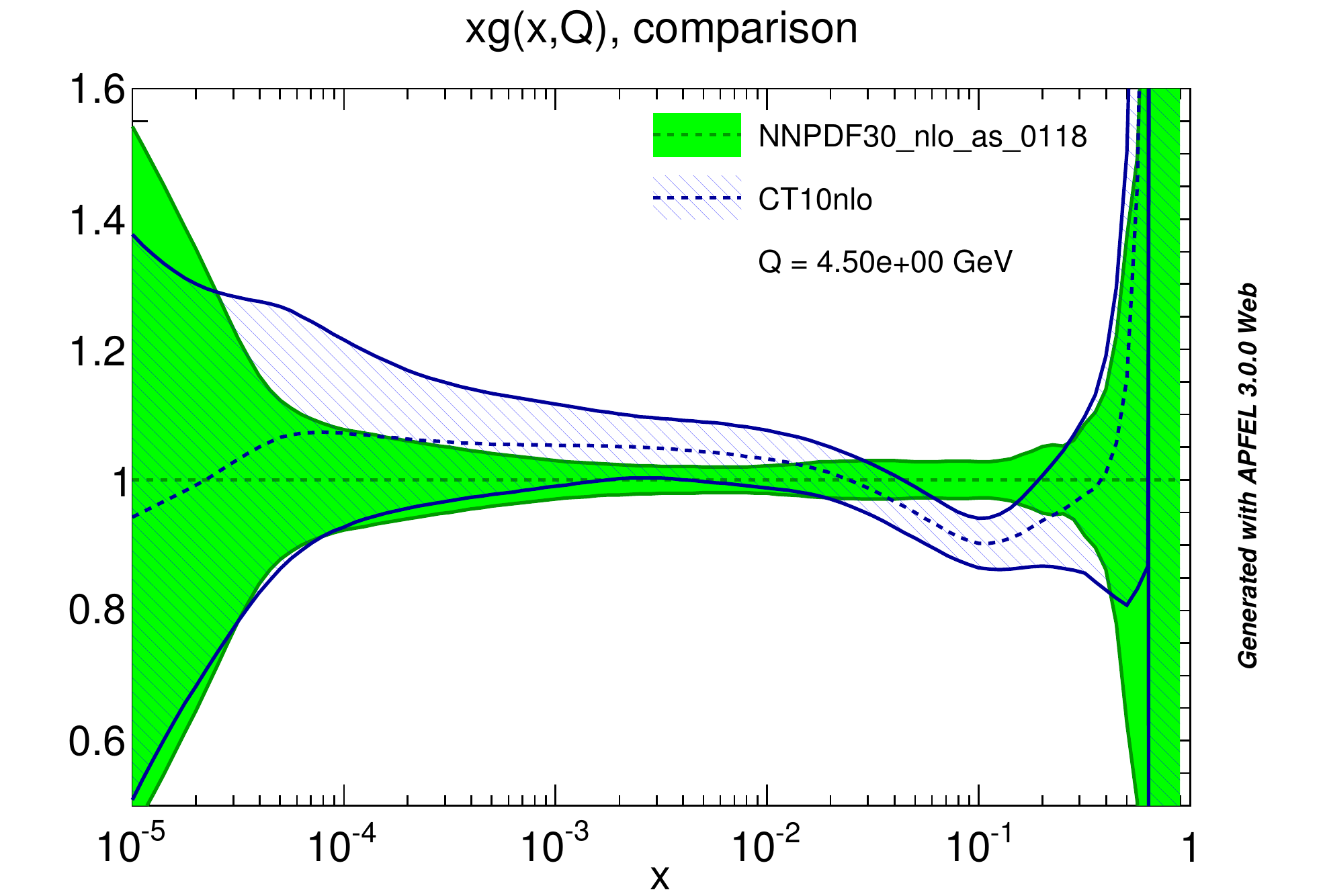}\\
\includegraphics[scale=0.39]{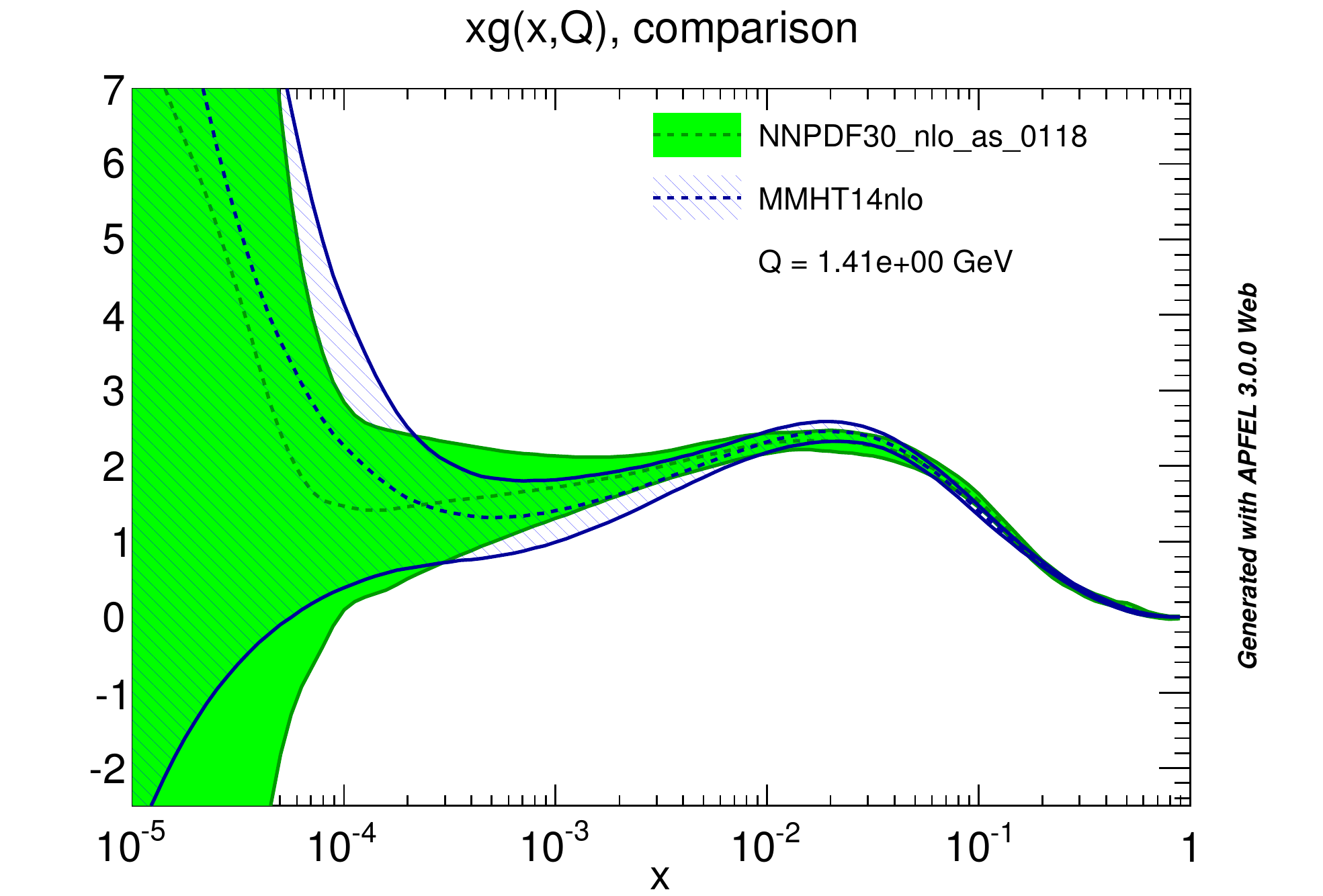}
\includegraphics[scale=0.39]{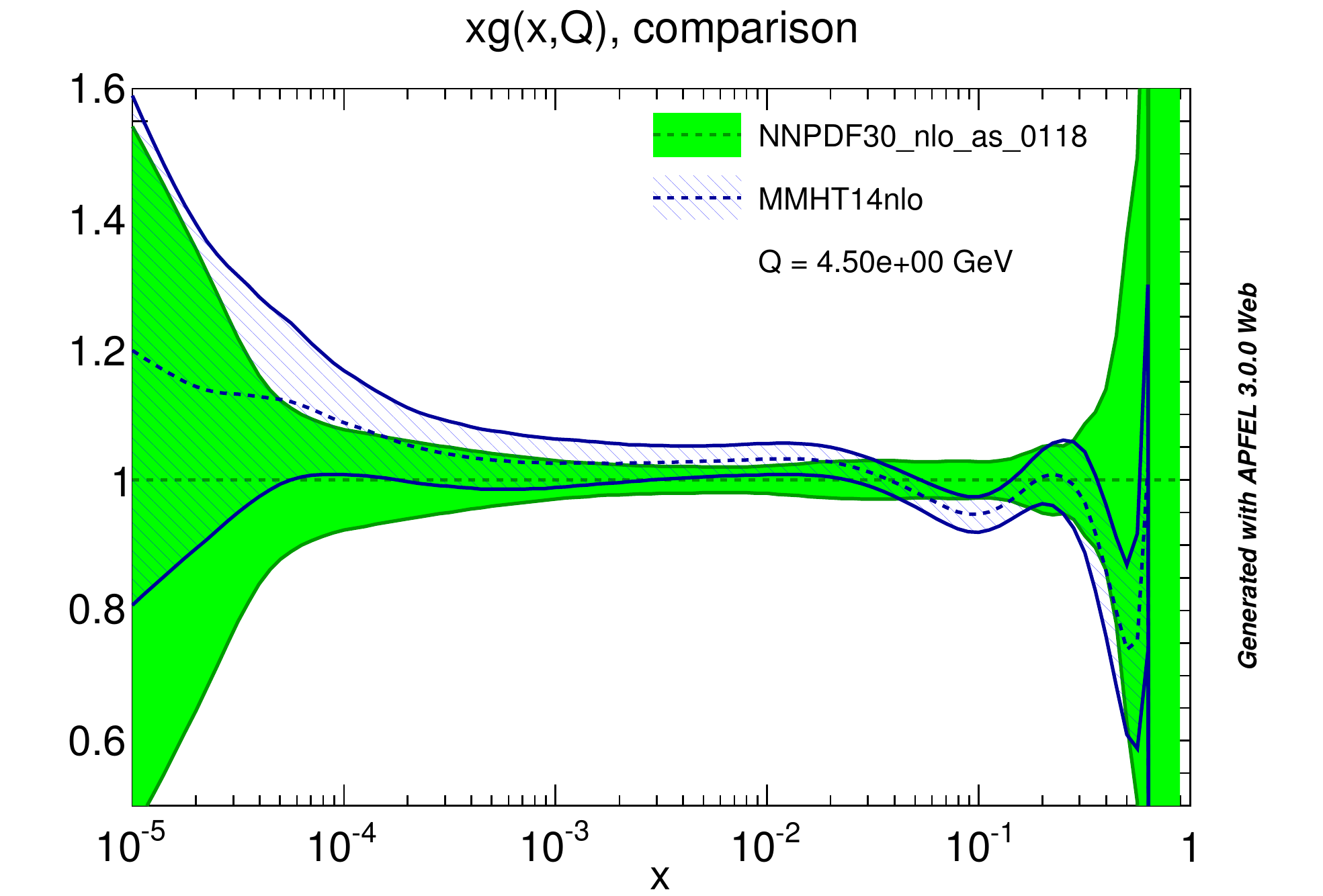}
\caption{\small Left plots: comparison of the small-$x$ gluon PDFs at
  $Q=1.4$ GeV 
  between NNPDF3.0 and (from top to bottom) CT10 and MMHT14.
  PDFs are compared in an absolute scale, and the bands
indicate
  the PDF uncertainties.
  Right plots: the same comparisons performed at $Q=4.5$ GeV,
  now shown as ratios
  with respect to the central NNPDF3.0 prediction.  }
\label{fig:pdfcomparison}
\end{figure}

  To better illustrate this point, and bearing in mind that heavy
  quark production at the LHC is driven by the $gg$ luminosity,
  it is useful to quantify the PDF uncertainties of the NNPDF3.0 gluon, 
  and compare this to other NLO PDF sets.
To ease these comparisons, we use the {\sc\small APFEL Web} on-line
PDF plotter~\cite{Bertone:2013vaa,Carrazza:2014gfa}.
In Fig.~\ref{fig:pdfcomparison} we show a comparison of the 
gluon PDFs evolved to the scale $Q=1.4$~GeV (corresponding to a typical value of the
charm mass) between the NNPDF3.0 and (from top to bottom) the
CT10~\cite{Gao:2013xoa} and
MMHT14~\cite{Harland-Lang:2014zoa} NLO PDF sets.
In each case, the bands correspond to the 68\% confidence level
for the PDF uncertainties.
The right plots of Fig.~\ref{fig:pdfcomparison} show the
same comparisons now performed at the scale $Q=4.5$ GeV, a value typical of
the bottom quark mass, shown as ratios with respect to the
central NNPDF3.0 prediction.

As can be seen, in the region relevant for charm production at LHCb,
with $x_2 \lsim \la x_2\ra\simeq 4.6\cdot10^{-5}$,
the gluon PDF uncertainties are
extremely large.
On the other hand, for the region relevant for bottom
production, with  $x_2 \lsim \la x_2\ra\simeq 1.3\cdot10^{-4}$, PDF
uncertainties are moderate, thanks to the constraints
from HERA data.
Importantly, as shown in Fig.~\ref{fig:pdfcomparison}, the description
of the gluon PDF at small-$x$ is quite similar, both in terms
of the central value and associated uncertainty --- particularly for the 
comparison between NNPDF3.0 and MMHT sets.
As will be shown explicitly, this agreement implies that
predictions for charm and bottom production at
LHCb obtained with NNPDF3.0 will be similar to those obtained
with CT10 or MMHT14 as input PDF sets.

\begin{figure}[t]
\centering 
\includegraphics[scale=0.39]{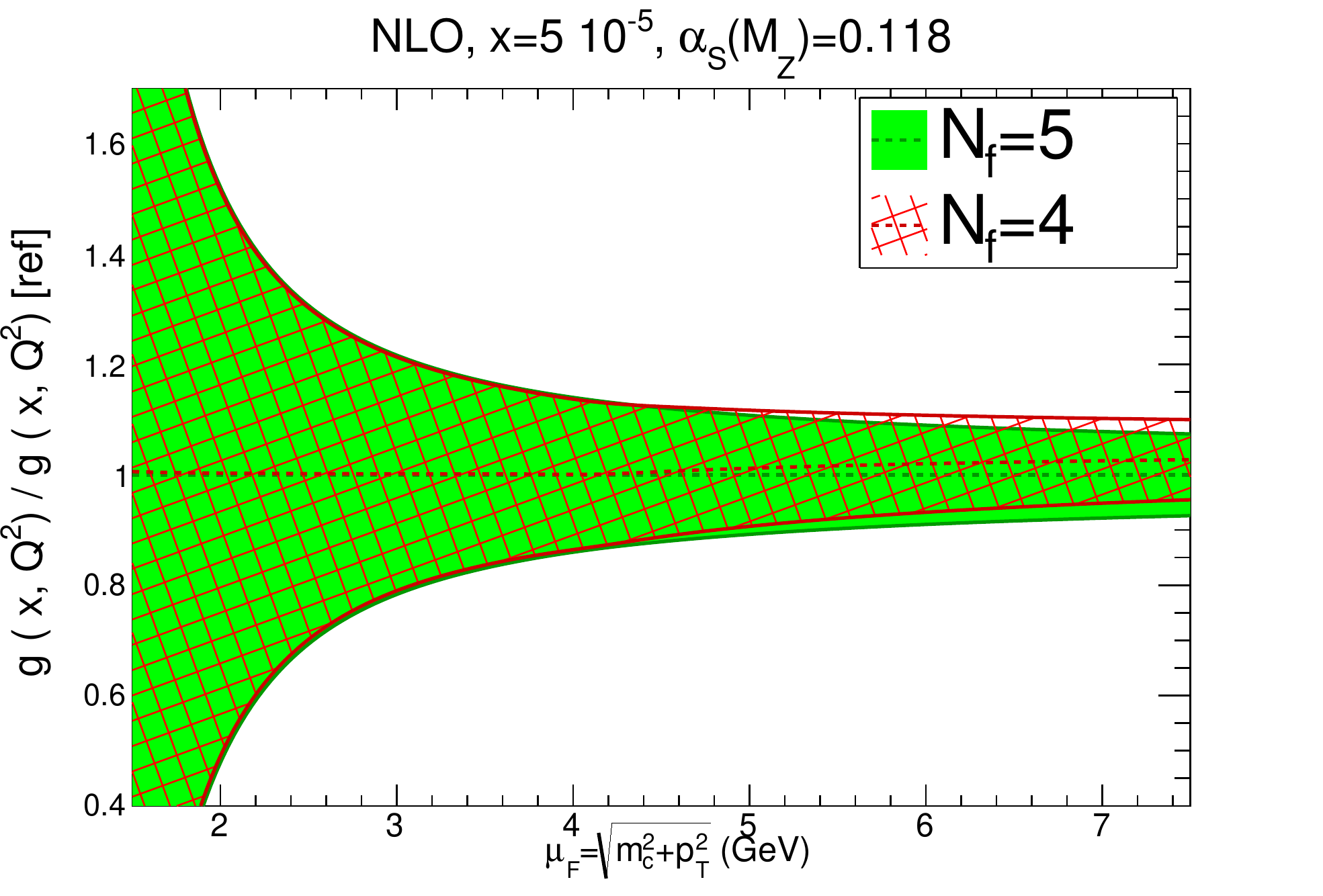}
\includegraphics[scale=0.39]{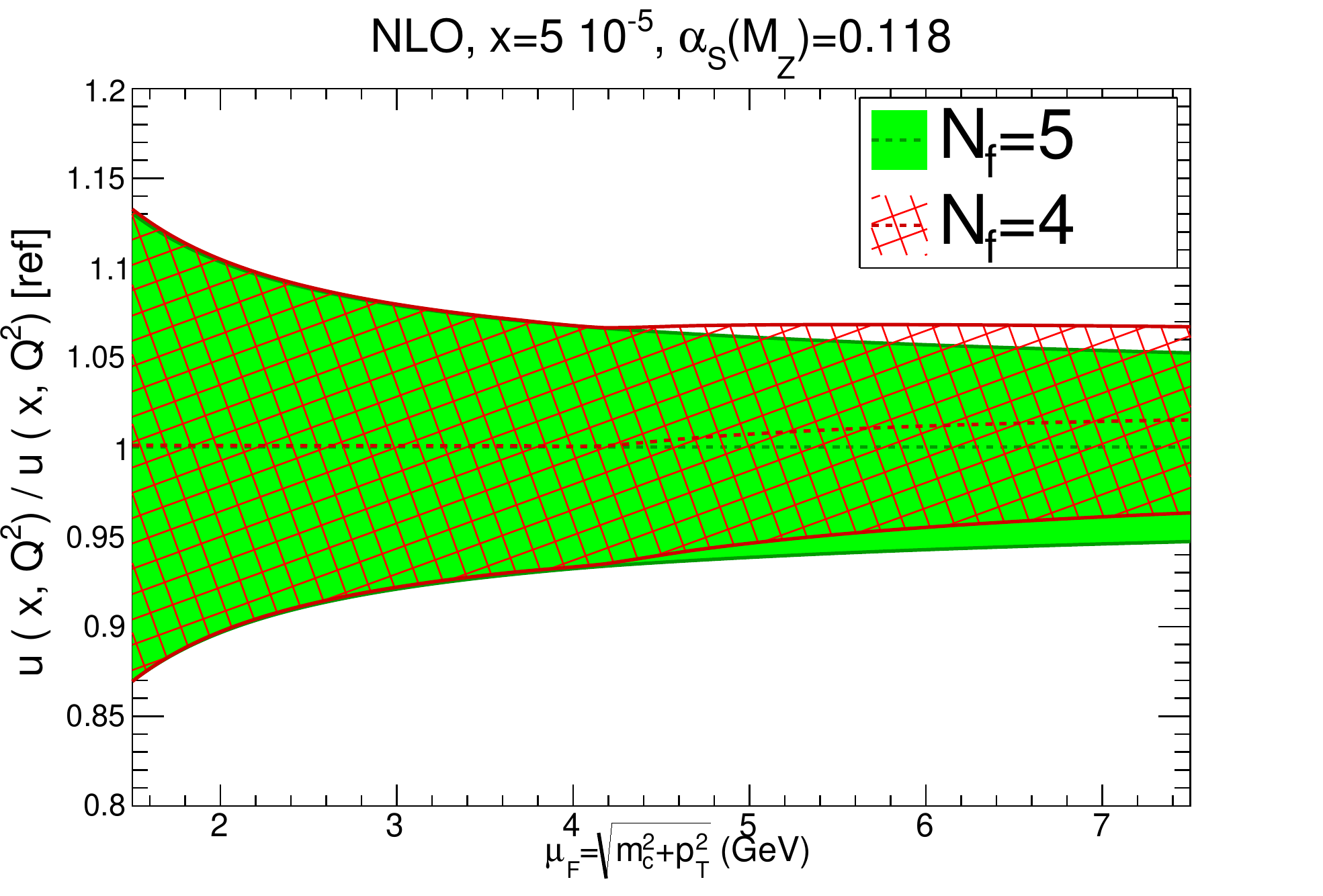}
\caption{\small Comparison between the $n_f=4$ and
  $n_f=5$ PDFs from the  NNPDF3.0 NLO set,
  as a function of $Q$ for $x=5\cdot 10^{-5}$,
  in the region relevant for forward charm production at LHCb.
  We show the gluon (left plot) and the up quark (right plot), normalized
  to the central value of the $n_f=5$ set.
}
\label{fig:pdfvfn}
\end{figure}

In Fig.~\ref{fig:pdfvfn} we show the comparison between the $n_f=4$
and $n_f=5$  gluon and up quark NNPDF3.0 NLO PDFs as a function of $Q$,
for a reference value $x=2\cdot 10^{-5}$, in the kinematical region relevant
for charm production at LHCb.
We see that the differences between the $n_f=4$ and $n_f=5$ schemes
are much smaller
than the associated PDF uncertainties.
We have also explicitly verified that either using
$n_f=4$ PDFs in the {\sc\small FONLL} calculations
or including the $n_f \to n_f+2$ scheme transformation terms
in {\sc\small POWHEG} 
leads to negligible modifications of our results.
These considerations justify our
choice of the NNPDF3.0 NLO $n_f=5$ set as baseline in our calculations.

The fact that gluon PDF uncertainties in the region
relevant for charm production at LHCb are large indicates that these 
measurements can be used to provide information on the poorly known
small-$x$ gluon.
This constraining potential has been recently verified by the
{\sc\small PROSA} analysis~\cite{Zenaiev:2015rfa}
based on the {\sc\small HERAfitter}
framework~\cite{Alekhin:2014irh}.
In Sect.~\ref{sec:rw} we will study the impact of the LHCb charm
data in the NNPDF3.0 NLO global analysis using the Bayesian
reweighting method.

\subsection{Comparison with the LHCb data}
\label{sec:lhcb}

We now perform a detailed comparison of the pQCD
calculations of charm and bottom production in the forward region
with the most recent LHCb data~\cite{Aaij:2013noa,Aaij:2013mga}.
The comparisons will be performed at the level of double
differential distributions,
\be
\frac{d^2\sigma(D)(y,p_T)}{dy^Ddp^D_T} \quad {\rm and}\quad \frac{d^2\sigma(B)(y,p_T)}{dy^Bdp^B_T} \, .
\ee
For all mesons, we have also
checked that good agreement is obtained for the total
cross-sections in the fiducial region.

For $D$ mesons, we restrict the comparison to the case of the
higher-statistics final states, namely $D^0$ and $D^{\pm}$, while for
the beauty mesons we will show results only for $B^0$ production.
For each calculation, we provide the central prediction
as well as the contribution arising from the various sources 
of theoretical uncertainty as outlined in Sect.~\ref{sec:thsettings}.

\begin{figure}[t]
\centering 
\includegraphics[scale=0.39]{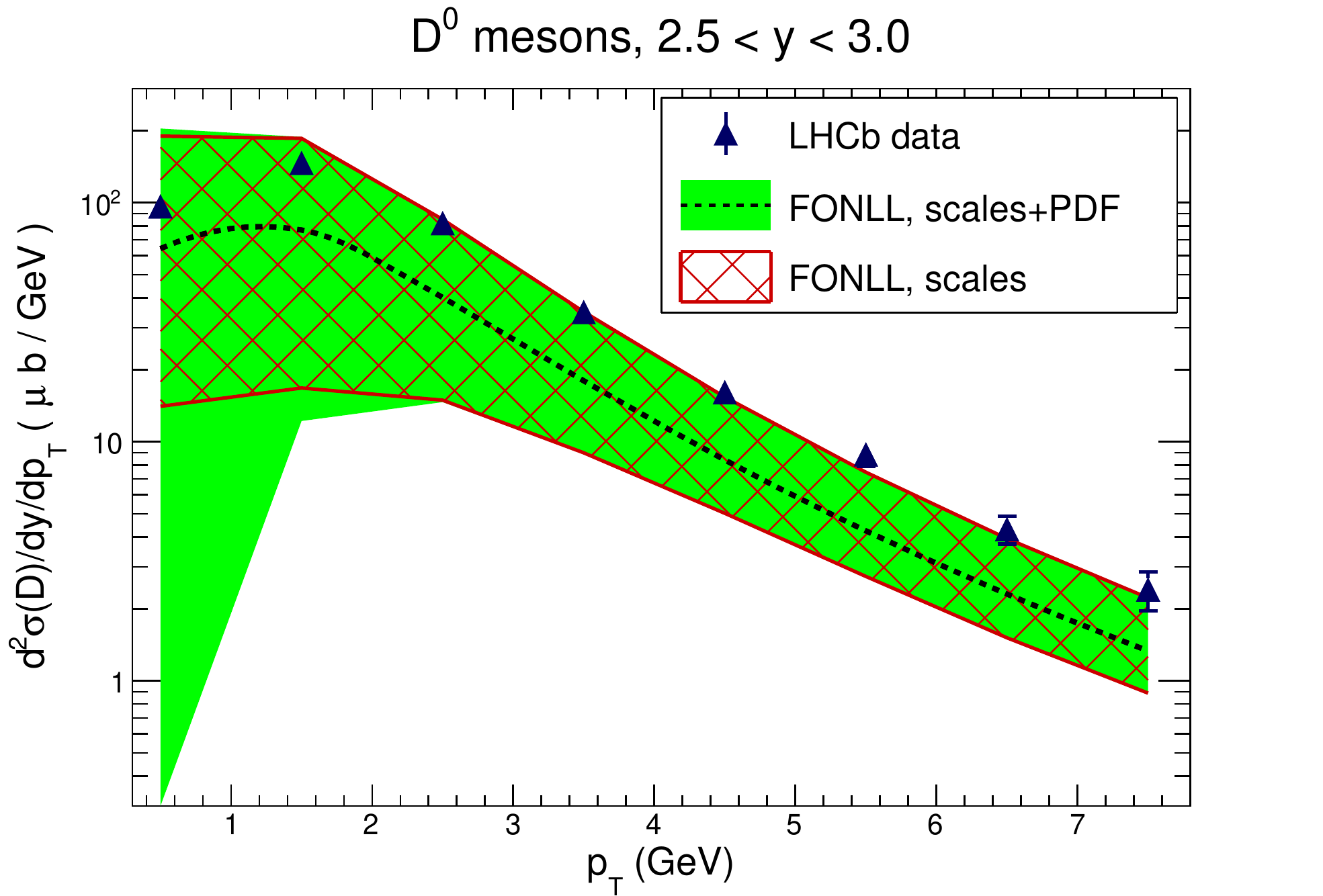}
\includegraphics[scale=0.39]{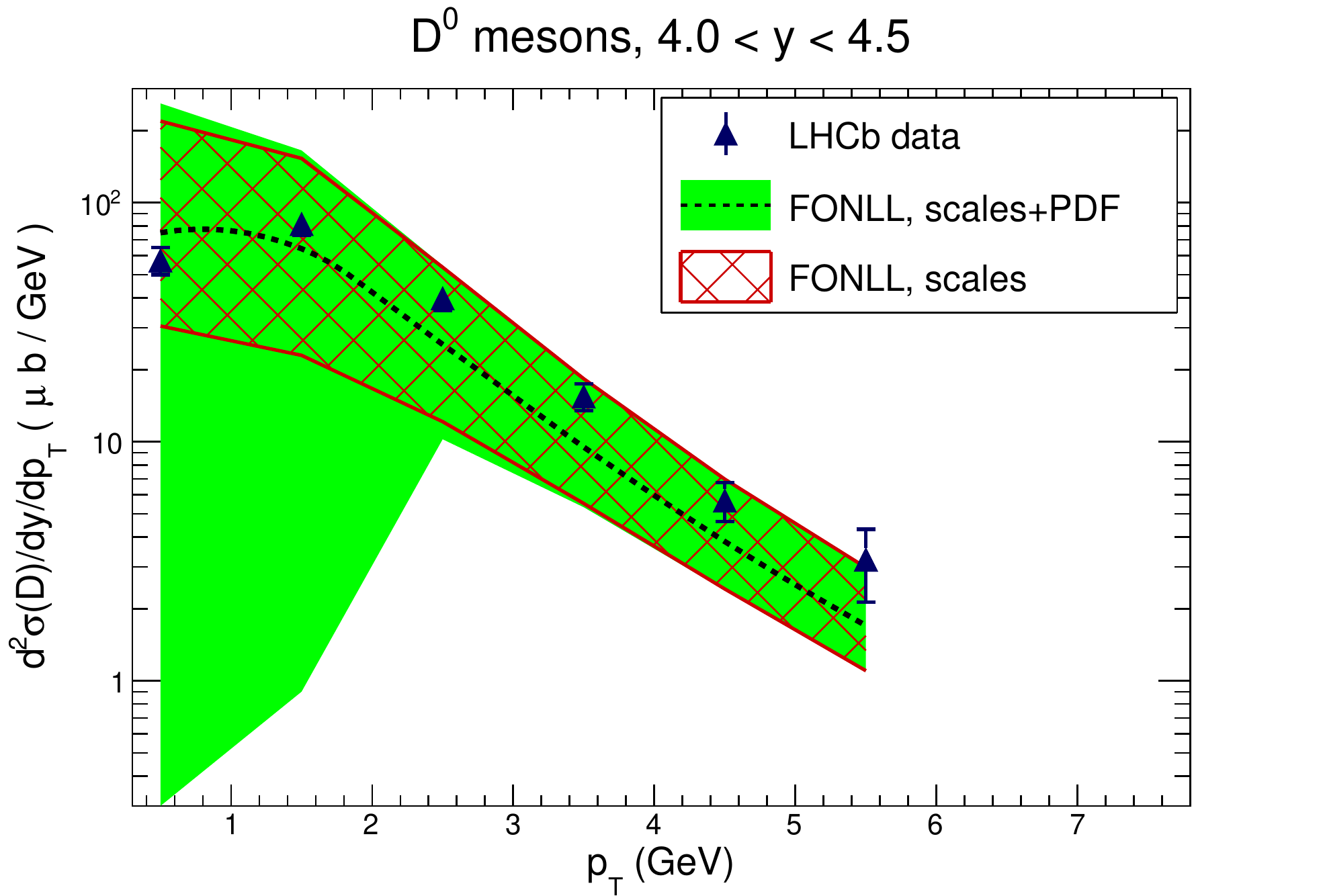}
\includegraphics[scale=0.39]{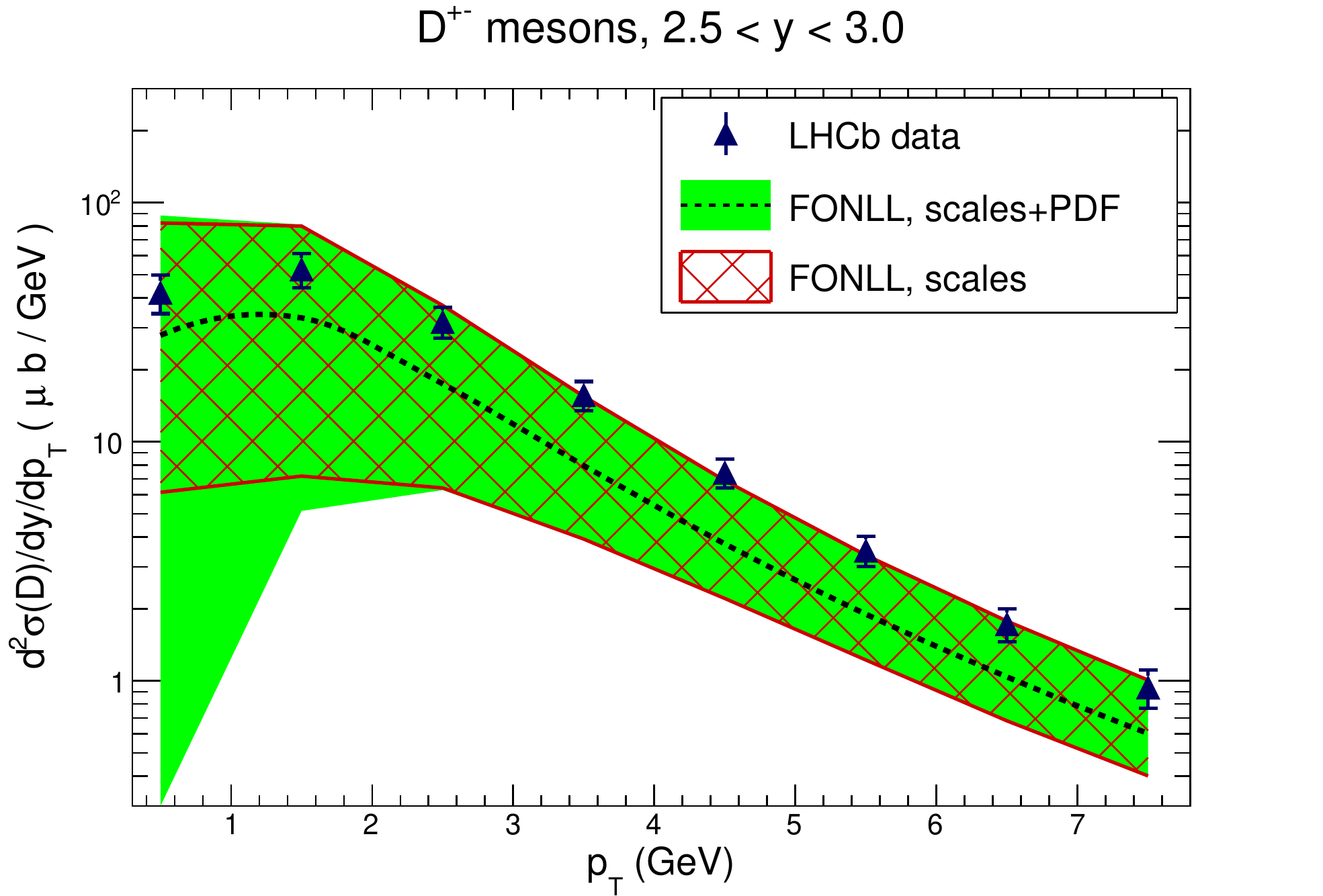}
\includegraphics[scale=0.39]{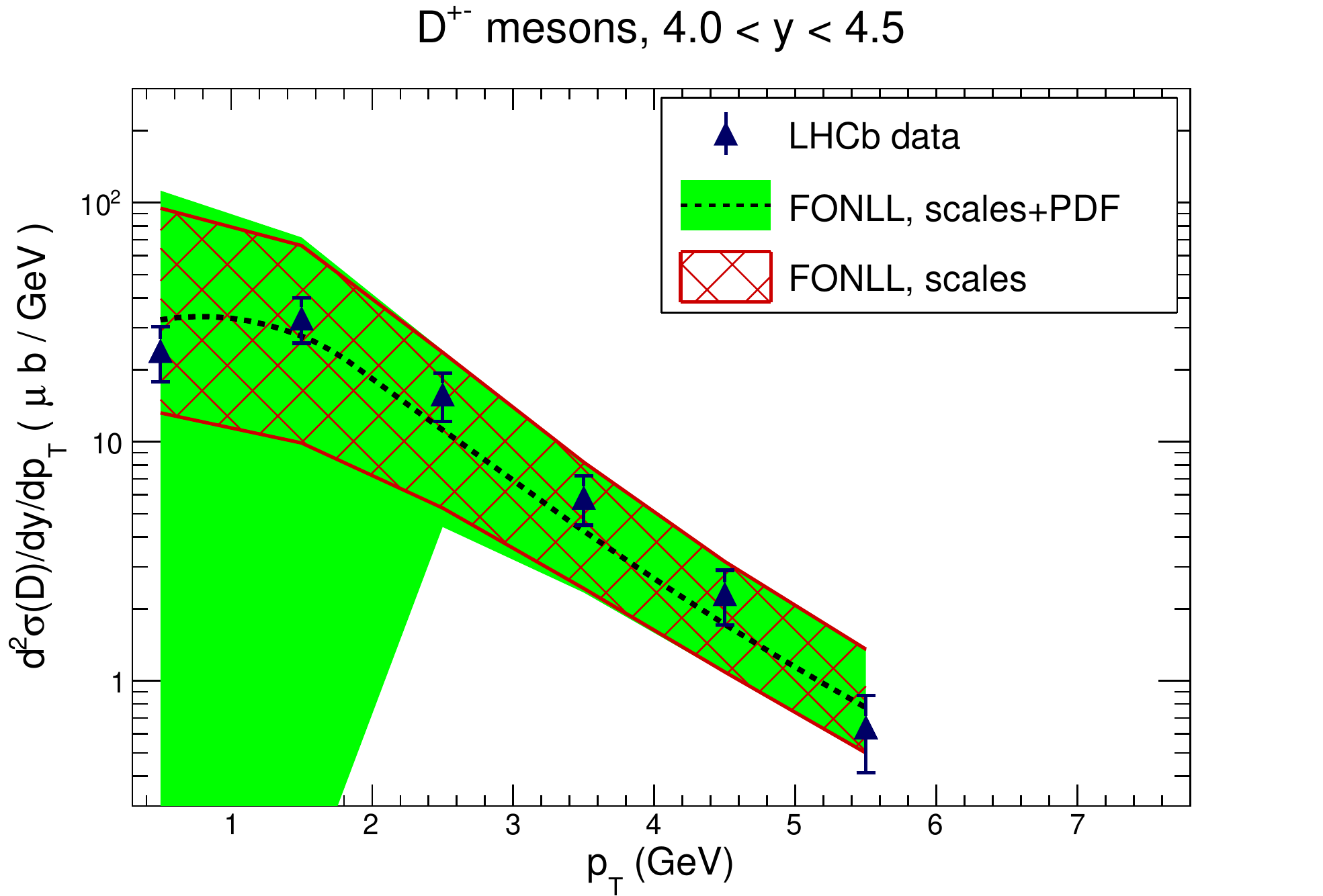}
\caption{\small Comparison between the LHCb data on $D$ meson
  production and the FONLL calculation using NNPDF3.0 as input.
  We show the results for the most central bin, $2.0 \le y \le 2.5$
  (left column) and a forward bin, $3.5 \le y \le 4.0$ (right
  column), both for $D^0$ data (upper row) and the $D^{\pm}$ data
  (lower row).
  The solid error band is obtained from the sum in quadrature of
  PDF and scale uncertainties, while the hatched band is only
  the scale variation component.
}
\label{fig:LHCbcharm}
\end{figure}

The comparison between the {\sc\small FONLL} calculation and the LHCb charm
production data is shown in Fig.~\ref{fig:LHCbcharm}.
We show the results for the most central bin, $2.0 \le y \le 2.5$  and
a  forward bin, $3.5 \le y \le 4.0$, both for the $D^0$
and the $D^{\pm}$  measurements.
 In Fig.~\ref{fig:LHCbcharm}, statistical and systematic uncertainties have
 been added in quadrature for the experimental data, while for the theory uncertainties
 we show both the scale uncertainty alone and also the sum in quadrature of scale and
 PDF uncertainties.

The agreement, within uncertainties, between the LHCb data and 
the NLO pQCD prediction across the entire kinematic range demonstrates the 
applicability of this approach to forward charm production.
The total theoretical uncertainty is dominated by scale variation, 
except in the low $p_T$ where the large
gluon PDF uncertainty at small-$x$
becomes comparable to the scale variation or even dominant.
Similar satisfactory agreement is found for the other data bins not
shown in Fig.~\ref{fig:LHCbcharm}.
%

\begin{figure}[t]
  \centering
  \includegraphics[scale=0.38]{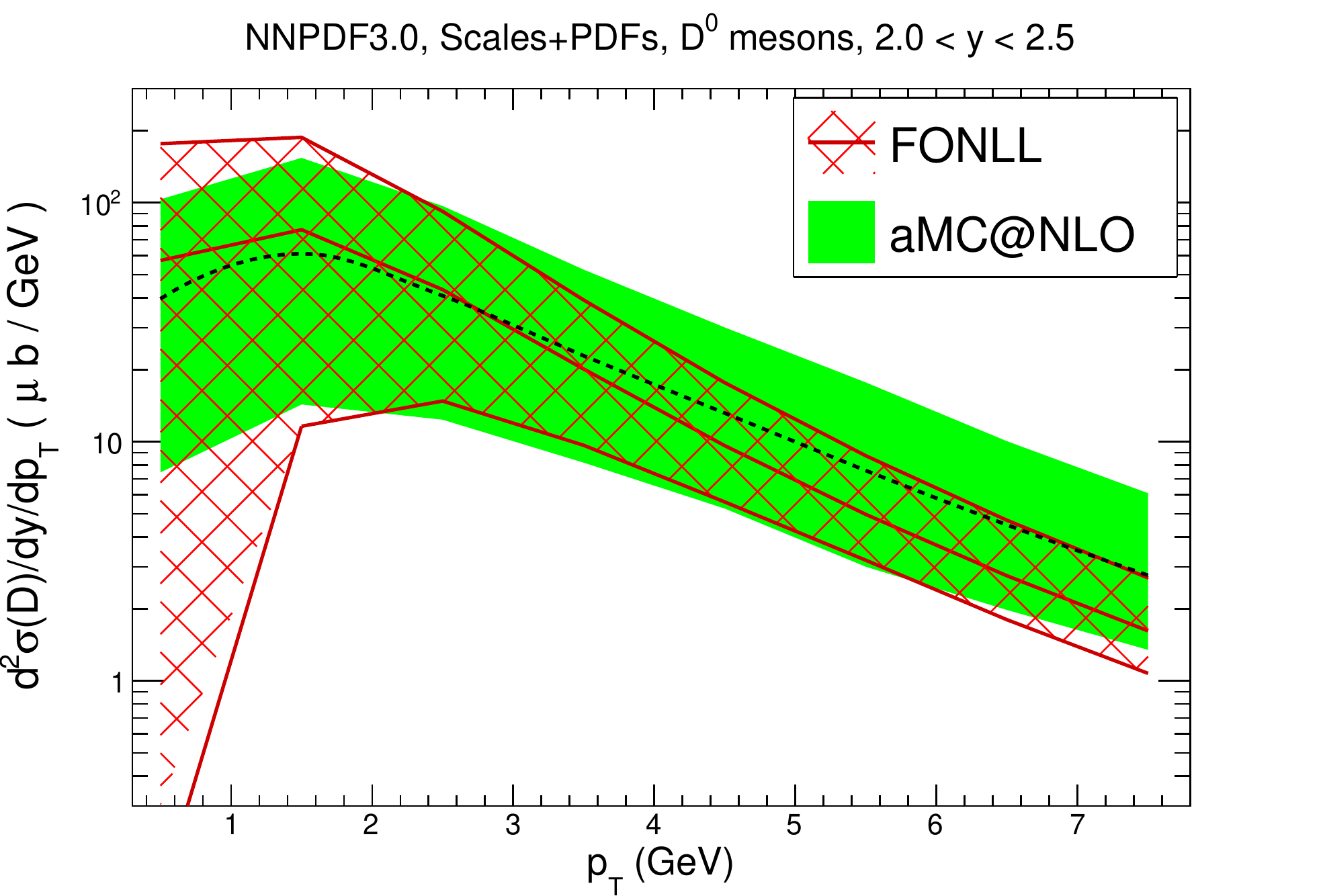}
\includegraphics[scale=0.38]{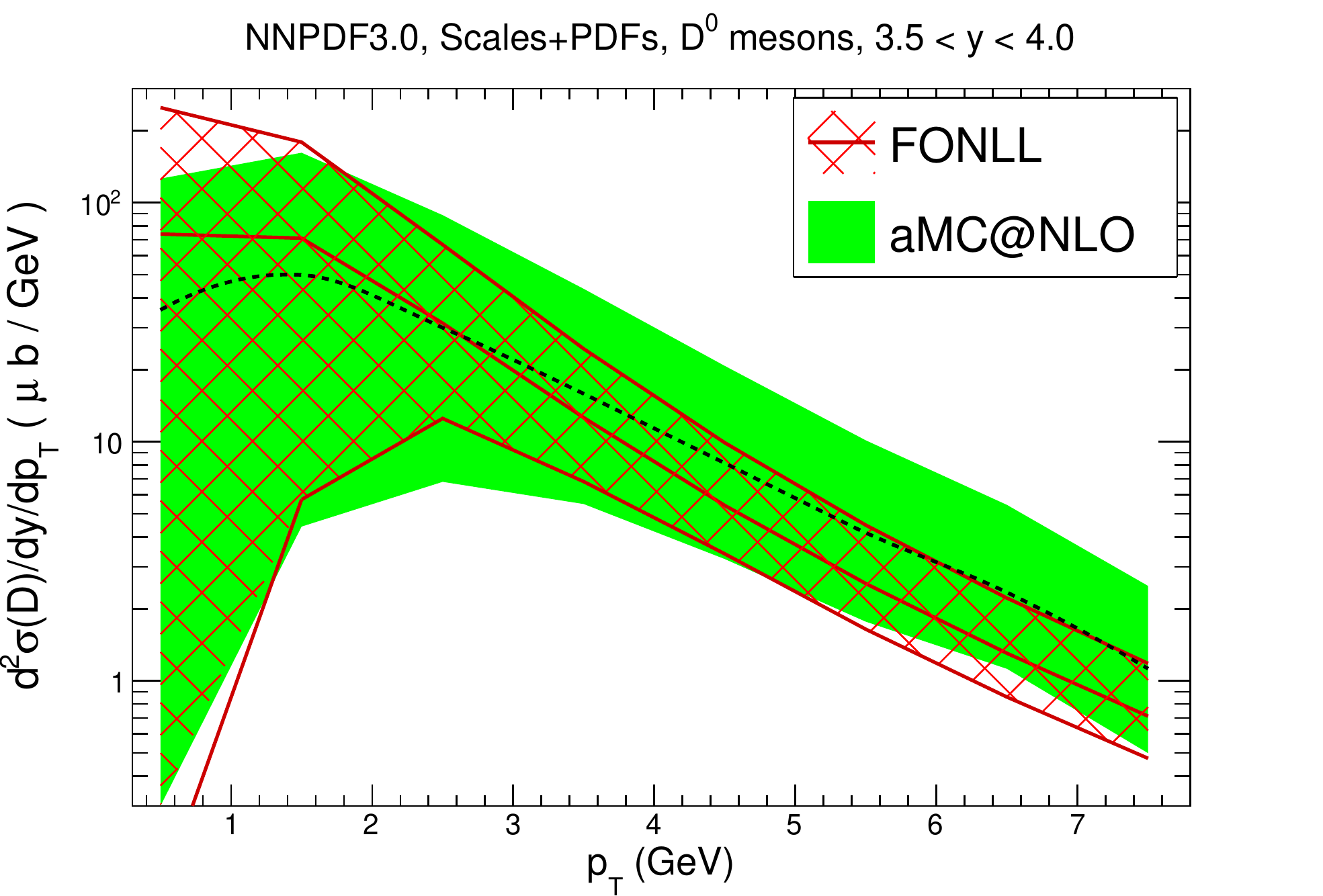}
  \includegraphics[scale=0.38]{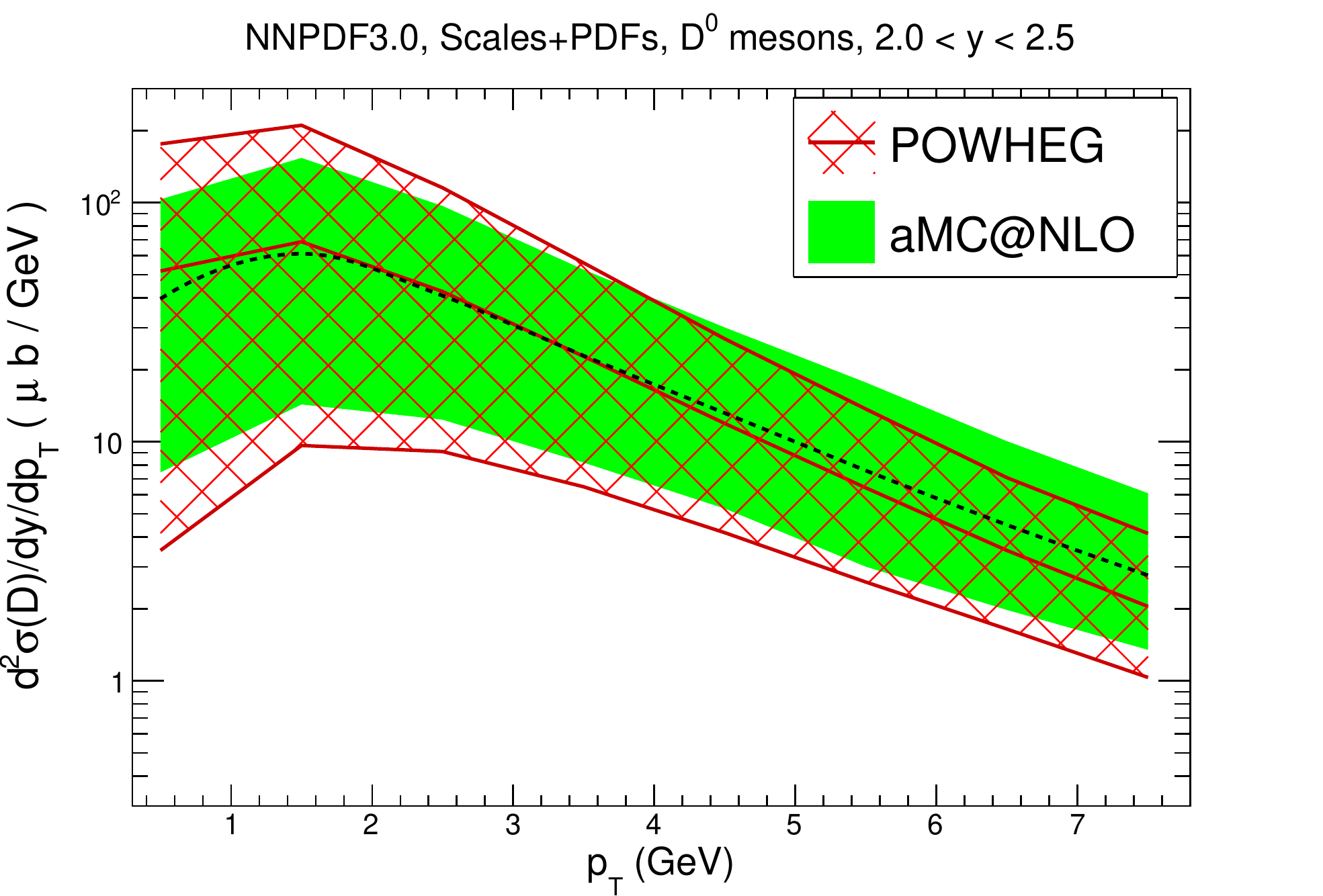}
\includegraphics[scale=0.38]{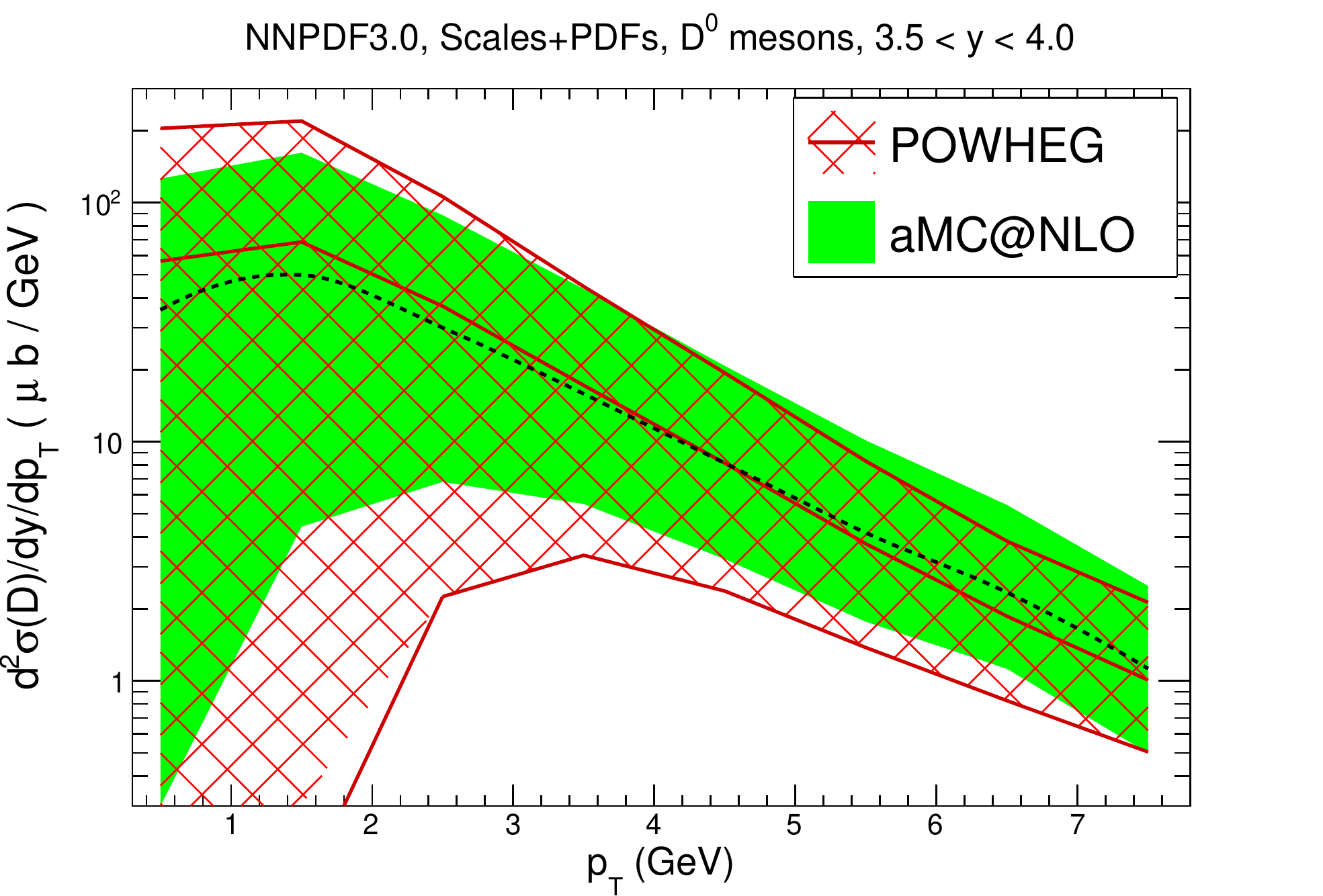}
\caption{\small Comparison between the
  {\sc\small FONLL} and {\sc\small {\rm a}MC@NLO} (upper plots)
  and between the {\sc\small POWHEG} and  {\sc\small {\rm a}MC@NLO}
   (lower plots) calculations for $D^0$ production in
  the same kinematics as the LHCb data of Fig.~\ref{fig:LHCbcharm}, using
  NNPDF3.0 NLO.
 The total
  theory uncertainly band is obtained by the addition in quadrature of
  scale and PDF uncertainties.
}
\label{fig:LHCbcharm2}
\end{figure}

Given the compatibility of the charm production data and theory prediction provided
by {\sc\small FONLL}, we now compare these predictions
to those obtained with the NLO Monte Carlo approaches,
{\sc\small
  {\rm a}MC@NLO} and {\sc\small POWHEG}.
First of all we compare the {\sc\small FONLL} results with the {\sc\small {\rm a}MC@NLO} calculation.
For simplicity, we only provide results for $D^0$ mesons.
The comparison is shown in Fig.~\ref{fig:LHCbcharm2}: clearly,
there is good agreement between the central values of the two
calculations.
For the total theory uncertainty band there is also reasonable
agreement, with the {\sc\small {\rm a}MC@NLO} band being typically
larger than, but still consistent, with the {\sc\small FONLL} result.
In this comparison the theory uncertainty band is obtained from adding scale
and PDF uncertainties in quadrature.
The corresponding comparison between the two
MC generators, {\sc\small
  {\rm a}MC@NLO} and {\sc\small POWHEG}, is shown in the lower
plots of  Fig.~\ref{fig:LHCbcharm2}.
Reasonable agreement is also found between the central predictions, well within the
uncertainty bands.
We note that scale uncertainties tend to be slightly larger
in the {\sc\small POWHEG} calculation.\footnote{This has
been traced back to a different solution of the RG equations for the running of $\alpha_s(Q)$
used in the {\sc\small POWHEG} calculation, leading to formally subleading corrections which are numerically
important at $Q \simeq m_c$.
As opposed to {\sc\small
    {\rm a}MC@NLO} and {\sc\small FONLL}, where $\alpha_s(Q)$ is consistently
  extracted from the PDF set that is being used via the {\sc\small LHAPDF6}~\cite{Buckley:2014ana}
  interface,
  {\sc\small POWHEG} uses its own internal routine for the running of $\alpha_s(Q)$.
  We thank Emanuele Re for clarifications about this point.
}

\begin{figure}[t]
\centering 
\includegraphics[scale=0.39]{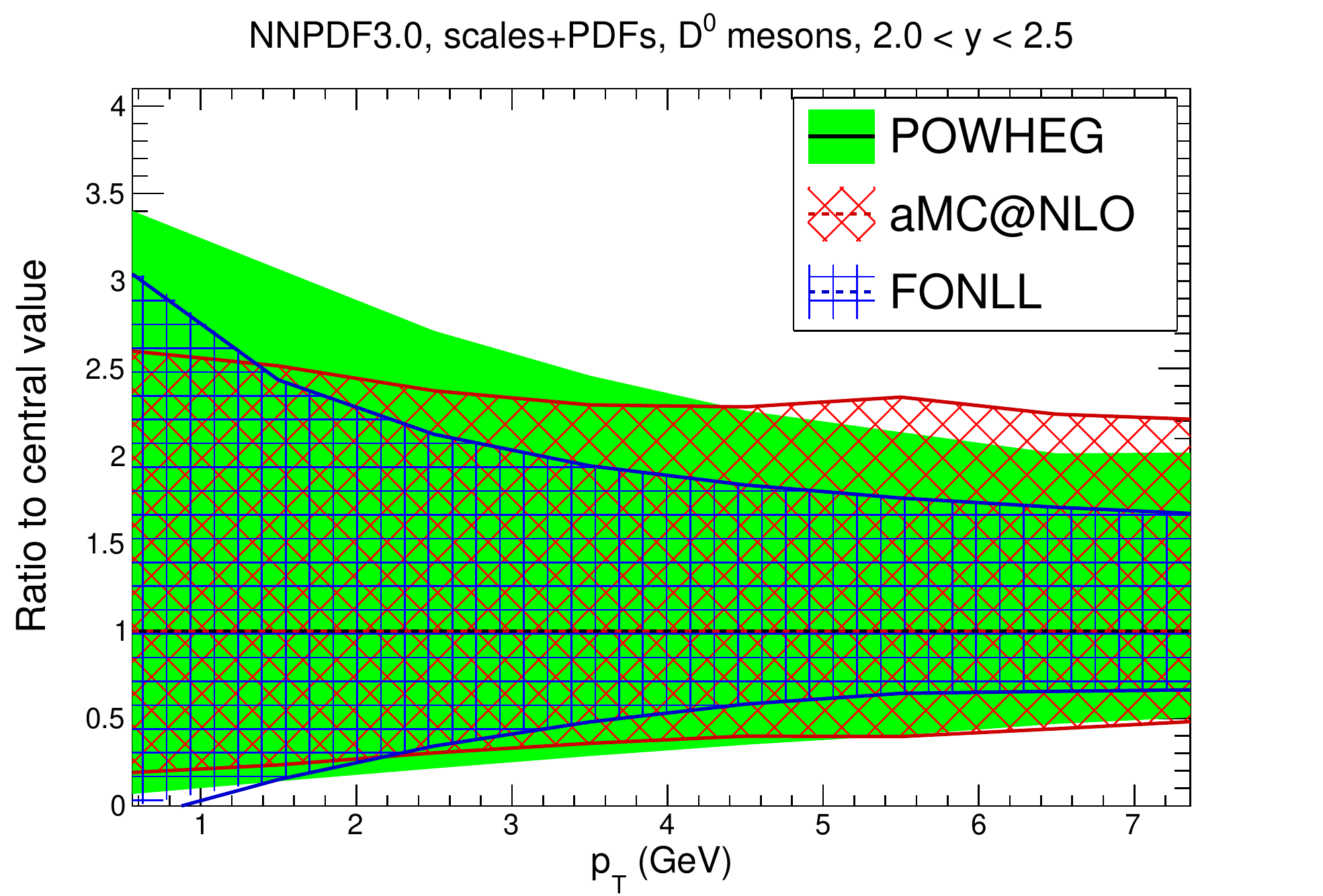}
\includegraphics[scale=0.39]{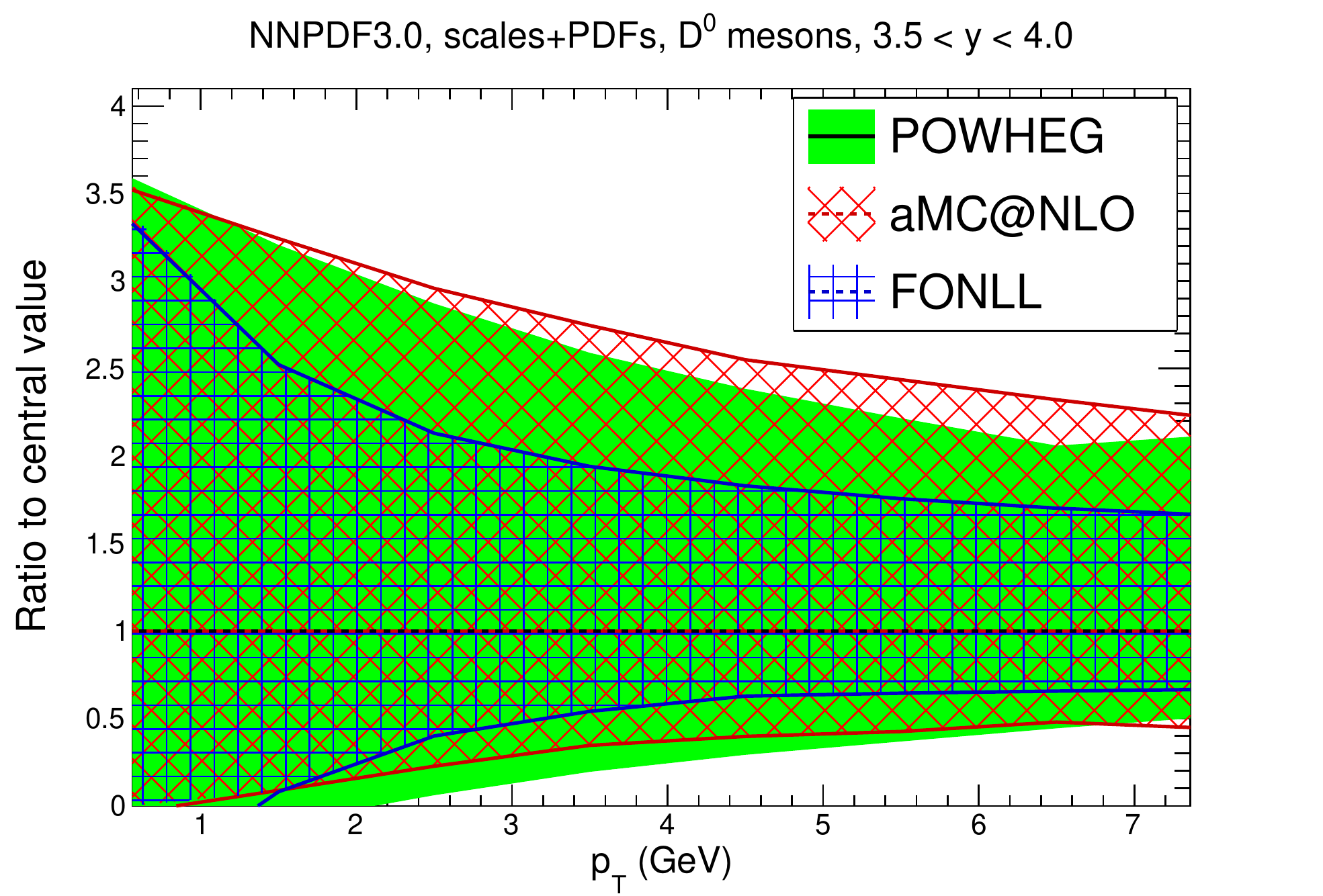}
\caption{\small Comparison between the
  total theoretical uncertainty (sum in quadrature of scale and
  PDF uncertainties) for the kinematics of $D^0$ production
  at LHCb.
  The results for the three calculations, 
  {\sc\small {\rm a}MC@NLO},
  {\sc\small POWHEG}, and {\sc\small FONLL} calculations,
  are normalised to the respective central values.
  }
\label{fig:LHCbcharm_thrat}
\end{figure}

In Fig.~\ref{fig:LHCbcharm_thrat} we perform the same comparison
between the three calculations 
as shown in Fig.~\ref{fig:LHCbcharm2}, but now normalising
each prediction to the corresponding central value.
This way we can gauge how the total theory uncertainty band
compares among the three calculations.
The total uncertainty is similar for {\sc\small POWHEG} and
 {\sc\small {\rm a}MC@NLO} calculations. Notably, the scale uncertainties of the
{\sc\small POWHEG} and  {\sc\small  {\rm a}MC@NLO} calculations tend to be larger
 than those of {\sc\small FONLL}, especially in the upper
 variations in the moderate and high $p_T$ region.
 While the origin of these differences remains to be understood,
 it might be related to the fact that {\sc\small FONLL} is a fixed-order calculation
 while {\sc\small POWHEG} and {\sc\small {\rm a}MC@NLO} are matched to 
 parton showers, and this matching may induce additional theoretical uncertainties.
 Indeed, we have verified that the scale uncertainties of the fixed-order NLO 
 computation of differential $c\bar{c}$ production (without fragmentation) 
 in {\sc\small {\rm a}MC@NLO} reproduces those of {\sc\small FONLL} to a few percent.

 From Fig.~\ref{fig:LHCbcharm_thrat} we see that the
 {\sc\small FONLL} semi-analytical calculation exhibits smaller
 theoretical uncertainty, and
 for this reason, in the following Section we will use the FONLL
 predictions to quantify the
 constraints of the LHCb charm production data on the NNPDF3.0 small-$x$ gluon PDF.

We now begin the comparison between the LHCb data
and the various theoretical calculations for the case of
$B$ meson production.
For simplicity, we show results only for $B^0$ mesons, though similar agreement
has been found for the other $B$ mesons.
As compared to the case of the $D$ mesons,
we expect a reduction of the theory
uncertainties for several reasons: the
calculation is performed at a higher scale $\sqrt{m_b^2+p_{T,b}^2}$, as compared to the
charm production case, $\sqrt{m_c^2+p_{T,c}^2}$, leading to an
improved convergence of the perturbative expansion; the relative uncertainty of the value of $m_b$ 
is smaller; and larger values of $x_{1,2}$ are probed within the proton,
a region well covered by HERA data as illustrated in Fig.~\ref{fig:kincovLO} and Fig.~\ref{fig:pdfcomparison}.

\begin{figure}[t]
\centering 
\includegraphics[scale=0.39]{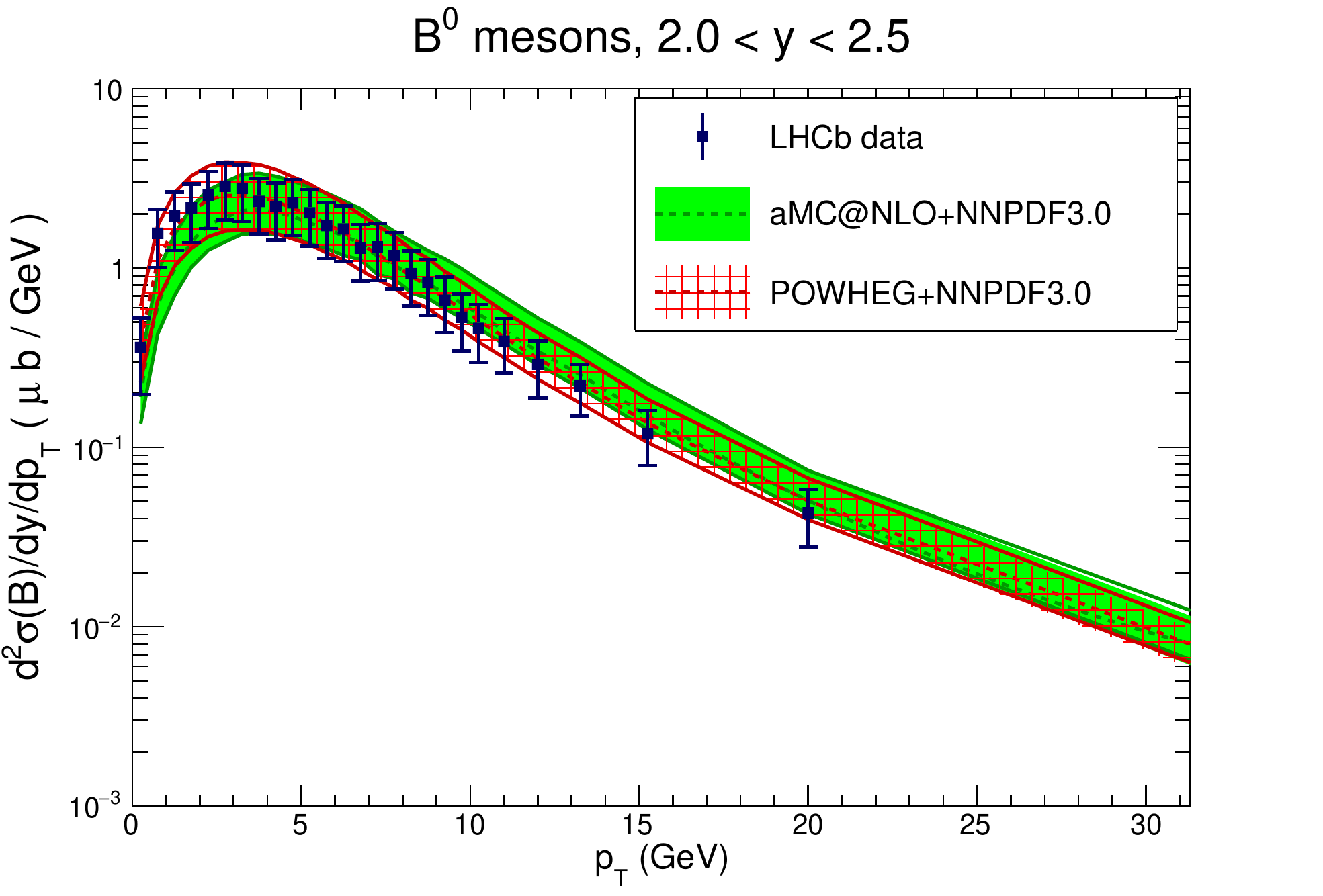}
\includegraphics[scale=0.39]{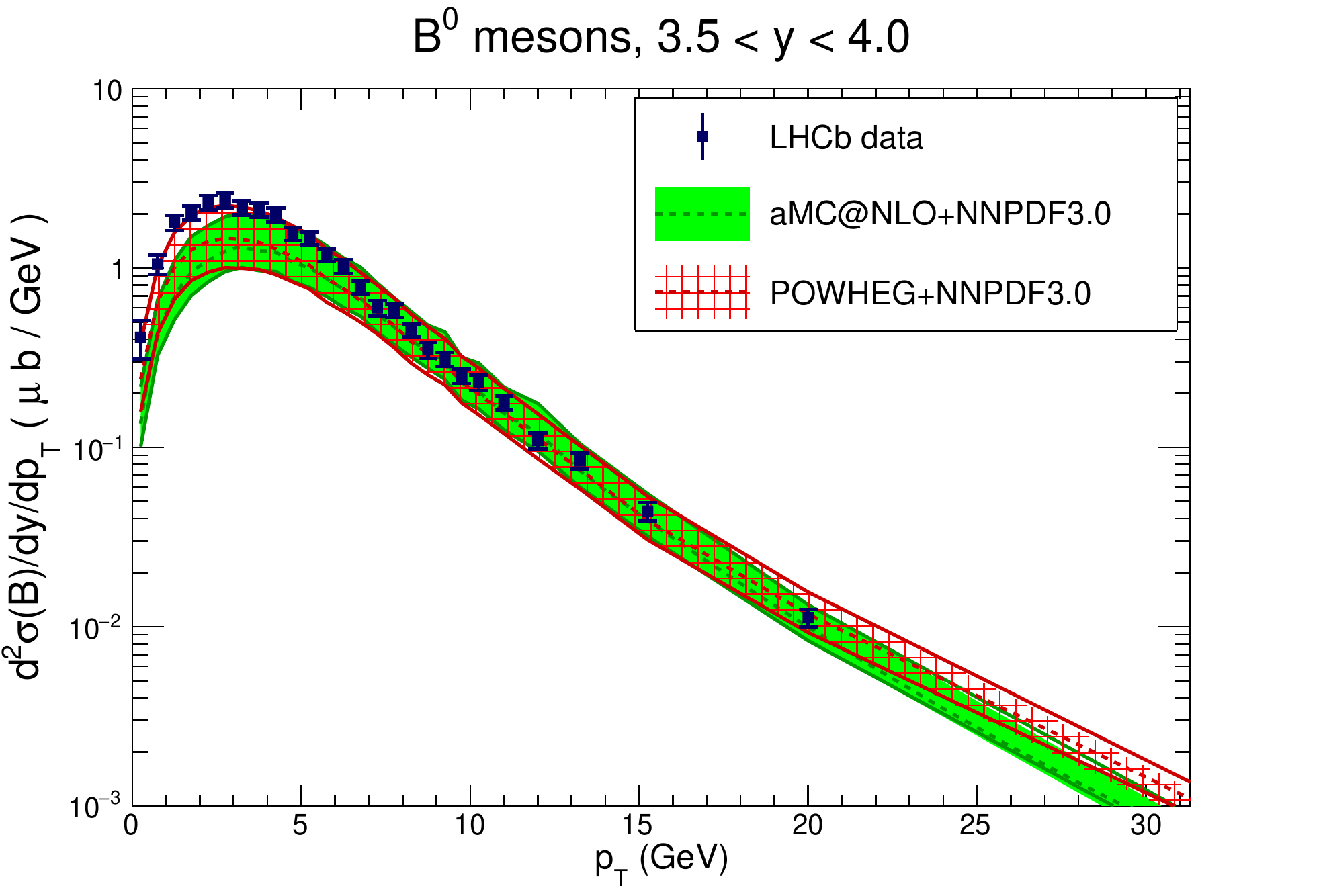}
\caption{\small Comparison of the LHCb data on $B^0$ meson production,
  both for central and for forward rapidities, with the theoretical
  predictions from {\sc\small POWHEG} and {\sc\small {\rm a}MC@NLO}.
  The theory uncertainty includes only scale uncertainties.
}
\label{fig:LHCbBottom}
\end{figure}

In Fig.~\ref{fig:LHCbBottom} we show the comparison of the LHCb data
for $B^0$ meson production, both for central and for forward
rapidities, with the corresponding {\sc\small POWHEG}
and {\sc\small {\rm a}MC@NLO} calculations.
The indicated theory uncertainty band includes only the scale uncertainties,
and we have verified that PDF uncertainties are not so relevant in this case.
As in the case of charm, satisfactory agreement between theory and data for $B$ meson production
in the forward region is found. There is also a substantial reduction of the theory uncertainty as 
compared to the $D$ meson case.
The {\sc\small POWHEG} and {\sc\small {\rm a}MC@NLO} predictions are in reasonable
agreement within the theory uncertainty band.

\begin{figure}[t]
\centering 
\includegraphics[scale=0.39]{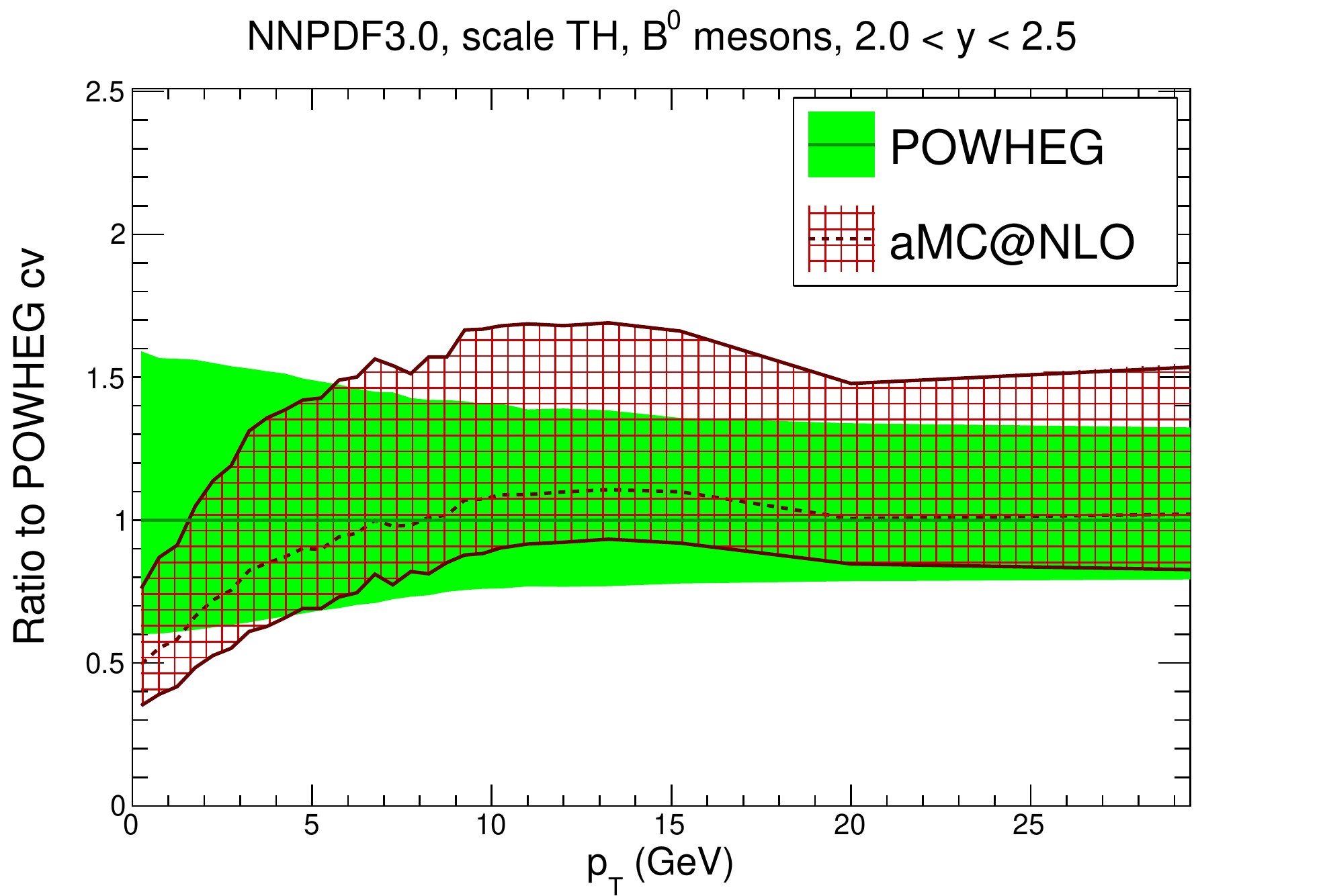}
\includegraphics[scale=0.39]{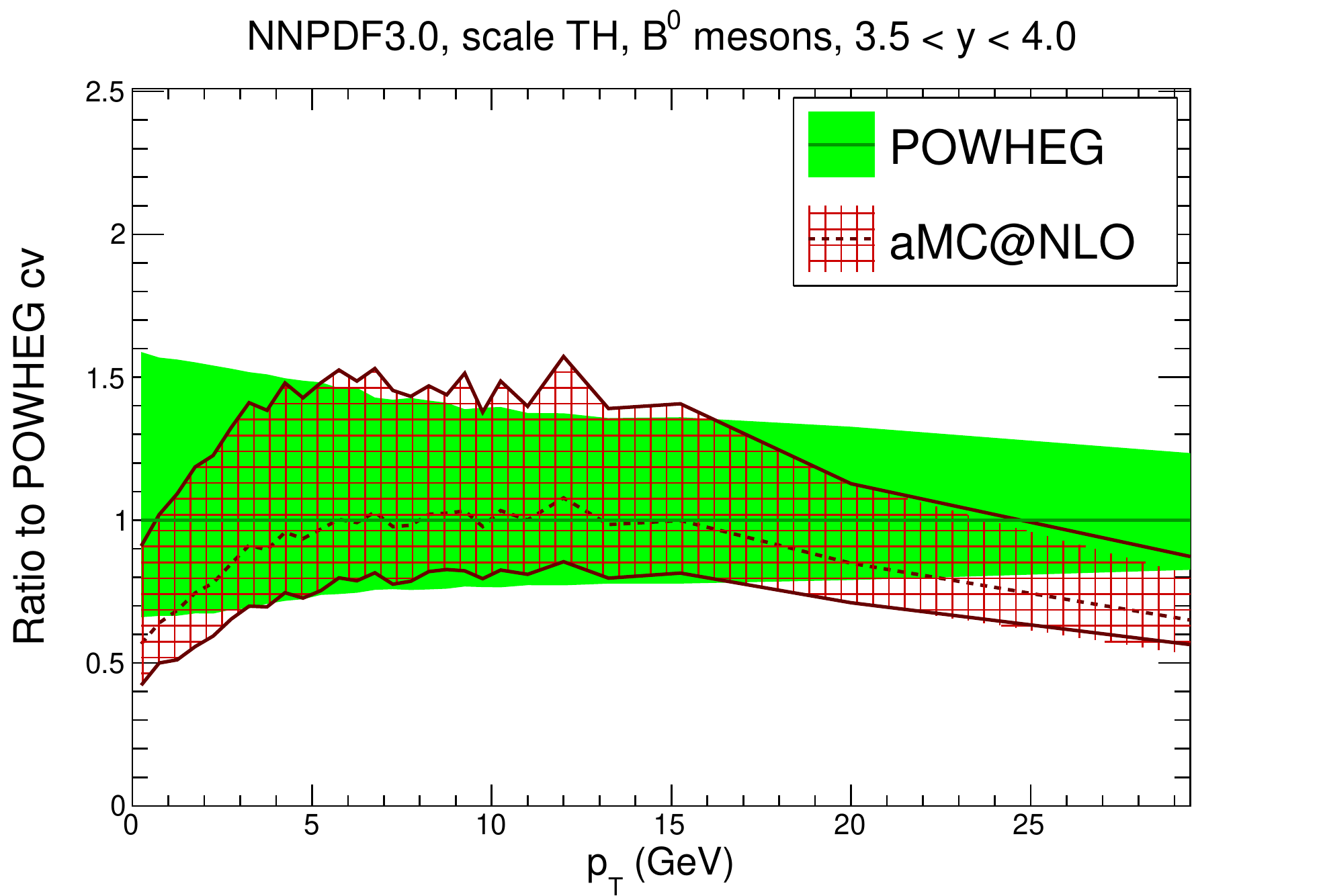}
\caption{\small Comparison of the theoretical predictions for $B^0$
  meson production at the LHCb kinematics between {\sc\small POWHEG}
  and {\sc\small {\rm a}MC@NLO}, shown in Fig.~\ref{fig:LHCbBottom}, but
  now where calculations are normalised to the  central
  value of the {\sc\small POWHEG} prediction.  }
\label{fig:LHCbBottom2}
\end{figure}

To better assess the differences between
the two NLO matched calculations, we compare them
again in Fig.~\ref{fig:LHCbBottom2}, this
time with the
distributions
normalised to the central {\sc\small POWHEG} prediction.
  The {\sc\small {\rm a}MC@NLO} and {\sc\small POWHEG} predictions agree across
  the considered kinematic range, with the {\sc\small POWHEG} prediction favouring
  a slightly larger cross section in the low $p_T$ range.
  In comparison to the charm results, Fig.~\ref{fig:LHCbcharm_thrat}, the reduction of
  scale uncertainties is evident, since now
  the scale variation amounts to an uncertainty of $\simeq 40\%$.
  We can conclude that the pQCD description of $B$ meson production
  in the forward region is completely satisfactory, and that theory uncertainties are
  substantially reduced as compared to charm production.

In this section we have restricted our study to 7~TeV, the only centre-of-mass
energy for which LHCb measurements are currently available.
Predictions for double differential distributions at 13~TeV, as well as for the ratio
of cross-sections at computed at 13 over 7~TeV, will be provided in 
Sect.~\ref{sec:appendix1} and~\ref{sec:appendix2}.

\subsection{PDF dependence of heavy quark production at LHCb}
\label{sec:deppdfset}

The results shown so far in this Section have been computed
using the NNPDF3.0 NLO set.
We have verified that the pQCD predictions for heavy quark production are
affected by a sizeable PDF uncertainty, which arises in turn from poor
knowledge of the small-$x$ gluon PDF due to a lack of direct
experimental constraints.
In this section we study the dependence of our predictions on the choice 
of input PDF set, in particular we compare those of the baseline NNPDF3.0 to 
CT10 and MMHT14 NLO sets.
The comparison of the small-$x$ gluon PDF
between these three sets shown in Fig.~\ref{fig:pdfcomparison}
indicates that predictions for
charm production cross-sections are expected to be
reasonably similar.

\begin{figure}[t]
\centering 
\includegraphics[scale=0.39]{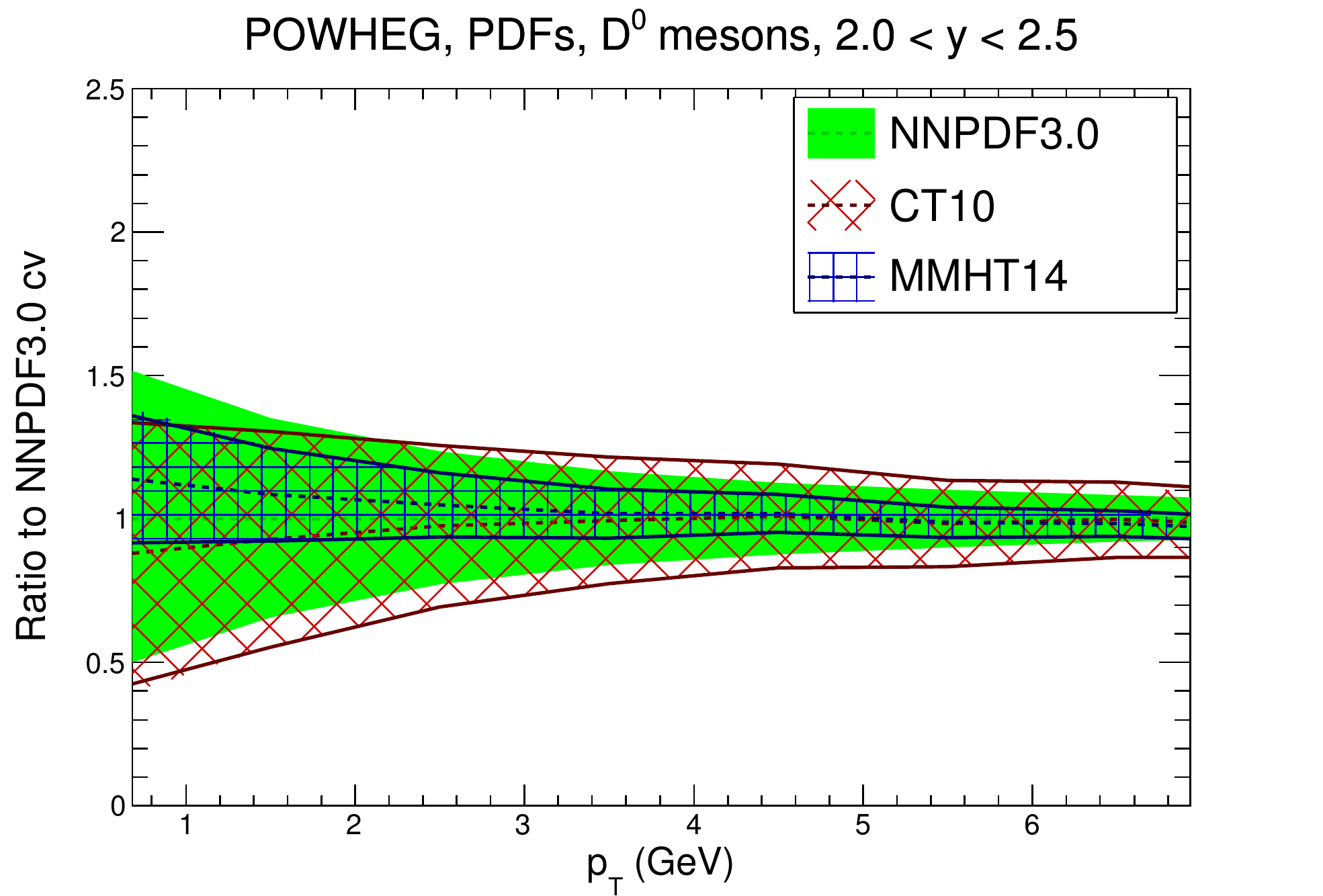}
\includegraphics[scale=0.39]{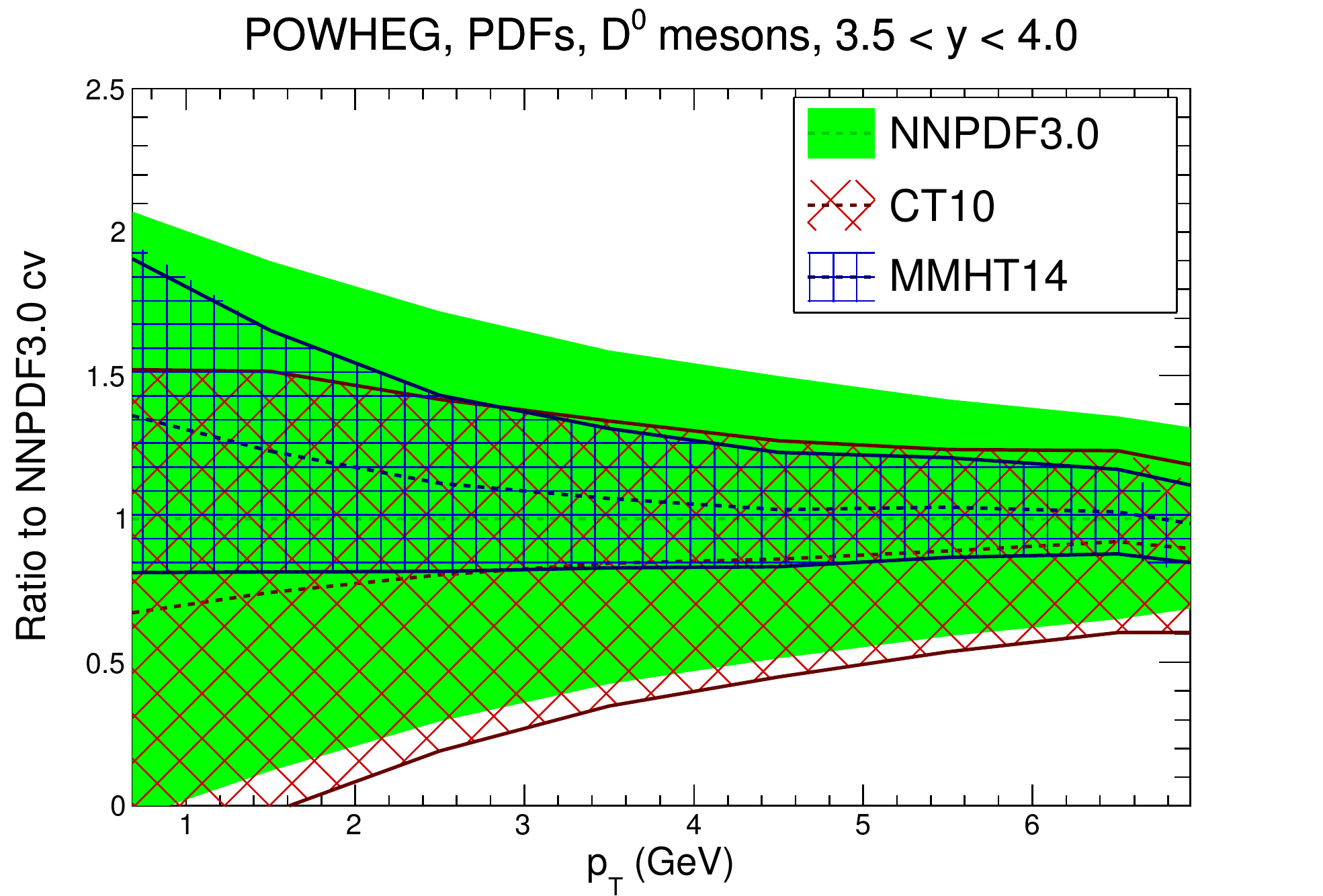}
\caption{\small Comparison of the theoretical predictions for
  $D^0$ meson production at $\sqrt{s}$=7 TeV in the LHCb kinematics with {\sc\small
    POWHEG}, for three different NLO sets of PDFs, NNPDF3.0, CT10 and
  MMHT14.
  The band corresponds to the respective one-sigma PDF uncertainty for
  each set.
  Results are shown normalised to the central value
  of the NNPDF3.0 prediction.  }
\label{fig:PDFdepLHC}
\end{figure}

In Fig.~\ref{fig:PDFdepLHC} we show the comparison of the
theoretical predictions for charm production at 7 TeV within the LHCb
acceptance found using the {\sc\small POWHEG} calculation with
NNPDF3.0, CT10 and MMHT14 PDFs.
  The uncertainty band corresponds to the 68\% confidence level for each PDF set, 
  and the shown results have been normalised to the central value
  of the NNPDF3.0 prediction.
  From this comparison, we see that the dependence of
  the charm cross-section on the choice of
  input PDF set is moderate, with the three central values
  consistent within large PDF uncertainties.
   Recall that at fixed rapidity, smaller values of the $D$ meson
   $p_T$ correspond to probing smaller $x$ values for the
   gluon PDF, and that, likewise, for a fixed value of $p_T$, forward rapidities
   corresponds to smaller $x$ values.
   It is therefore reasonable that PDF uncertainties are
   largest at small $p_T$ and forward
   rapidities, as shown in Fig.~\ref{fig:PDFdepLHC}.

  Even though predictions suffer from large PDF uncertainties, the central value 
  of these three PDF sets
  are reasonably consistent. This agreement can in part 
  be explained by the fact that at small-$x$ PDF constraints
  in the three sets come from the same dataset, the combined
  HERA-I measurements~\cite{Aaron:2009aa}.
  We note that the relative size of the PDF uncertainties is similar for NNPDF3.0 
  and CT10, while the MMHT14 uncertainty is about a factor of two smaller. Another 
  feature of these predictions is the preference for the CT10 and MMHT14 
  central values towards relatively smaller and larger differential cross sections
  for small $p_T$ values, respectively. This can be traced to the relatively softer and harder gluon
  PDF at small-$x$ preferred by the CT10 and MMHT14 respectively as compared 
  to NNPDF3.0 --- see Fig.~\ref{fig:pdfcomparison}.

We conclude from Fig.~\ref{fig:PDFdepLHC} that although there is some dependence 
on the choice of input PDF set, these differences are small within
the large intrinsic PDF uncertainties, and therefore it
is sufficient to use a single PDF set,  NNPDF3.0, as baseline in
our calculations.

%% file: sec-rw.tex
\section{Constraints on the small-$x$ gluon PDF
  from forward charm production data}
\label{sec:rw}

As demonstrated in the previous section, the production of
charmed hadrons in the forward region and the associated theoretical uncertainty 
depends on the description of the gluon PDF at small-$x$.
We now use the charm production data from
LHCb to substantially reduce the small-$x$ gluon PDF uncertainties. 
This will allow a more reliable prediction for both forward charm production 
at the LHC Run II and the prompt neutrino cross section arising from high energy 
cosmic rays --- an important input for calculating the background
neutrino flux at IceCube.

The basic idea is similar to the study performed by the
{\tt PROSA} Collaboration~\cite{Zenaiev:2015rfa}, where 
the impact of forward $B$ and $D$ LHCb data on the low-$x$
PDFs is studied\footnote{We would like to stress that
  preliminary results for our work were presented already in February
  2015, \url{http://benasque.org/2015lhc/talks_contr/179_BenasqueGauld.pdf}, before
  the publication of the {\tt PROSA} paper.
  Preliminary results of the {\tt PROSA} study were also
  presented in~\cite{PROSAurl}.
}.
The {\tt PROSA} study is based on the
{\sc\small  HERAfitter} framework~\cite{Alekhin:2014irh}, and quantifies
the error reduction in a HERA-only PDF fit when the LHCb $B$ and $D$ meson
production data is included using the MNR code~\cite{Mangano:1991jk} in a FFN $N_f=3$ scheme.
Similarly as will be done here, theoretical uncertainties can be reduced by suitable
normalisations.
This said, there are important methodological differences in the two analysis
(global fit versus HERA-only fit, theory calculations, data normalisation strategies),
and so the two approaches complement one another.

The starting point is the NNPDF3.0 NLO set, with
$\alpha_s(m_Z)=0.118$, supplemented by the LHCb measurements of the 7~TeV 
differential distributions for $D^0$ and $D^{\pm}$
production~\cite{Aaij:2013mga}.
The LHCb data will be added to the NNPDF3.0 global dataset by means of
the Bayesian reweighting technique~\cite{Ball:2010gb,Ball:2011gg}.
This method allows to quantify the impact of new data in a set of
Monte Carlo PDFs without the need of redoing the full global QCD
analysis, and has been used before in a number of related applications
in order to quantify the impact on PDF fits from data for isolated
photon production~\cite{d'Enterria:2012yj,Carminati:2012mm}, top quark pair
production~\cite{Czakon:2013tha}, and polarised
$W^{\pm}$ and jet production~\cite{Nocera:2014gqa}.
As an alternative to the reweighting, it should also have been
possible to use the {\sc\small aMCfast}~\cite{amcfast}
interface to construct
an {\sc\small APPLgrid}~\cite{Carli:2010rw}
fast implementation of the
{\sc\small {\rm a}MC@NLO} calculations presented in the previous section.

As input to the reweighting, we consider the $\lp y, p_T\rp$ double
differential distributions for $D^0$ and $D^{\pm}$ production at
LHCb, but exclude the data from other final states
such as $D^{*\pm}$ and $D^{\pm_s}$ which are affected by larger
experimental uncertainties, and therefore have reduced impact on the
fit.
These data cover a range in rapidity of $[2.0,4.5]$ and
in $p_T$ of $[0,8]$ GeV.
In total, we are adding $N_{\rm dat}=75$ new data points into the
NNPDF3.0 analysis.

\begin{figure}[t]
\centering 
\includegraphics[scale=0.39]{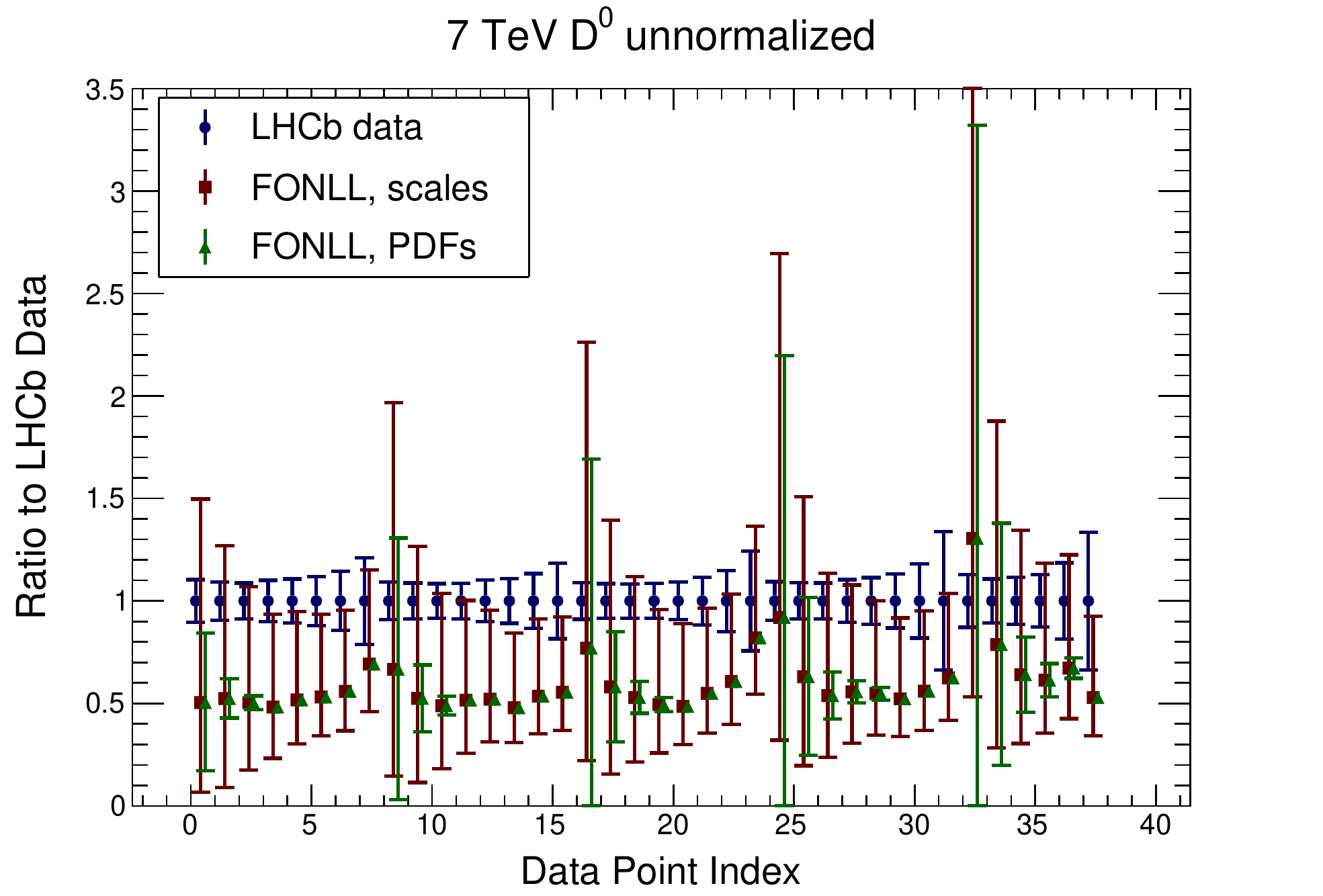}
\includegraphics[scale=0.39]{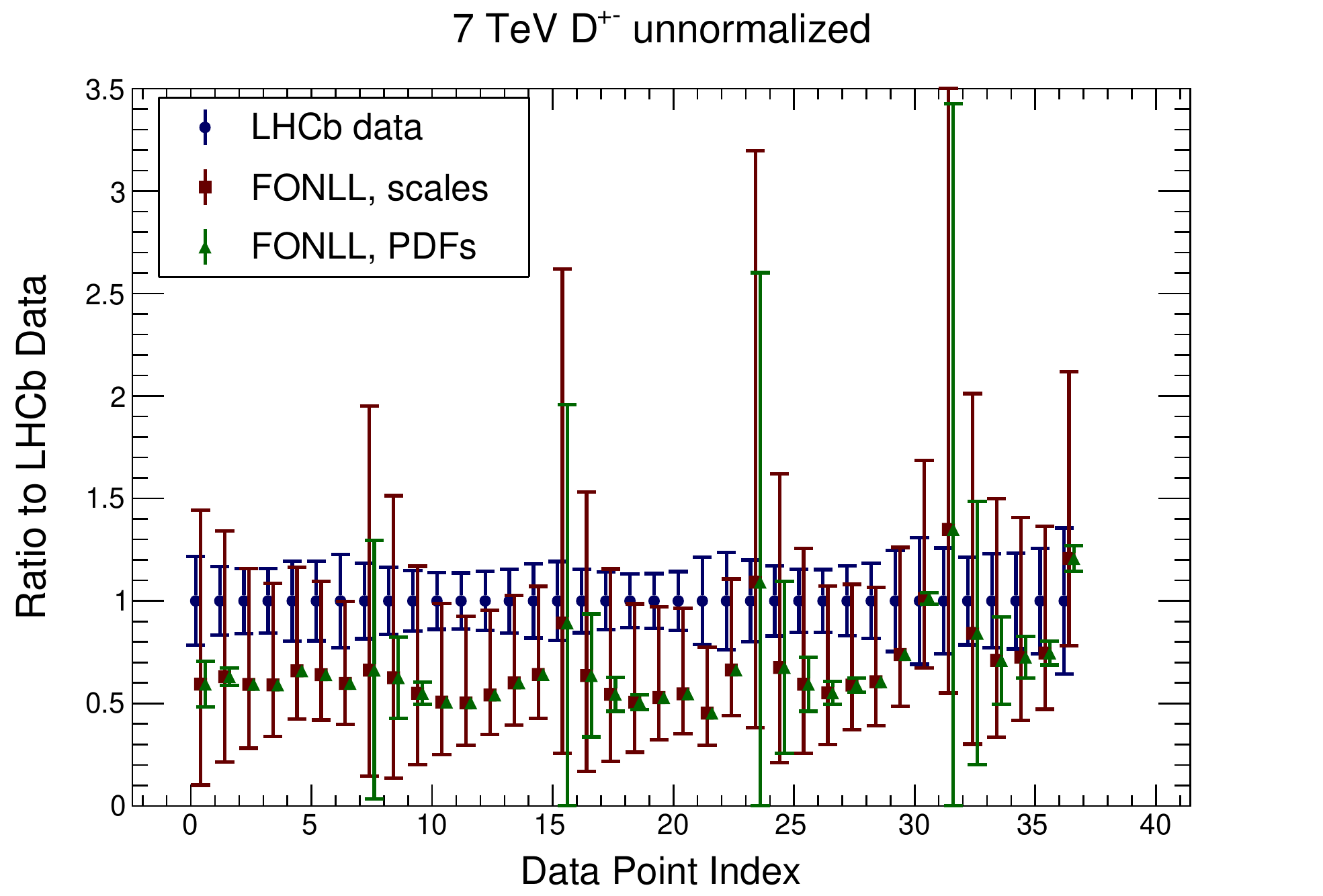}\\
\caption{\small Comparison between the LHCb charm production
  data and the {\sc\small FONLL} calculation with NNPDF3.0 NLO at the level
  of unnormalized (absolute) cross-sections.
  The left plot shows the $D^0$ data while the right plot corresponds
  to the $D^{\pm}$ data.
  Results are shown normalised to the central value of the LHCb data.
  For the  {\sc\small FONLL} calculation we show separately the scale
  and the PDF uncertainties.
  The data is ordered in increasing rapidity bins, an within each of
  these, in increasing $p_T$ bins.
}
\label{fig:fonll1}
\end{figure}

For the theoretical calculations, we use the {\sc\small FONLL} predictions, with
the settings discussed in the previous Section.
In Fig.~\ref{fig:fonll1} we compare the LHCb charm production data and
the {\sc\small FONLL} prediction for the $D^0$ and $D^{\pm}$ data.
Results are shown normalised to the central value of the respective experimental data point.
The experimental statistical and systematic uncertainties have been
added in quadrature, and both scale and PDF uncertainties are independently 
shown for the {\sc\small FONLL} theoretical prediction.

It is clear by inspection of Fig.~\ref{fig:fonll1} that the scale
uncertainties in the NLO calculation are large, by as much as a factor
of two for some bins.
In general, they are reduced when going towards higher $p_T$ bins,
thanks to the improved convergence of the perturbative expansion
in this region.
Although PDF uncertainties are also large, especially at low $p_T$ and
forward rapidities where the small-$x$ gluon is being probed, 
they are sub-dominant as compared to the scale uncertainties.
This is
concerning from the point of view of a PDF analysis, in which a scale
choice for the central value of the theory prediction must be made.

To bypass this problem, the strategy that will be adopted in this work
is to normalise all the data bins to that with highest $p_T^D$, $\lc
7,8\rc$ GeV, and central rapidity $y^D$, $\lc 2.0,2.5\rc$.
The rationale for this choice is that scale uncertainties will
partially cancel in the ratio, while the cancellation of PDF 
uncertainties will not be as severe, given
that different bins in $\lp y^D,p_T^D\rp$ probe different values of
$(x,Q^2)$ of the gluon PDF.
The reference bin has been chosen precisely for this reason, as PDF 
uncertainties for this particular bin are the smallest.
Note that this is strategy is different as compared
to the {\tt PROSA} analysis~\cite{Zenaiev:2015rfa}, where, separately for each bin
in $p_T^D$, the rapidity bin $3.0 \le y^D \le 3.5$ was used to normalize
the data and the theory calculations.

\begin{figure}[t]
\centering 
\includegraphics[scale=0.39]{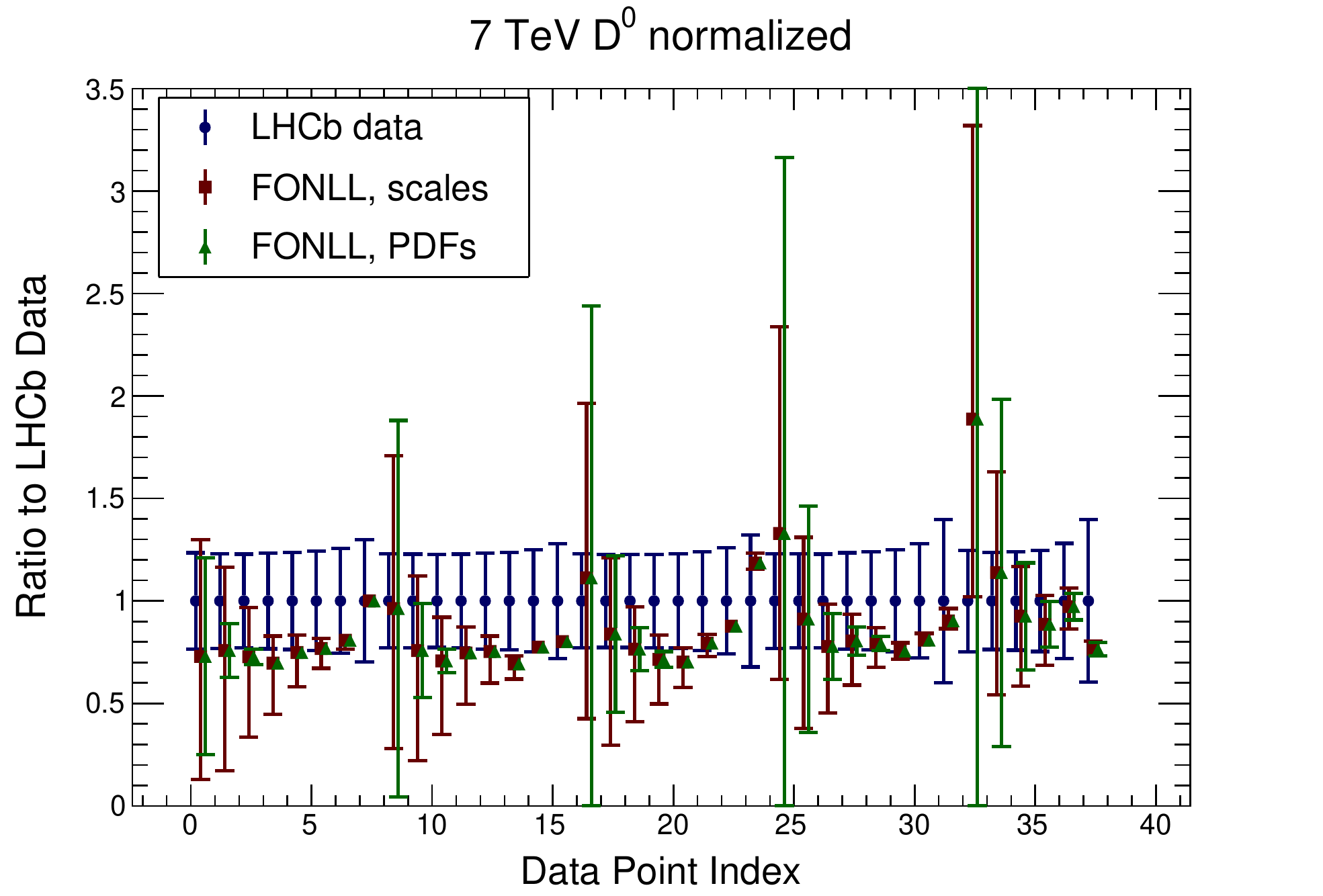}
\includegraphics[scale=0.39]{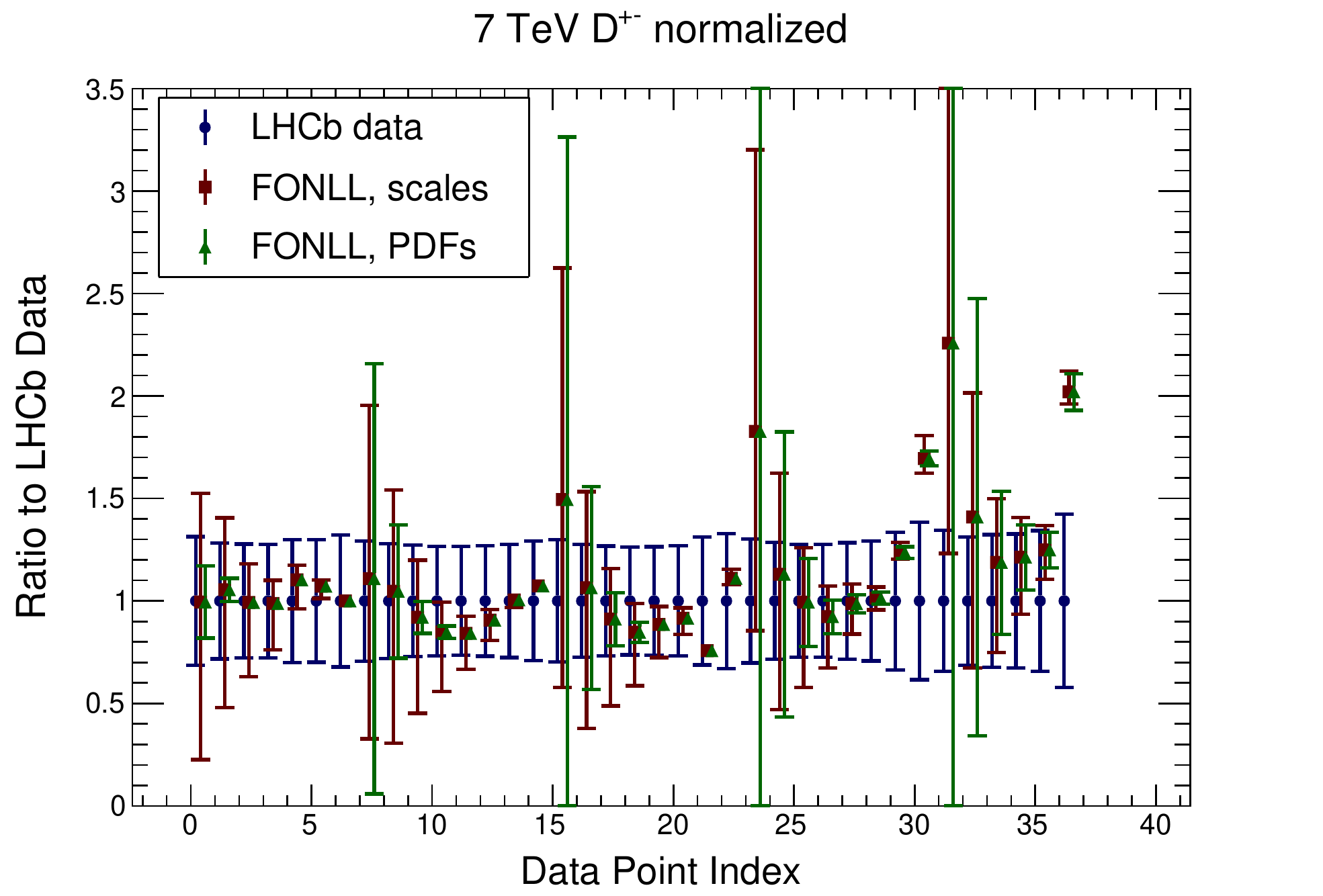}\\
\caption{\small Same as Fig.~\ref{fig:fonll1}, where now both data and
  theory have been respectively normalised with respect to the bin
  with $7~{\rm GeV} \le p_T \le 8$ GeV and $2.0 \le y \le 2.5$ (the bin
  with data point index 8).
  }
\label{fig:fonll2}
\end{figure}

In Fig.~\ref{fig:fonll2} we provide the same comparison of
Fig.~\ref{fig:fonll1}, but this time at the level of normalised
distributions.
In Fig.~\ref{fig:fonll2} we have added in quadrature the experimental
uncertainties in the numerator and the denominator, this being the
only option since the full experimental covariance matrix with the information
of correlations between bins is not
available.
Theoretical uncertainties are taken to be fully correlated among all the data bins.
%

The comparison between Fig.~\ref{fig:fonll1} and Fig.~\ref{fig:fonll2}
illustrates how after the normalisation procedure has been applied, 
scale uncertainties are substantially reduced in the low-$p_T$ and large-$y$ bins.
Importantly, the PDF uncertainties are now larger than the corresponding scale and experimental 
uncertainties in these bins, which justifies the inclusion of the normalised charm
production cross-sections into NNPDF3.0 using Bayesian reweighting.
In this respect, after the normalisation, the theoretical
status of forward charm production
becomes similar to that of other hadronic processes
routinely included in global NLO
fits, such as jet production.

The results of the reweighting are summarised in
Table~\ref{tab:rw}.
The breakdown of the $\chi^2$ per data point of the $D^0$ and $D^{\pm}$
data before and after reweighting, as well as the number of effective replicas
left out of the original $N_{\rm rep}=100$ replicas, is provided.
The description of the normalised LHCb charm
data turns out to be excellent
even using the original NNPDF3.0 set,
with a value of $\chi^2/N_{\rm
  dat}=1.10$.
This is certainly reassuring, since it shows that both
NNPDF3.0 and the {\sc\small FONLL} calculation provide a good description
of charm production in the LHCb acceptance.
Once the data is included by the reweighting, the $\chi^2_{\rm rw}/N_{\rm
  dat}=0.74$ is even better, and the effective number of replicas is
$N_{\rm eff}=50$, confirming that this data is indeed very constraining on
the small-$x$ gluon PDF.
Note that since we are neglecting the correlations between
systematics, we are underestimating the impact of these data. Future
measurements with the full systematic breakdown should be even more
powerful.

\begin{table}[h]
  \centering
  \begin{tabular}{c|c|c}
    \hline
    &   NNPDF3.0  &  NNPDF3.0+LHCb data \\
    \hline
    \hline
   $\chi^2/N_{\rm dat}$ for $D^{0}$ + c.c. & 1.13    &  1.05    \\
   $\chi^2/N_{\rm dat}$ for $D^{+}$ + c.c. & 1.06    &  0.40  \\
    $\chi^2/N_{\rm dat}$ for $D^{0}$ + $D^{\pm}$ &  1.10   & 0.74    \\
    \hline
       \hline
    $N_{\rm eff}$ for $D^{0}$ + $D^{\pm}$  & - & 50 \\
    \hline
  \end{tabular}
  \caption{\small Results of the reweighting of NNPDF3.0 with the LHCb
    charm production data.
    We give the value of the $\chi^2/N_{\rm dat}$, both for the original
    NNPDF3.0 set, and for the reweighted NNPDF3.0+LHCb set, as well
    as the effective number of replicas left, $N_{\rm eff}$.
    \label{tab:rw}}
\end{table}
    
\begin{figure}[t]
\centering 
\includegraphics[scale=0.39]{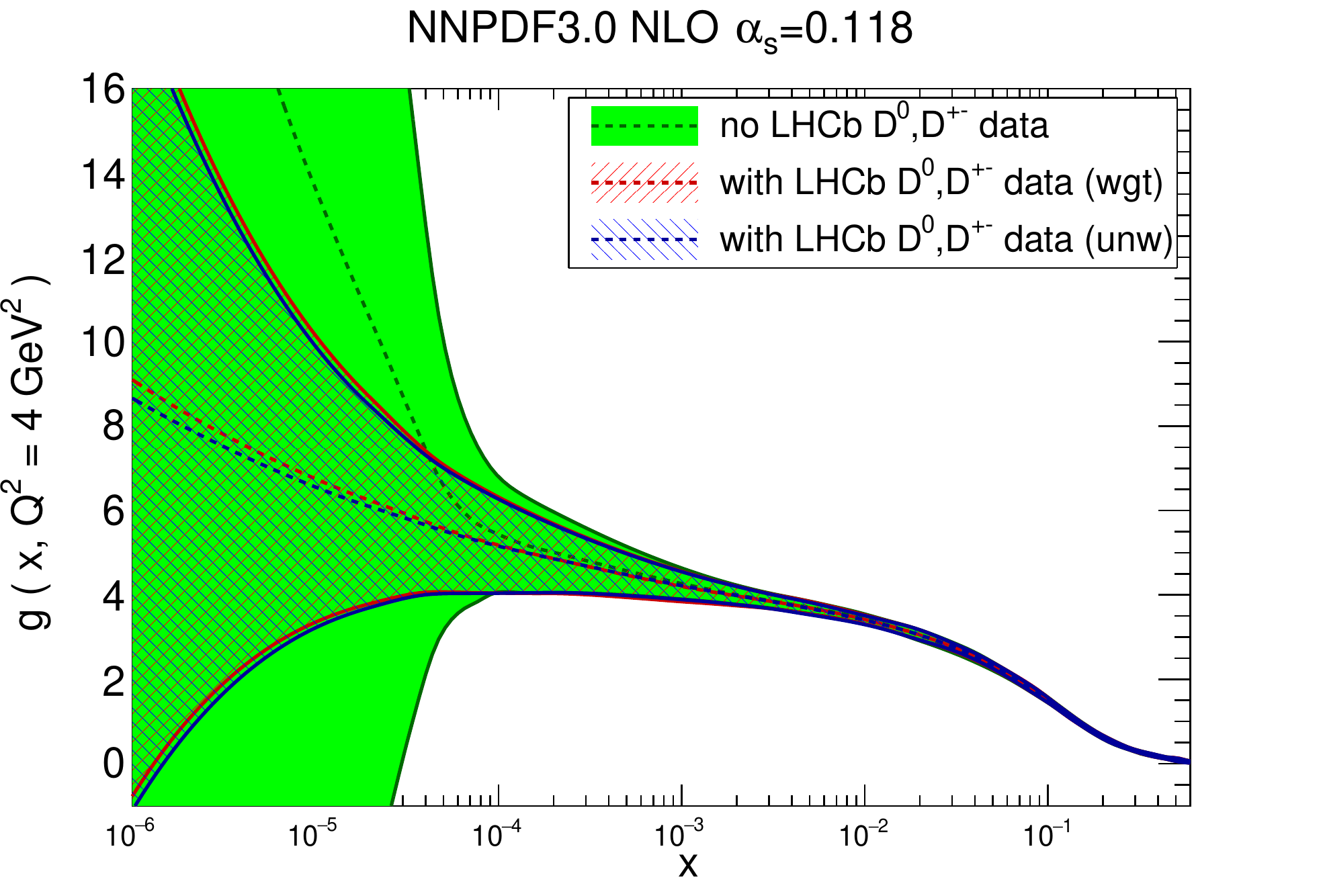}
\includegraphics[scale=0.39]{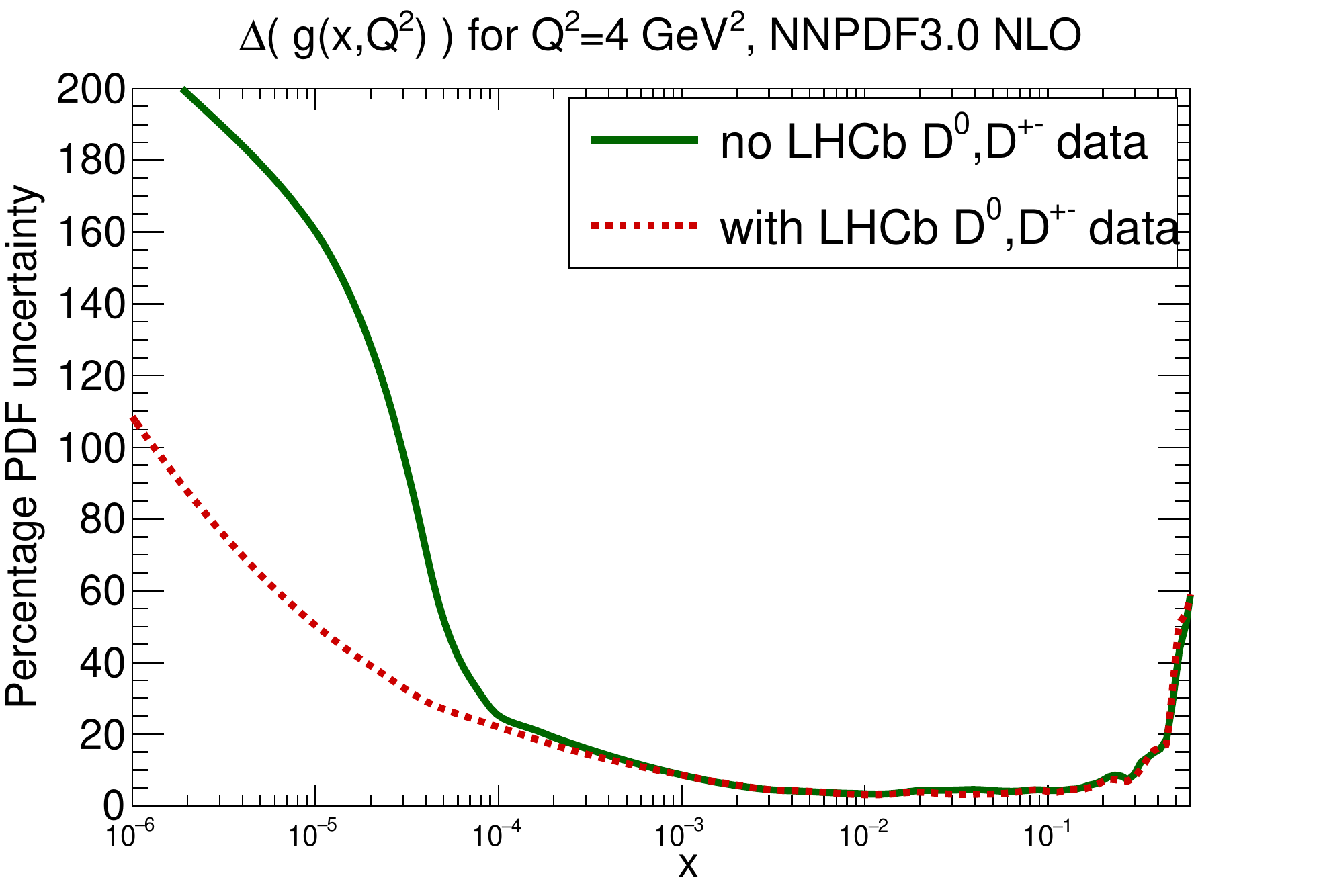}
\caption{\small Left: The NNPDF3.0 NLO small-$x$ gluon, evaluated at
  $Q=2$ GeV, comparing the global fit result with 
  with the new gluon obtained
  from the inclusion of the LHCb charm production data.
  In the latter case, we show both the reweighted ({\it rwg})
  and the unweighted ({\it unw}) results.
  Right: comparison of percentage PDF uncertainties
  for the NNPDF3.0 gluon with and without the inclusion of the LHCb data, computed
  also at $Q=2$ GeV,
  that illustrate the reduction of PDF uncertainties for $x\lsim 10^{-4}$.
}
\label{fig:smallxgluon}
\end{figure}

The impact of the LHCb charm production data into the small-$x$ gluon
PDF can be seen in Fig.~\ref{fig:smallxgluon}.
We show the NNPDF3.0 small-$x$ gluon, evaluated at $Q=2$ GeV,
compared with the new gluon obtained after the
inclusion in the fit of the normalised LHCb charm data.
  As a cross-check, we have also verified that it is possible to
  unweight the results to produce a stand-alone {\sc\small LHAPDF6}
  grid for the combined NNPDF3.0+LHCb fit (indicated as ``(unw)'' in the
  plot legend).
  In Fig.~\ref{fig:smallxgluon} we also compare the
  percentage PDF uncertainties
  for the NNPDF3.0 gluon with and without the inclusion of the LHCb data,
  which quantify the reduction of PDF uncertainties at small-$x$.

  We see that the impact of LHCb data is negligible for $x\gsim
  10^{-4}$, where most of the HERA data is available, but becomes 
  substantial for $x\lsim 10^{-4}$, where the previously large PDF
  uncertainties are dramatically reduced.
  For instance, for $x\sim 10^{-5}$, the PDF uncertainties in the
  gluon PDF are reduced by more than a factor three.
  We also note that the central value at small-$x$ of the gluon PDF
  preferred by the LHCb charm data is less steep than that of the global
  fit, although fully consistent within uncertainties.
  The quark PDFs are essentially unaffected by the inclusion of the
  LHCb charm data and are thus not shown here.

Since the resulting PDF set from the inclusion of the LHCb data into
NNPDF3.0 has been
unweighted to a
a {\sc\small LHAPDF6} grid, 
it can be easily used both for the predictions of
heavy quark production at 13 TeV at LHCb,
presented in Sect.~\ref{sec:results13tev},
and for the prompt neutrino cross-sections relevant
for IceCube in Sect~\ref{sec:results}.

\begin{figure}[t]
\centering 
\includegraphics[scale=0.50]{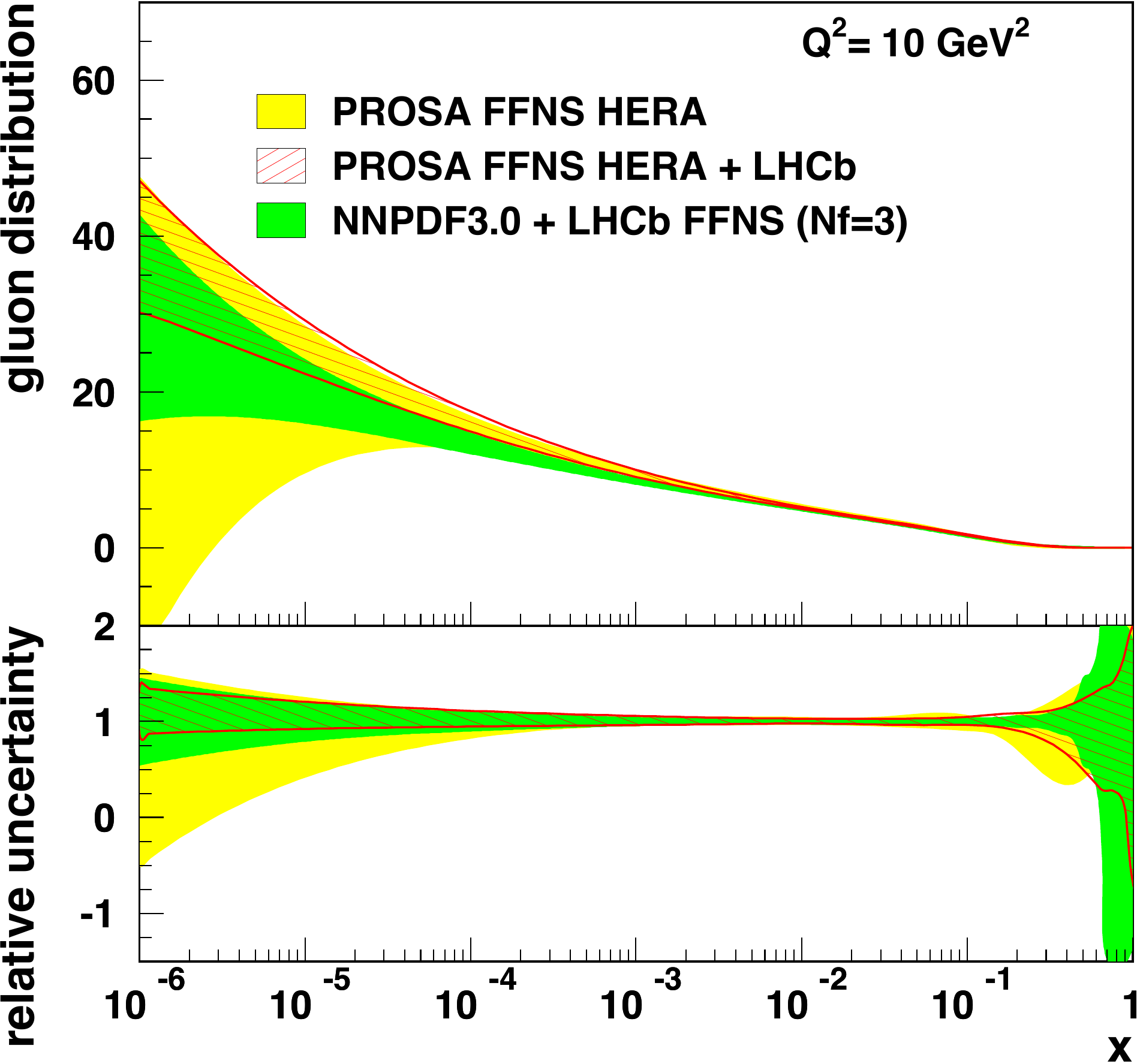}
\caption{\small The gluon PDF at $Q^2=10$ GeV$^2$ in the FFN scheme
  with $N_{f}=3$, comparing the results of this work with those
  of the {\tt PROSA} analysis.
  In the latter case, we show the results both in the HERA-only fit
  and in the HERA+LHCb fit.
  The lower panel compares the relative PDF uncertainties in each case.
}
\label{fig:prosa}
\end{figure}

It is interesting to assess how the results of this analysis compare
to those of the {\tt PROSA} study~\cite{Zenaiev:2015rfa}.
Note that the two analysis use rather different methodologies
(HERA-only fit versus global fit, {\sc\small HERAfitter} versus NNPDF reweighting),
and given that this is the first time that forward charm data is used
in a PDF fit, it is important assess the robustness of the results
by performing a cross-check.
Since the {\tt PROSA} analysis is performed in the FFN $n_f=3$ scheme,
we have constructed a FFN $n_f=3$ version of the NNPDF3.0+LHCb NLO set
using {\tt APFEL}~\cite{Bertone:2013vaa}.
The results of this comparison are shown in
Fig.~\ref{fig:prosa}, where we show the
gluon PDF at $Q^2=10$ GeV$^2$ in the FFN scheme
  with $N_{f}=3$, 
  In the {\tt PROSA} case, we show the results both in the HERA-only fit
  and in the HERA+LHCb fit.\footnote{We thank Katerina Lipka for
    providing us this plot, which compares the
  {\tt PROSA} and NNPDF results.}
  The lower panel compares the relative PDF uncertainties in each case.
  As can be seen, there is good agreement both between central values
 (the two gluons agree within their one-sigma band)
  and especially between PDF uncertainties, which is a non-trivial
  verification of the two analyses.

\begin{figure}[t]
\centering 
\includegraphics[scale=0.39]{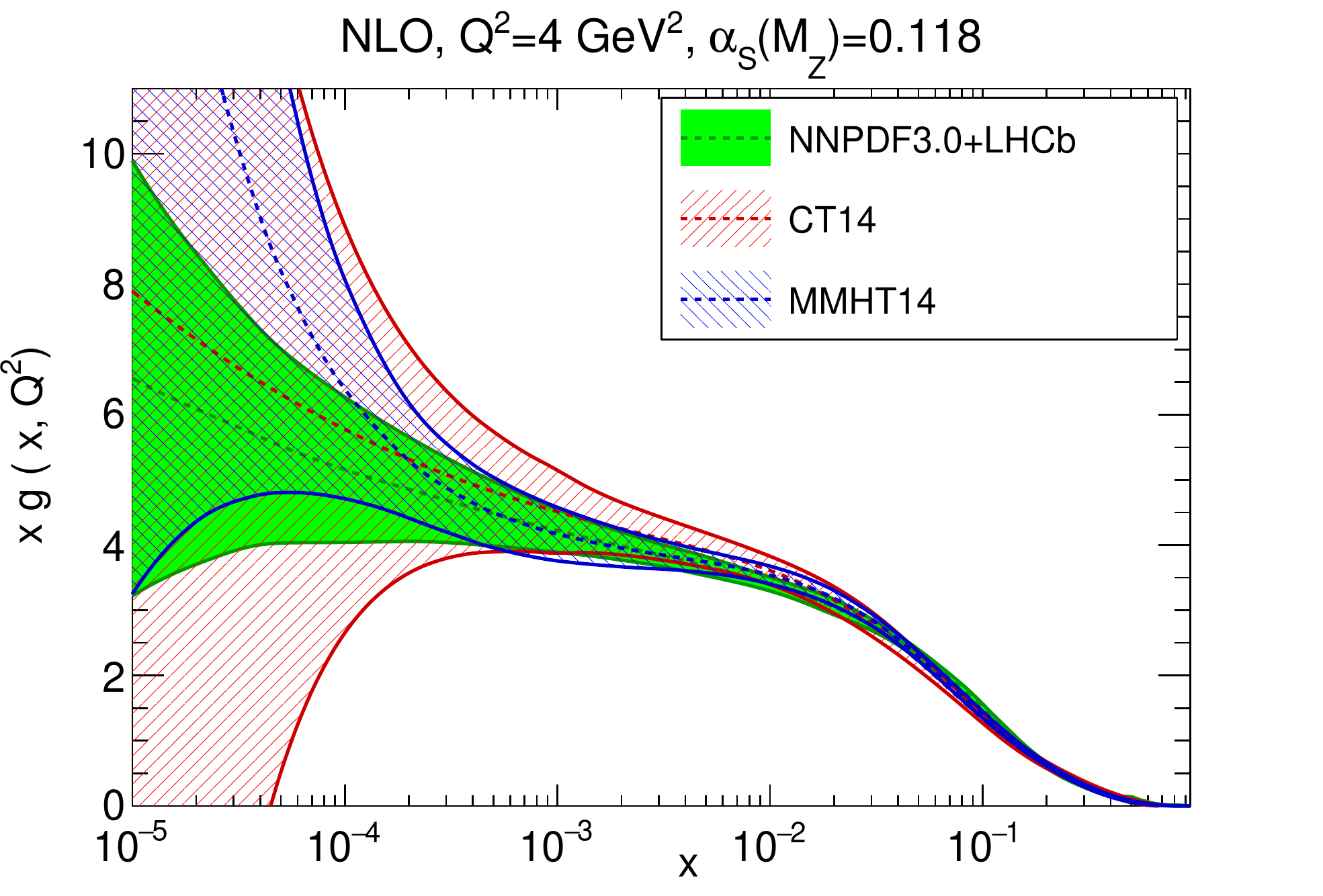}
\includegraphics[scale=0.39]{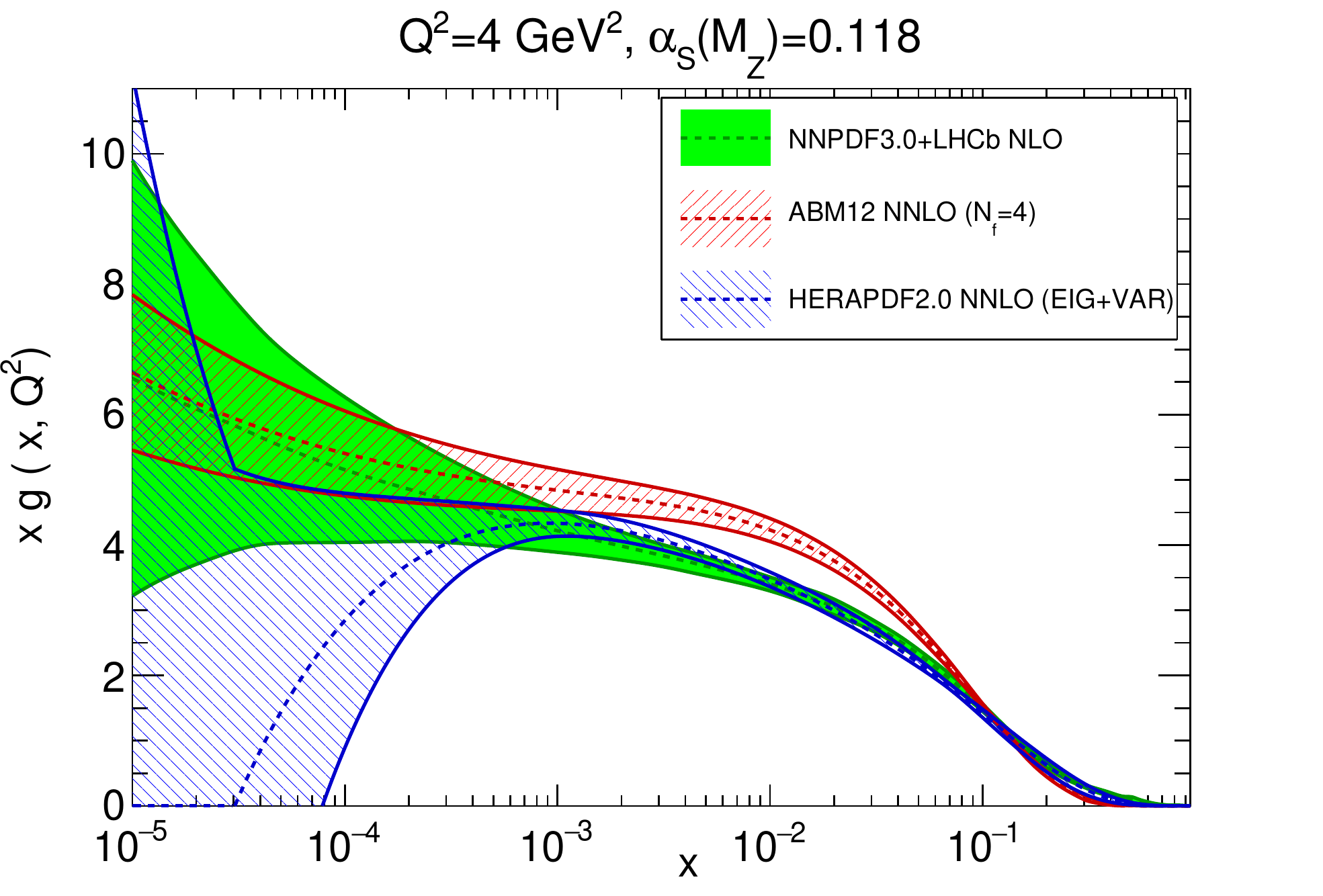}
\caption{\small The NNPDF3.0+LHCb gluon PDF at $Q^2=4$ GeV$^2$
  compared with CT14 and MMHT14 (left plot), and to ABM12
  and HERAPDF2.0 (right plot).
  In the case of HERAPDF2.0, both the experimental, model
  and parametrization uncertainties are included.
}
\label{fig:gluon-comparison}
\end{figure}

Finally, let us compare the resulting gluon PDF in this analysis
with those of other recent PDF fits.
In Fig.~\ref{fig:gluon-comparison}
we compare the NNPDF3.0+LHCb gluon PDF at $Q^2=4$ GeV$^2$
  with the CT14~\cite{Dulat:2015mca} and MMHT14 results (left plot), and to the ABM12~\cite{Alekhin:2013nda}
  and HERAPDF2.0~\cite{Abramowicz:2015mha}  results (right plot).
  In the case of HERAPDF2.0, both the experimental, model
  and parametrization uncertainties are included.
  In the case of ABM12, the $n_f=4$ set has been adopted.
  From Fig.~\ref{fig:gluon-comparison} we note that the
  NNPDF3.0+LHCb central value is close to the CT14 result, but
  with much smaller uncertainties, while the MMHT14 gluon is
  substantially larger at small-$x$.
  From the comparison with ABM12 we find reasonable agreement
  for $x\le 10^{-4}$, while HERAPDF2.0 predicts a much smaller
  (negative gluon), though consistent with the NNPDF3.0+LHCb result within
  the PDF large uncertainties.

%% file: sec-13tev.tex
\section{Predictions for 13 TeV and for the 13/7 TeV ratio}
\label{sec:results13tev}

In this section we provide predictions
for $D$ and $B$  production within the LHCb acceptance 
at 13~TeV.
We also provide 
predictions for the ratio
of differential cross-sections between 13 and 7~TeV.
Our predictions are have been computed using the {\sc\small POWHEG}
and {\sc\small {\rm a}MC@NLO} calculations with the improved 
NNPDF3.0+LHCb PDF set constructed in Sect.~\ref{sec:rw}, 
and can be used to compare with the upcoming
Run II measurements at LHCb.
Using the theoretical value of the ratio between inclusive
fiducial cross-sections at 13 and 7 TeV, and the LHCb 7 TeV data ($R_{13/7}$),
we also provide predictions for $B$ and $D$ mesons in fiducial cross-sections at 13 TeV.
A tabulation of our results is provided
in Appendix~\ref{appendix}, and predictions for different
binning choices and other meson species are available from
the authors on request.\footnote{Very recently, the
  LHCb 13 TeV charm production measurements have been
  presented~\cite{Aaij:2015bpa}.
  The LHCb publication
  includes a detailed comparison between data and the 
  theoretical predictions presented in this work, showing
  good agreement within
  uncertainties.
  This agreement for the 13 TeV
  data provides further validation of the robustness
  of our approach.
}

\subsection{Forward heavy quark production at 13~TeV}
\label{sec:appendix1}

First of all,
we provide theory predictions required to compare
with the upcoming LHCb data on charm and bottom production
which will be collected at 13~TeV. 
Our results are presented according to the binning scheme adopted in
the 7 TeV measurements~\cite{Aaij:2013mga,Aaij:2013noa}, with the exception
that a slightly finer binning for the charm predictions is chosen at low $p_T$ and 
the high $p_T$ range is slightly extended. For all predictions, the uncertainty due to 
scales, PDFs, and the heavy quark mass is provided as a sum in quadrature.

%
  
\begin{figure}[t]
\centering 
\includegraphics[scale=0.39]{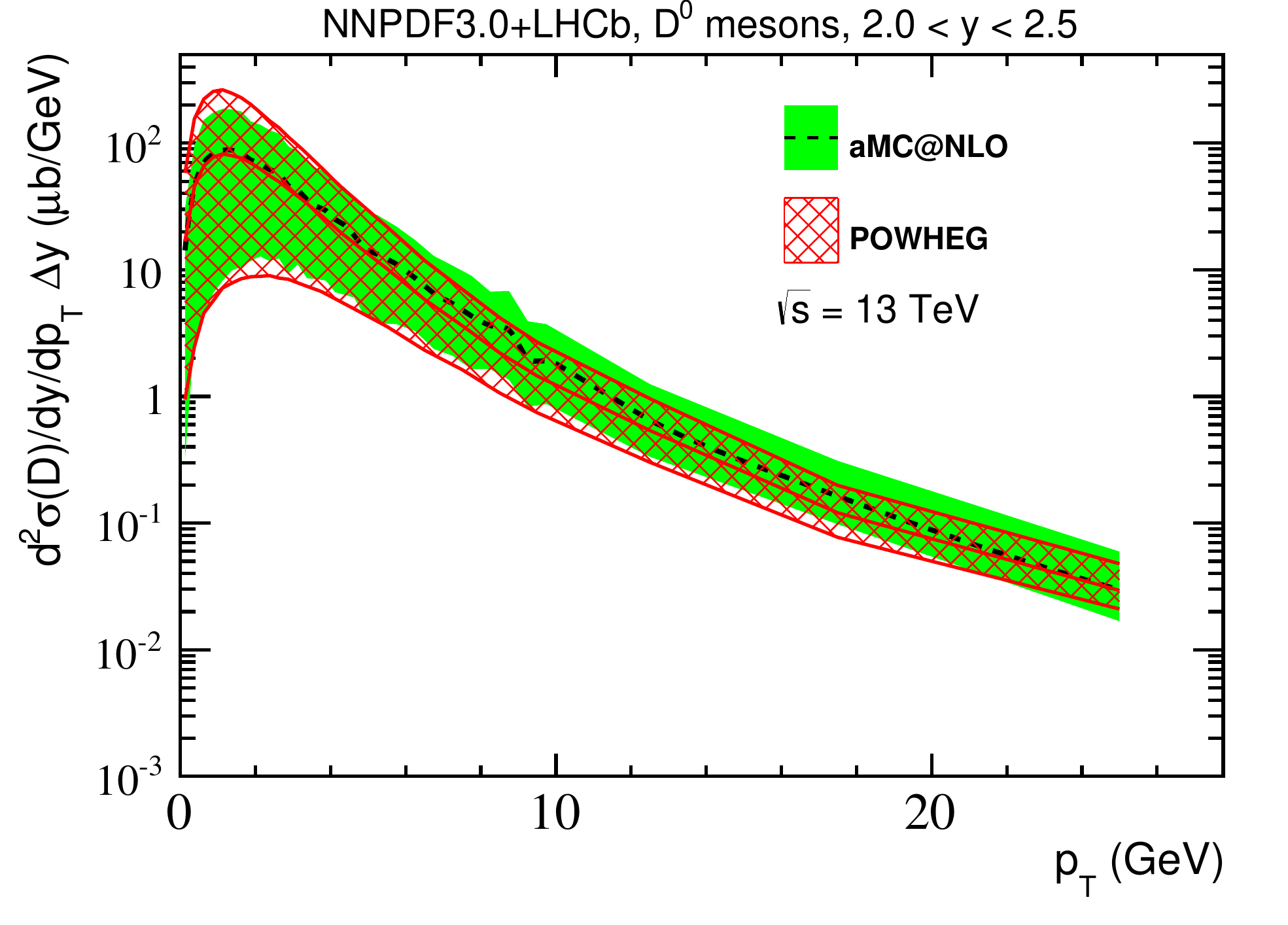}
\includegraphics[scale=0.39]{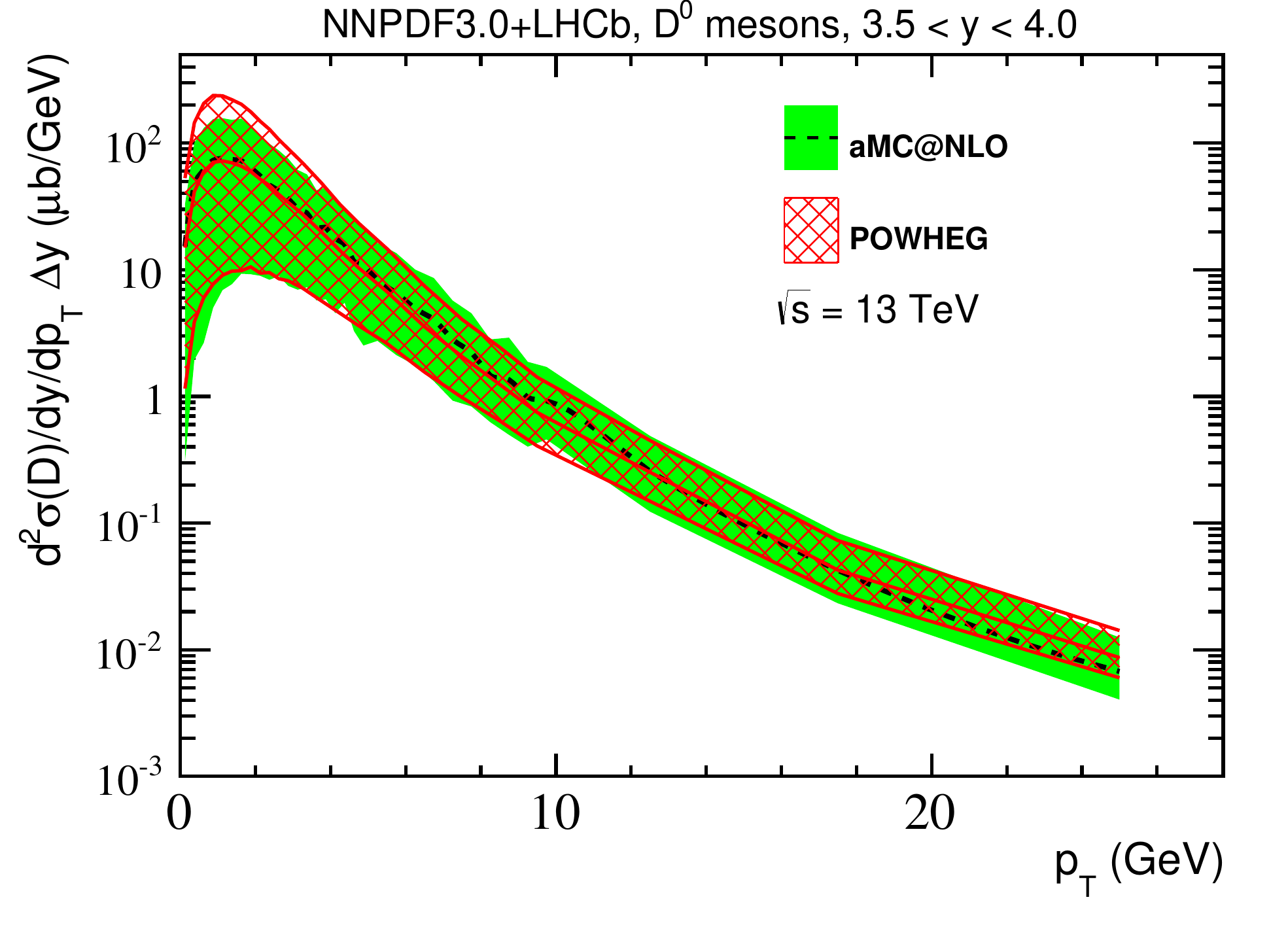}
\caption{\small The double-differential distribution, $d^2\sigma(D)/dydp_T$, for
  the production of $D^0$ mesons at LHCb for a centre-of-mass energy of
  13 TeV.
  We show representative results for the central ($2.0 \le y \le 2.5$)
  and forward ($3.5 \le y \le 4.0$) regions.
  We compare the {\sc\small POWHEG} and  {\sc\small {\rm a}MC@NLO} calculations,
  using the NNPDF3.0+LHCb PDF set.
  For both calculations, the theory uncertainty band is computed adding
  in quadrature 
  scales, PDF and charm mass uncertainties.
}
\label{fig:LHCbcharm13}
\end{figure}

In Fig.~\ref{fig:LHCbcharm13}, the double differential distributions for
$D^0$ mesons at 13~TeV are shown for both a central and a forward
rapidity bin within the LHCb acceptance. The central value and total uncertainty of both 
 {\sc\small POWHEG} and {\sc\small {\rm a}MC@NLO} calculations are provided.
  This comparison demonstrates that there is good agreement between the two
  calculations, both in terms of central values and in terms of the total
  uncertainty band --- agreement also holds for other $D$ mesons and rapidity regions, which 
  are not shown here.
  Thanks to using the improved NNPDF3.0 PDFs with 7~TeV LHCb data,
  PDF uncertainties turn out to be moderate even at 13~TeV,
  with scale variations being the dominant source of theoretical
  uncertainty.

\begin{figure}[t]
\centering 
\includegraphics[scale=0.39]{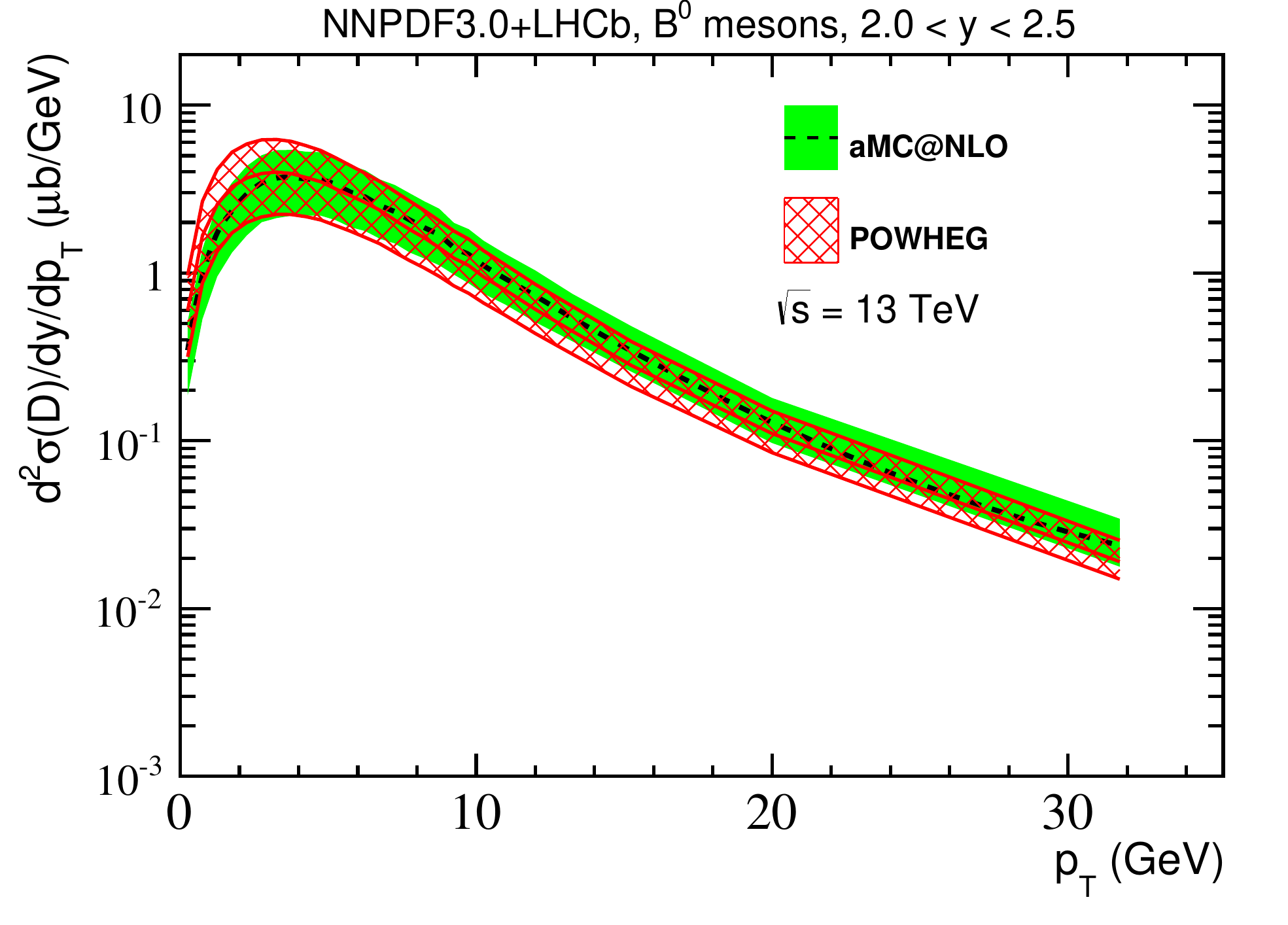}
\includegraphics[scale=0.39]{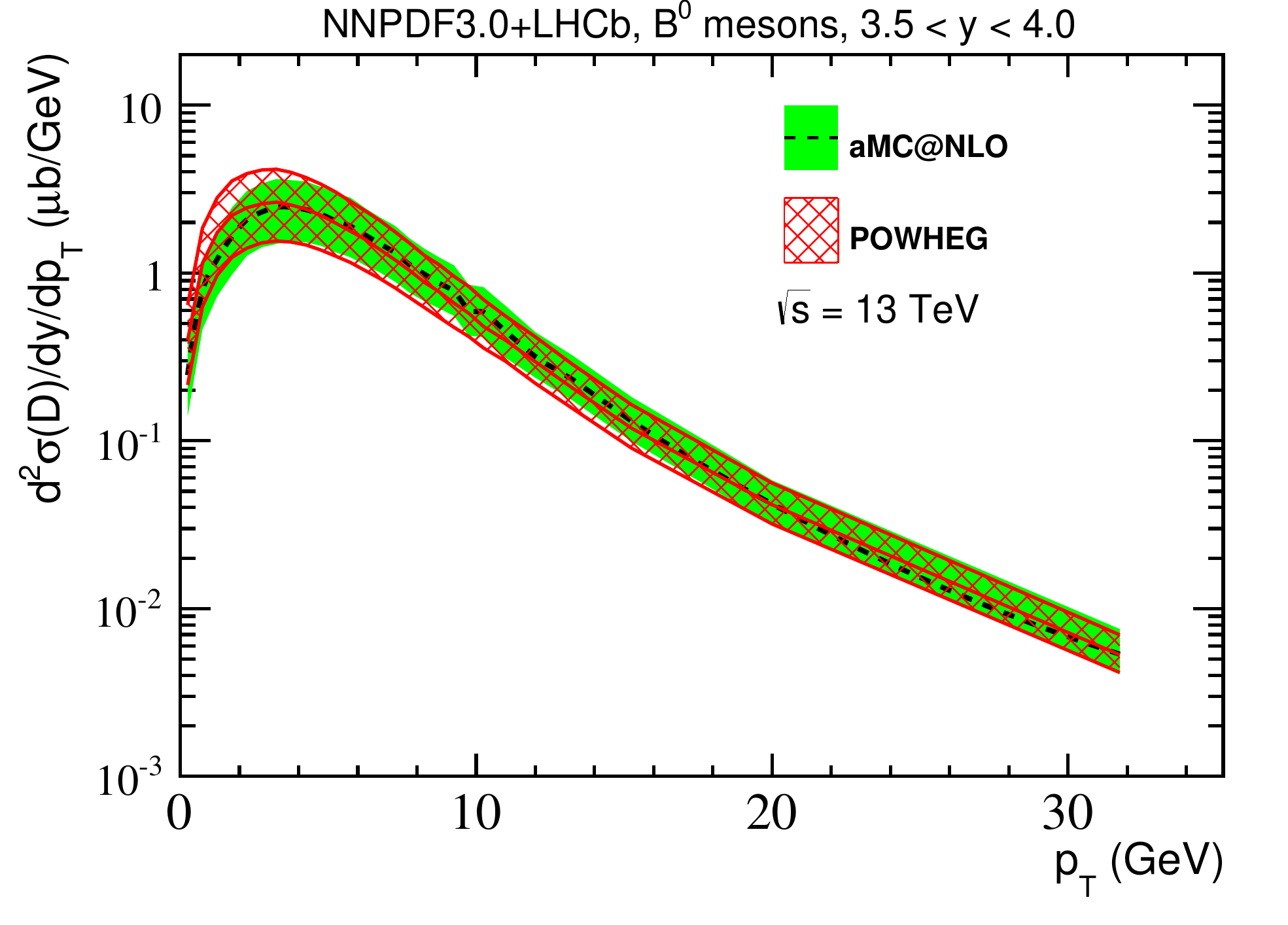}
\caption{\small Same as Fig.~\ref{fig:LHCbcharm13} for $B^0$ mesons.}
\label{fig:LHCbbeauty13}
\end{figure}

The corresponding comparison for $B^0$ mesons is shown
in Fig.~\ref{fig:LHCbbeauty13}.
As in the case of the charm, there is excellent agreement between
the {\sc\small POWHEG} and {\sc\small {\rm a}MC@NLO} calculations 
within the LHCb acceptance.
%

The tabulation of the results shown in Figs.~\ref{fig:LHCbcharm13}
and~\ref{fig:LHCbbeauty13} are provided in Appendix~\ref{appendix},
in particular in Tables~\ref{tab:Ddiff} (for $D^0$ mesons)
and~\ref{tab:Bdiff} (for $B^0$ mesons).

\subsection{Predictions for the ratio between the 13 and 7 TeV cross-sections}
\label{sec:appendix2}
In addition to differential cross section measurements, it will also become possible 
to measure the ratio of differential cross sections performed at 13 and 7~TeV when
the 13~TeV data is available. 
As discussed in Ref.~\cite{Mangano:2012mh},
measurements of the ratio of cross-sections at different centre-of-mass 
energies are well motivated as many theoretical uncertainties, 
such as scale uncertainties, mass dependence, and 
fragmentation/branching fractions cancel in the ratio to 
a good approximation. In addition, many experimental uncertainties
also cancel in such ratios which allows stringent tests of the
Standard Model to be performed.
The relevance of the ratio of 13 over 7 TeV heavy quark production cross-sections
at LHCb for PDF studies has also been recently emphasised in Ref.~\cite{CMN},
in a study of the various theoretical uncertainties associated to charm and bottom
production in the forward region.

On the other hand, PDF uncertainties do not cancel completely, because of the different
kinematical range covered by the measurements at the two centre-of-mass energies,
and thus these ratio measurements provide in principle useful
PDF discrimination power.
This idea has been implemented already by a number of LHC analyses, like the ATLAS measurement
of the ratio of 7 TeV over 2.76 TeV jet cross-sections~\cite{Aad:2013lpa}
and the CMS measurement of the ratio
of 8 TeV over 7 TeV Drell-Yan distributions~\cite{CMS:2014jea}.

\begin{figure}[t]
\centering 
\includegraphics[scale=0.39]{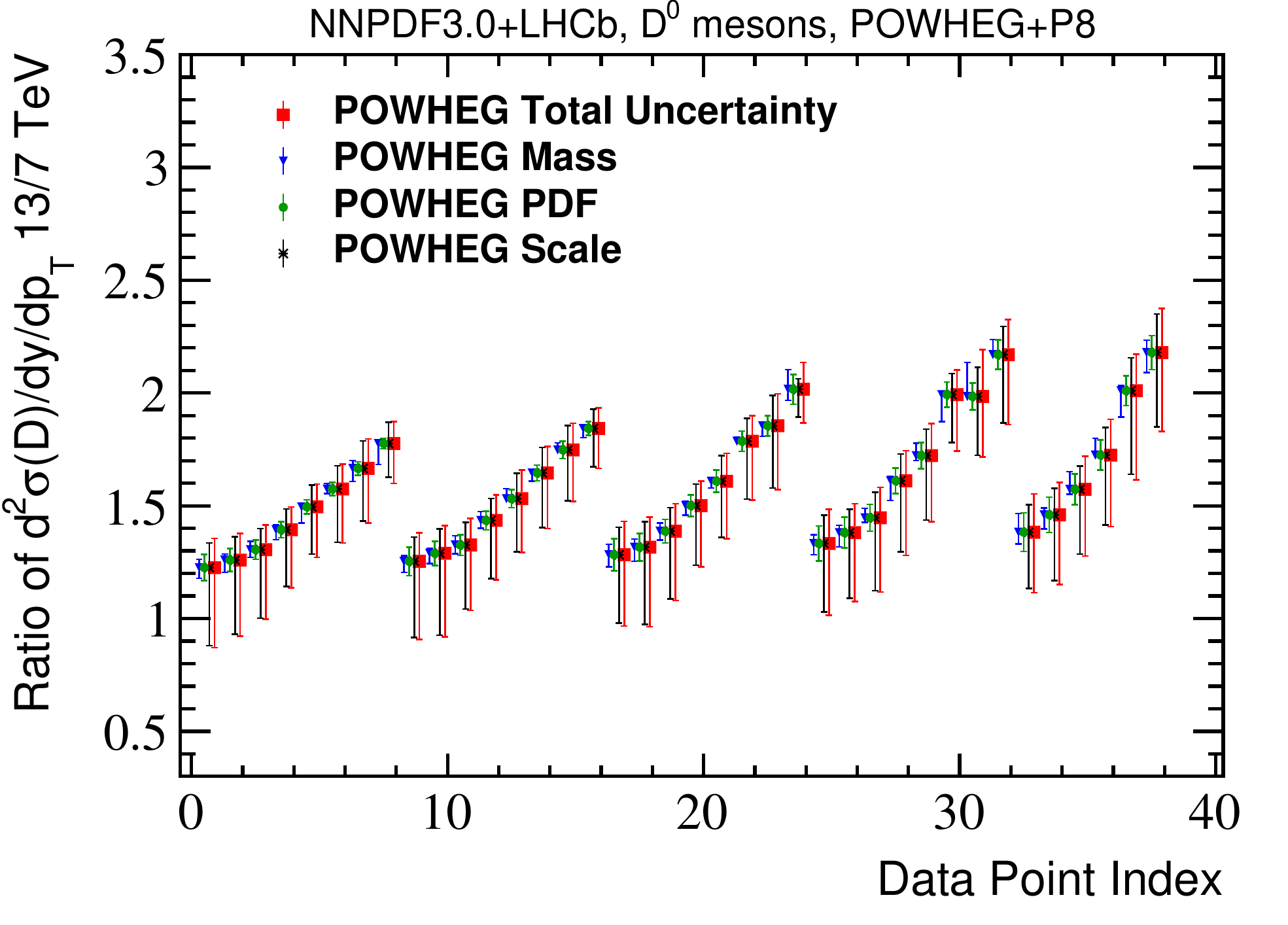}
\includegraphics[scale=0.39]{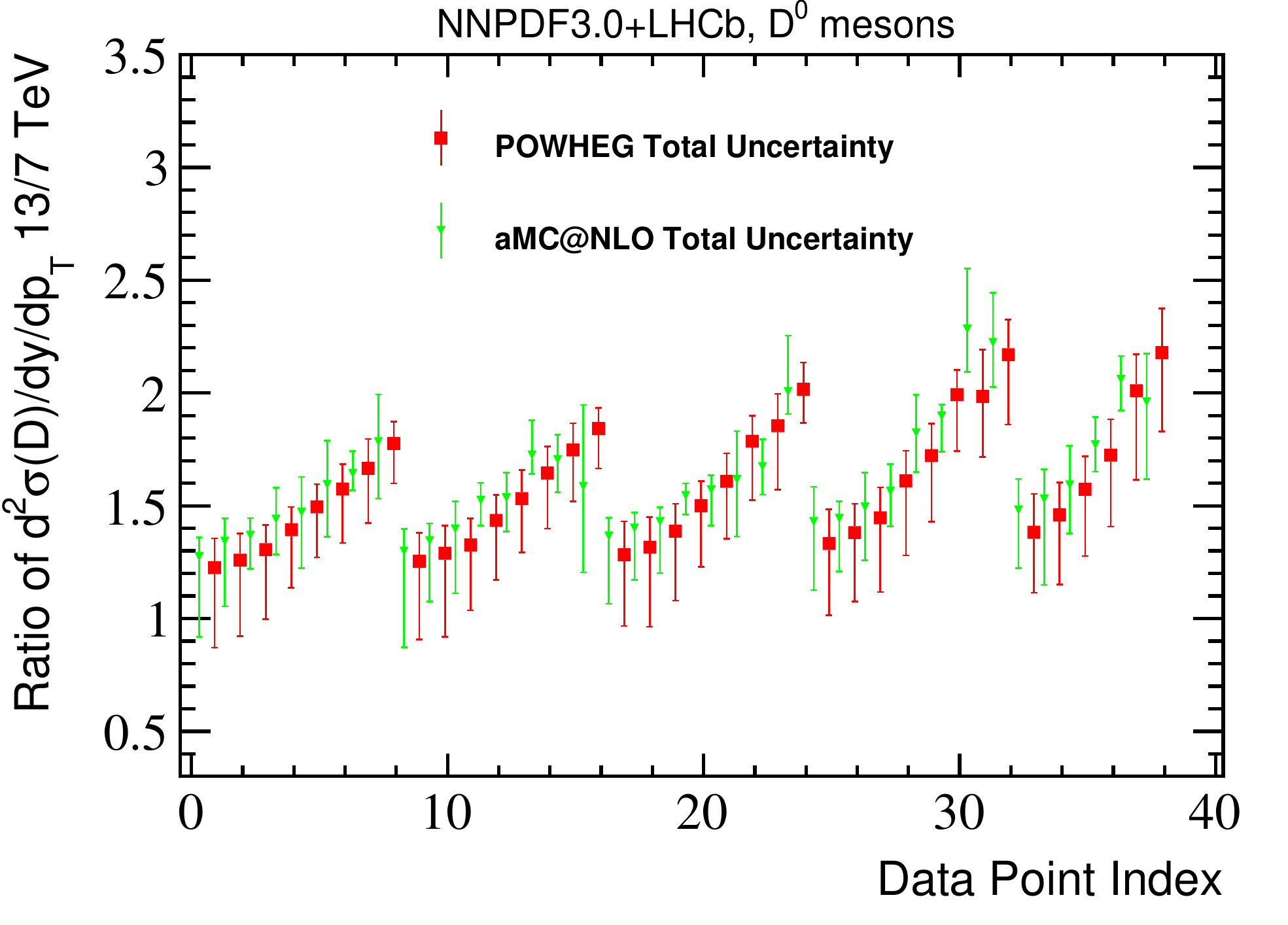}
\caption{\small Left plot: predictions for the ratio of differential cross-sections
  $R_{13/7}^{D^0}$, Eq.~(\ref{eq:ratioD0}), for the production of $D^0$ mesons between 13 TeV and 7 TeV,
  computed using {\sc\small POWHEG} and NNPDF3.0+LHCb.
  Results are ordered in increasing bins in rapidity, in within each, in increasing
  bins of $p_T$.
  The total theoretical uncertainty in the ratio is decomposed into its various sources: scale, PDF
  and charm quark mass variations.
  Right plot: comparison of the predictions for $R_{13/7}^{D^0}$ between {\sc\small POWHEG}
  and {\sc\small {\rm a}MC@NLO}, for central values and for the total theory uncertainties.
}
\label{fig:LHCbcharmRatio}
\end{figure}

In Fig.~\ref{fig:LHCbcharmRatio} we show the predictions for the ratio of differential
cross-sections for $D^0$ production between 13 TeV and 7 TeV, defined as
\be
\label{eq:ratioD0}
R_{13/7}^{D^0}(y^D,p_T^D)\equiv\frac{d^2\sigma(D^0)(y^D,p_T^D,13~{\rm TeV})}{dy^Ddp_T^D}\Bigg/
\frac{d^2\sigma(D^0)(y^D,p_T^D,7~{\rm TeV})}{dy^Ddp_T^D} \, ,
\ee
where the same binning as in the 7 TeV LHCb measurement has been assumed.
In the left plot we show the results computed with {\sc\small POWHEG} and
the NNPDF3.0+LHCb PDF set, for each of the bins of the 7 TeV measurement
(data points are ordered in increasing bins of rapidity, and within each of these five
rapidity bins, in increasing bins of $p_T$). 
 The central value of the ratio $R_{13/7}^{D^0}$ varies between 
 1.20 and 2.2 for increasing values of $p_T$ and more forward 
 rapidity bins, where the opening of phase space between 13 TeV 
 with respect to 7 TeV is more important.

In the left plot of Fig.~\ref{fig:LHCbcharmRatio} we have separated the total theory uncertainty into
the individual contributions from scales, PDFs and charm mass to highlight their importance.
We see that the total uncertainty in  $R_{13/7}^{D^0}$ varies between 10\% and 30\%,
depending on the specific bin, and that scale variation is found to dominate the total 
uncertainty in $R_{13/7}^{D^0}$. Note however the
substantial cancellation of scale uncertainties as compared to the absolute differential cross-sections shown in
Fig.~\ref{fig:LHCbcharm13}.
In Appendix~\ref{appendix} we provide a tabulation of the results of Fig.~\ref{fig:LHCbcharmRatio},
which will be useful for comparison if the ratio $R_{13/7}^{D^0}$ is measured in the upcoming
LHCb 13 TeV analysis.
In the same appendix we also quantify the reduction of PDF uncertainties in $R_{13/7}^{D^0}$
by comparing the predictions using the original NNPDF3.0 set with our baseline
predictions obtained with NNPDF3.0+LHCb.
The substantial reduction of PDF uncertainties in $R_{13/7}^{D^0}$, thanks to
the constraints from the 7 TeV normalised charm cross-sections derived in Sect.~\ref{sec:rw},
improve the robustness of our theory prediction for $R_{13/7}^{D^0}$.
Conversely, the measurement of $R_{13/7}^{D^0}$ should provide important PDF discrimination
power, and it would be interesting to verify the consistency of the constraints
on the small-$x$ gluon from $R_{13/7}^{D^0}$ from those that we have derived
from the normalised 7 TeV data.

To validate the cancellation of the theoretical systematics in the {\sc\small POWHEG} calculation, we 
have also computed the ratio with the {\sc\small {\rm a}MC@NLO} calculation.
The comparison of these two calculations, including their total uncertainties, is shown
in the right plot of Fig.~\ref{fig:LHCbcharmRatio}.
Reasonable agreement is found, both for the central values and for the uncertainties.
In particular, for most of the bins, the central predictions for $R_{13/7}^{D^0}$ agree within 10\% at most.
This agreement should be considered satisfactory especially taking into account the very large
theory uncertainties in the absolute distributions.

\begin{figure}[t]
\centering 
\includegraphics[scale=0.39]{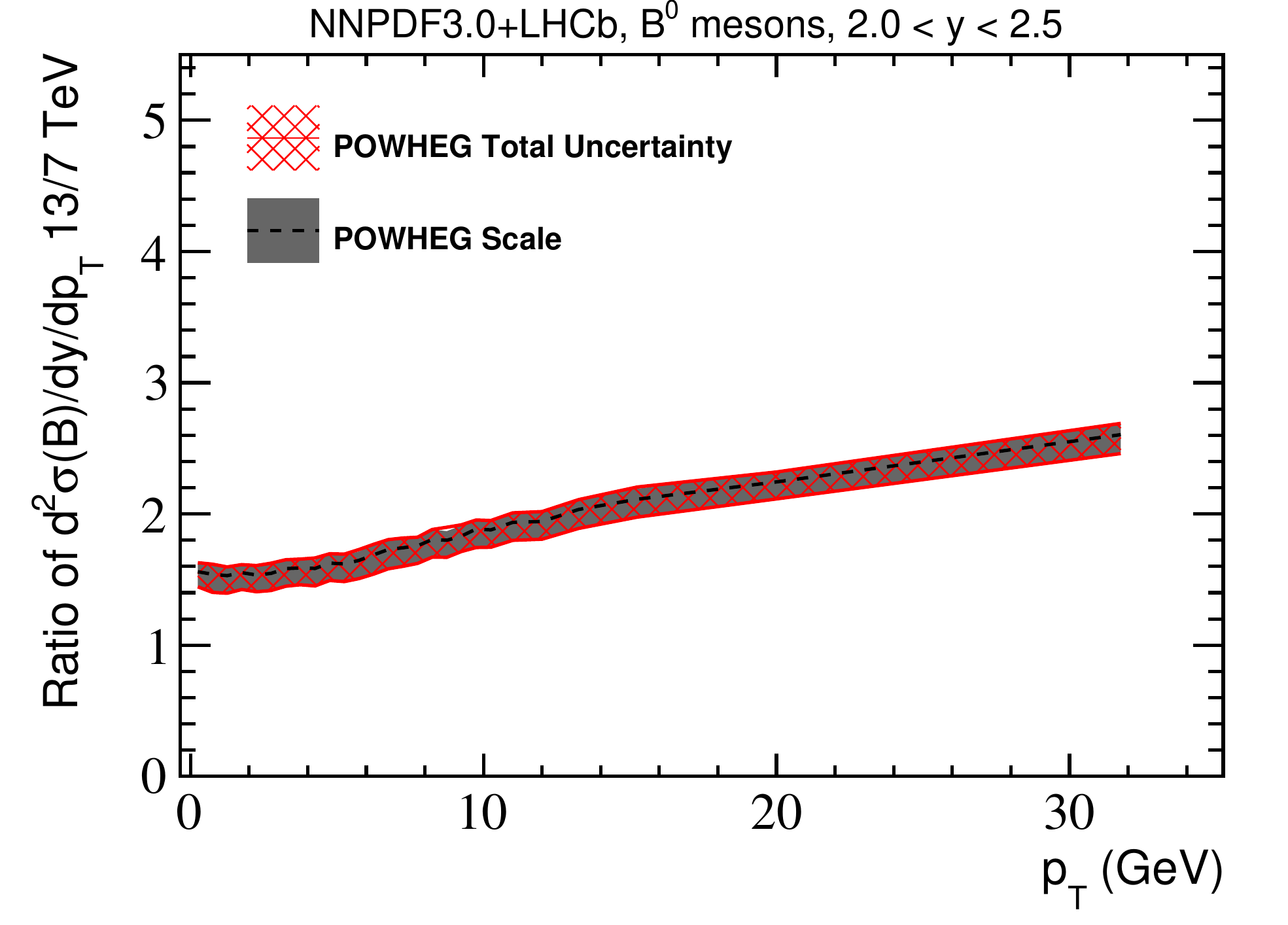}
\includegraphics[scale=0.39]{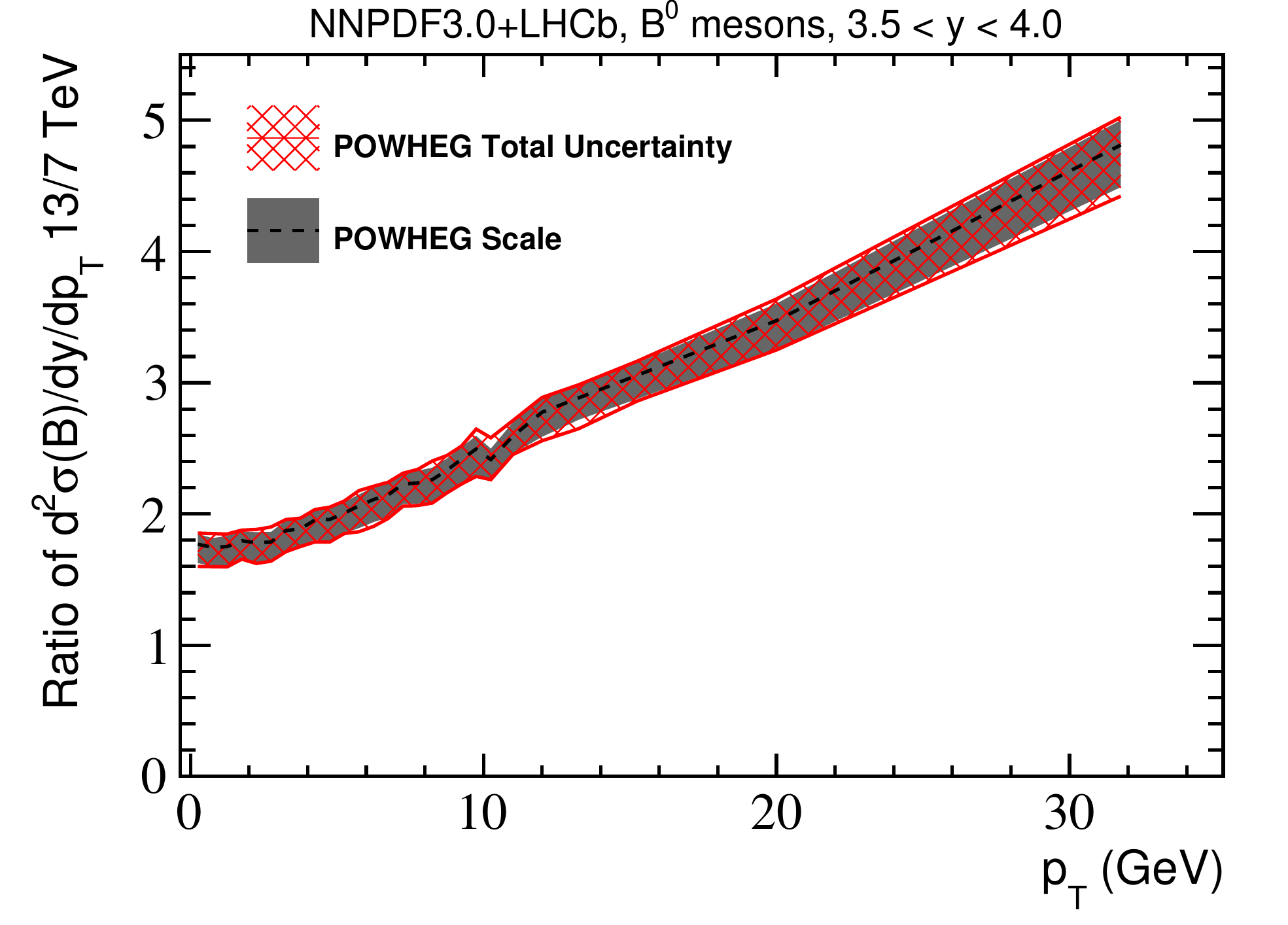}
\caption{\small Theoretical predictions for the ratio $R_{13/7}^{B^0}$ Eq.~(\ref{eq:ratioB0})
  between $B^0$ meson distributions between 13 and 7 TeV.
  Results have been computed with {\sc\small POWHEG} using NNPDF3.0+LHCb.
  We show the predictions for two representative bins in rapidity, one central
  (left plot) and the other forward (right plot), as a function of $p_T^B$.
  The total theory uncertainty (hatched band) is compared with the scale
  uncertainty (solid band).
}
\label{fig:LHCbbeautyRatio}
\end{figure}

Next we provide the corresponding predictions for the ratio of $B$ meson differential
distributions between 13 TeV and 7 TeV, defined as
\be
\label{eq:ratioB0}
R_{13/7}^{B^0}(y^B,p_T^B)\equiv\frac{d^2\sigma(B^0)(y^B,p_T^B,13~{\rm TeV})}{dy^Bdp_T^B}\Bigg/
\frac{d^2\sigma(B^0)(y^B,p_T^B,7~{\rm TeV})}{dy^Bdp_T^B} \, ,
\ee
for the case of $B^0$ mesons, which we choose for illustrative purposes.
In Fig.~\ref{fig:LHCbbeautyRatio}
we show the theoretical predictions for the ratio $R_{13/7}^{B^0}$
computed with {\sc\small POWHEG} using NNPDF3.0+LHCb for two representative bins in rapidity, one central
  (left plot) and one forward (right plot), as a function of $p_T^B$.
  The total theory uncertainty (hatched band) is compared with the scale
  uncertainty (solid band).
  We have verified that the results for $R_{13/7}^{B^0}$ obtained with {\sc\small {\rm a}MC@NLO} are fully
  consistent the {\sc\small POWHEG} calculation.
  In the results of Fig.~\ref{fig:LHCbbeautyRatio}, the same binning as in the 7 TeV measurement
  has been used~\cite{Aaij:2013mga}.

  From Fig.~\ref{fig:LHCbbeautyRatio} we see that $R_{13/7}^{B^0}$ varies between 1.3 at central rapidities
  at low $p_T$ to almost 5 at forward rapidities and large $p_T$, for the same reasons as
  $R_{13/7}^{D^0}$.
  The total uncertainty in  $R_{13/7}^{B^0}$ ranges between 5 and 10\%, depending on the specific bin,
  and is dominated by the scale uncertainty (but only due to using the improved NNPDF3.0+LHCb set).
  As in the case of charm production, in Appendix~\ref{appendix} we tabulate our predictions for $R_{13/7}^{B^0}$,
  that can be used to compare the the upcoming LHCb measurement.
  The corresponding predictions for other $B$ meson species are available upon request.

\subsection{Predictions for inclusive fiducial cross-sections at 13 TeV}
\label{sec:predictionsforratio}

In addition to the double differential distributions,
it is also useful to provide predictions for the charm and bottom inclusive cross-section, that is,
the cross-sections measured
within the full LHCb fiducial region.
In the case of $D$ mesons, the fiducial region is defined as
\be
\label{eq:fiducial}
 0 \le p_T^D \le 8~{\rm GeV} \, , \qquad 2.0 \le y^D \le 4.5 \, ,
 \ee
 while the corresponding fiducial region for the production of $B$ mesons is
 defined by
 \be
\label{eq:fiducialB}
 0 \le p_T^B \le 40~{\rm GeV} \, , \qquad 2.0 \le y^B \le 4.5 \, .
 \ee
 In order to compute the 13 TeV predictions for the charm and bottom inclusive cross-sections
 in the fiducial region, there are two possible strategies that can be adopted, namely
 \begin{itemize}
 \item integrating the {\sc\small POWHEG} calculation for the absolute double differential cross-sections,
   shown in Fig.~\ref{fig:LHCbcharm13}, for the acceptance in Eq.~(\ref{eq:fiducial}), or instead
\item using the theoretical predictions for the ratios $R_{13/7}^D$ and $R_{13/7}^B$ to rescale the corresponding
  7 TeV  LHCb inclusive measurements reported in~\cite{Aaij:2013mga,Aaij:2013noa}.
\end{itemize}
The main advantage of the second option is that theoretical uncertainties are substantially reduced
in the ratios $R_{13/7}^D$ and $R_{13/7}^B$ as compared to
the absolute cross-sections,
allowing a reasonably accurate extrapolation
for the 13 TeV inclusive cross-sections, with precision comparable to that 
expected for the corresponding experimental measurement.

Let us illustrate how the two strategies compare in the case of $D$ meson production.
For simplicity, we will show the results for $D^0$ mesons but the same ideas apply to the other $D$
mesons.
In this case, the prediction for the inclusive ratio, with the total associated theory uncertainty, is given by
\beq
\label{eq:r137}
R_{13/7}^{D^0}({\rm th,incl}) = 1.39 \,~ ^{+0.12 \,(8.3\%)}_{-0.29 \,(20.5\%)} \,.
\eeq
This can be combined with the 7 TeV LHCb inclusive measurement~\cite{Aaij:2013mga} in the fiducial
region for
$D^0$ mesons
\be
\sigma^{D^0}_{\rm 7TeV}({\rm LHCb,incl}) = 1661 \pm 129\,~(\pm 7.8\%) \,~{\rm \mu b} \ ,
\ee
to obtain an accurate prediction for the corresponding 13 TeV inclusive cross-section in the same fiducial region.
This leads to
\be
\label{eq:10}
\sigma^{D^0}_{\rm 13TeV}({\rm th,incl}) = \sigma^{D^0}_{\rm 7TeV}({\rm LHCb,incl})\cdot
R_{13/7}^{D^0}({\rm th,incl})  = 2236^{+308 \, (14\%)}_{-521 \, (23\%)} \,~{\rm \mu b} \ ,
\ee
where the theoretical uncertainty from $R_{13/7}^{D^0}$ is slightly larger than
that of the 7 TeV measurement, and dominates the precision of the prediction for $\sigma^{D^0}_{\rm 13TeV}$
performed in this way.
In Eq.~(\ref{eq:10}) we have added in quadrature the theory uncertainties from
$R_{13/7}^{D^0}$ with the experimental uncertainties of the LHCb measurement.

In Table~\ref{tab:Dsigma} the prediction for the inclusive cross-section
$\sigma^{D^0}_{\rm 13TeV}$ obtained using the 7 TeV measurement and the calculation of $R_{13/7}^D$ is compared
to the corresponding result computed from the integral of the absolute differential distributions.
The advantage of the ratio strategy is apparent: when integrating the absolute distributions, the prediction is affected by
large theory uncertainties up to 200\% which render the comparison with the much more accurate
experimental measurement not very informative.
On the other hand, our prediction obtained using $R_{13/7}^D$ has a 10-20\% accuracy, comparable to that of
the upcoming Run II LHCb measurement, and therefore should provide interesting information for the comparison between
data and theory in a hitherto unexplored kinematical region.
In Table~\ref{tab:Dsigma} we also provide the predictions for the inclusive charm pair production
cross-section using the two methods, obtained from rescaling the meson-level result by the branching
fraction of charm into $D^0$ mesons,
\be
\label{eq:rescaling}
\sigma^{c\bar{c}}_{\rm 13TeV}({\rm th,incl}) = \sigma^{D^0}_{\rm 13TeV}({\rm th,incl}) / \left(2f\lp c\to D^0\rp\right) \, .
\ee
This prediction is useful to compare with parton-level predictions of charm production, which do not account
for the fragmentation of charm quarks into $D$ mesons.

\renewcommand*{\arraystretch}{1.5}
\begin{table}[h]
\centering
\begin{tabular}{@{} c|rcl|rcl@{}}
  \hline
  13~TeV			&	\multicolumn{3}{c|}{$D^0$}			& \multicolumn{3}{c}{$c\bar{c}$} \\ \hline\hline
  	$\sigma_{\rm 13TeV}({\rm th,incl}) (\mu b)$ (from ratio)	&	2236 & \hspace{-0.7cm} & $^{+308 \, (14\%)}_{-521 \, (23\%)}$ &1979 & \hspace{-0.7cm} & $^{+249 \, (13\%)}_{-447 \, (23\%)}$ \\ \hline
	$\sigma_{\rm 13TeV}({\rm th,incl})	(\mu b)$ (from abs)	&	1097 & \hspace{-0.7cm} & $^{+2082 \, (190\%)}_{-896 \, (82\%)}$ & 970 & \hspace{-0.7cm} & $^{+1843 \, (190\%)}_{-793 \, (82\%)}$ \\ \hline
     \end{tabular}
  \caption{\small Predictions for the inclusive $D^0$ production cross-section in the fiducial region Eq.~(\ref{eq:fiducial}) at
    13 TeV
    using the two methods discussed in the text (integrating the absolute distributions and rescaling the 7 TeV
    LHCb measurement with the ratio $R^D_{13/7}$).
    Predictions are also provided for the corresponding $c\bar{c}$ cross-sections using
    Eq.~\ref{eq:rescaling}.
  }
  \label{tab:Dsigma}
\end{table}

The same strategies can be applied to obtain accurate predictions for the inclusive
$B$ meson production cross-sections at 13 TeV in the fiducial region
defined by Eq.~(\ref{eq:fiducialB}).
For simplicity we restrict ourselves to $B^0$ mesons, though
the same method also applies to all other  $B$ mesons that will
be measured at Run II.
The first method, integrating the absolute differential cross-sections from
Fig.~\ref{fig:LHCbbeauty13} in this fiducial region leads to the following
prediction
\beq
\label{eq:intB}
\sigma^{B^0}_{\rm 13TeV}({\rm th,incl})	(\mu b) ({\rm from~abs}) = 55.07 \,~ ^{+28.77 \,(52.3\%)}_{-20.76 \,(37.7\%)} \,{\rm \mu b} \,.
\eeq
Now, using the prediction for the  ratio of inclusive cross-sections between
13 and 7 TeV for $B^0$ mesons,
\beq
\label{rescalingB}
R^{B^0}_{13/7}({\rm th,incl}) = 1.84 \,~^{+0.08 \,(4.1\%)}_{-0.12 \,(6.8\%)} \,,
\eeq
to rescale the 7 TeV LHCb measurements~\cite{Aaij:2013noa} in this fiducial region,
\be
\sigma^{B^0}_{\rm 7TeV}({\rm LHCb,incl}) = 38.1 \pm 6.0\,~(\pm 15.6\%) \,~{\rm \mu b} \, ,
\ee
we obtain the following prediction for the 13 TeV fiducial $B^0$ production cross-section
\beq
\sigma_{\rm 13TeV}({\rm th,incl}) ({\rm from~rat}) =
\sigma^{B^0}_{\rm 7TeV}({\rm LHCb,incl}) \cdot R^{B^0}_{13/7}({\rm th,incl})=
70.02\,^{+11.42 \,(16.3\%)}_{-12.03 \,(17.2\%)} \,\mu b \, .
\eeq
In the above procedure, the theoretical uncertainties from $R_{13/7}^B$ and the experimental
uncertainties from the 7 TeV measurement have been added in quadrature.
In this case, the advantage of using  $R_{13/7}^B$ are even more marked: as the theoretical uncertainties
of the ratio are smaller than those of the 7 TeV LHCb inclusive measurement, the extrapolation from
7 to 13~TeV is essentially limited by the precision of the 7 TeV cross-section, with very small
theoretical uncertainty in the procedure.
Note that using the ratio strategy our theoretical prediction for $\sigma_{\rm 13TeV}$ leads to a prediction 
with uncertainties
which are around three times smaller as compared to the prediction obtained from the integration of the
absolute distributions, Eq.~(\ref{eq:intB}).
Similar improvements can be observed for other meson species.
Note also that in the case of $B$ mesons, the fragmentation is essentially the same for all the meson
types, and thus the same rescaling Eq.~(\ref{rescalingB}) can be applied to all the $B$ meson species. 
For example, we find $R^{B^{\pm}}_{13/7}({\rm th,incl}) = R^{B^0}_{13/7}({\rm th,incl})$ to the precision 
provided in Eq.~(\ref{rescalingB}).

In summary, in this section we have provided accurate predictions for the 13 TeV fiducial cross-sections
for the production of $D$ and $B$ mesons at LHCb, using the ratios $R^{D}_{13/7}$ and $R^{B}_{13/7}$ to
extrapolate the 7 TeV measurements.
The robustness of this extrapolation is illustrated by the fact that, upon rescaling by the ratio,
the corresponding 13 TeV prediction has uncertainties which are at most two times larger than than the 
precision of the 7~TeV data.
Note that the predictions from the absolute distributions have significantly larger uncertainties as compared
to the foreseen prediction of the 13 TeV uncertainties, particularly in the case of charm, where theory uncertainties
for the fiducial cross-section can be as large as $\sim 200\%$
(see Table~\ref{tab:Dsigma}).

%% file: sec-results.tex
\section{QCD predictions for charm-induced neutrino production}
\label{sec:results}

The dominant background for
the detection of ultra-high-energy neutrinos from astrophysical
sources in experiments like IceCube arises from the flux of neutrinos
originating from the prompt decay of energetic charmed mesons produced
in cosmic ray collisions in the upper atmosphere.
We now
provide state-of-the-art pQCD predictions for
the cross-sections of charm-induced neutrino
production.
These cross-sections are an important ingredient of the
full calculation of prompt neutrino event rates at
IceCube, which is beyond
the scope of this paper.

As compared to previous works~\cite{Lipari:1993hd,Pasquali:1998ji,Enberg:2008te,Gondolo:1995fq,Martin:2003us,Gelmini:1999ve,Bhattacharya:2015jpa},
here we want to fully exploit the
flexibility of our approach for the computation of the charm
production cross-sections, based on NLO Monte Carlo event generators.
We can derive a robust prediction for the primary neutrino
flux arising from the decays of charmed mesons produced in cosmic ray
collisions from pQCD, eliminating the need of
model assumptions, and being able to 
estimate all the associated sources of theoretical uncertainties in our
calculation.
 Being fully differential,
      our calculation of the prompt neutrino flux can be processed in cascade
      codes and in neutrino telescopes detector simulation software with
    arbitrary selection cuts.

To achieve this goal, using the results of Sects.~\ref{sec:pqcd}
and~\ref{sec:rw}
we have computed
\be
\label{eq:dsigmadEnu}
\frac{d\sigma(pN \to \nu X;E;E_{\nu})}{dE_{\nu}} \, , 
\ee 
that is, the differential cross-section for the production of
neutrinos from the decays of charmed hadrons in proton-nucleon
collisions, as a
function of the neutrino energy $E_{\nu}$, for different values of the
incoming cosmic ray energy $E$.\footnote{Eq.~(\ref{eq:dsigmadEnu})
  accounts only for the flux of {\it primary}  prompt neutrinos, those
  produced in the first interaction of the cosmic ray with
  air nuclei.
  To compute the complete flux one should also include the contribution
  from secondary production solving the cascade equations.
}

To compute the neutrino energy distribution, Eq.~(\ref{eq:dsigmadEnu}),
using {\sc\small POWHEG} and {\sc\small {\rm a}MC@NLO}, charmed
hadrons are first decayed using the {\sc\small Pythia8} shower,
summing over all hadron species and neutrino flavours.
Subsequently,
a Lorentz boost is applied for the conversion of the neutrino
energy distribution from the centre-of-mass frame, where the
prediction of MC event generators is provided, to the laboratory
frame.
The magnitude of this
boost is determined by the incoming cosmic ray energy.

Results have been computed for a number of values of the
incoming cosmic ray energy $E$ between
$E=10^3$ GeV to $E=100$ PeV, corresponding to centre of mass energies
$\sqrt{s}=\sqrt{2m_NE}$ ranging from 44 GeV to 14 TeV.
As discussed before, we emphasize the overlap between
the kinematic region
crucial for neutrino telescopes
and that of the LHCb charm production data.

The fact that cosmic rays collide
with air nucleus rather than with isolated (isoscalar) nucleons can
be accounted for by rescaling the cross-section for
$pN$ collisions with the mean atomic number
of air nuclei $\la A\ra\simeq 14.5$, that is, to  good
approximation we can write
\be
\label{eq:nuclear}
\sigma(pA \rightarrow c\bar{c}X) \simeq \la A \ra \cdot \sigma(pN \rightarrow
c\bar{c}X) \, .  
\ee
Eq.~(\ref{eq:nuclear}) assumes that nuclei can be treated
as an incoherent sum of their protons and neutrons, and that nuclear
corrections to the nucleon PDFs can be
neglected as compared
other theoretical uncertainties in the calculation.

The assumption of neglecting
nuclear shadowing
in charm production is justified by the recent CMS measurements of $B$
mesons in proton-lead collisions at $\sqrt{s_{NN}}=5$
TeV~\cite{CMS-PAS-HIN-14-004}, which cover a similar kinematical range
as for charm production in cosmic rays, and that show no evidence for
suppression induced by nuclear PDFs.
Moreover, available sets of nuclear
PDFs~\cite{Eskola:2009uj,deFlorian:2003qf,deFlorian:2011fp}
are unconstrained at 
small-$x$ due to the absence of experimental data, and thus
cannot be used reliably in our calculation.
In addition,
a recent calculation of forward $D$ production at $\sqrt{s_{NN}}=5$~TeV
incorporating the EPS09 nuclear PDF modifications~\cite{Gauld:2015lxa}
indicates that a cross section suppression of at most $\simeq 10\%$ can expected 
in proton-lead collisions, within substantial
uncertainties.
%

\begin{figure}[t]
  \centering
  \includegraphics[scale=.38]{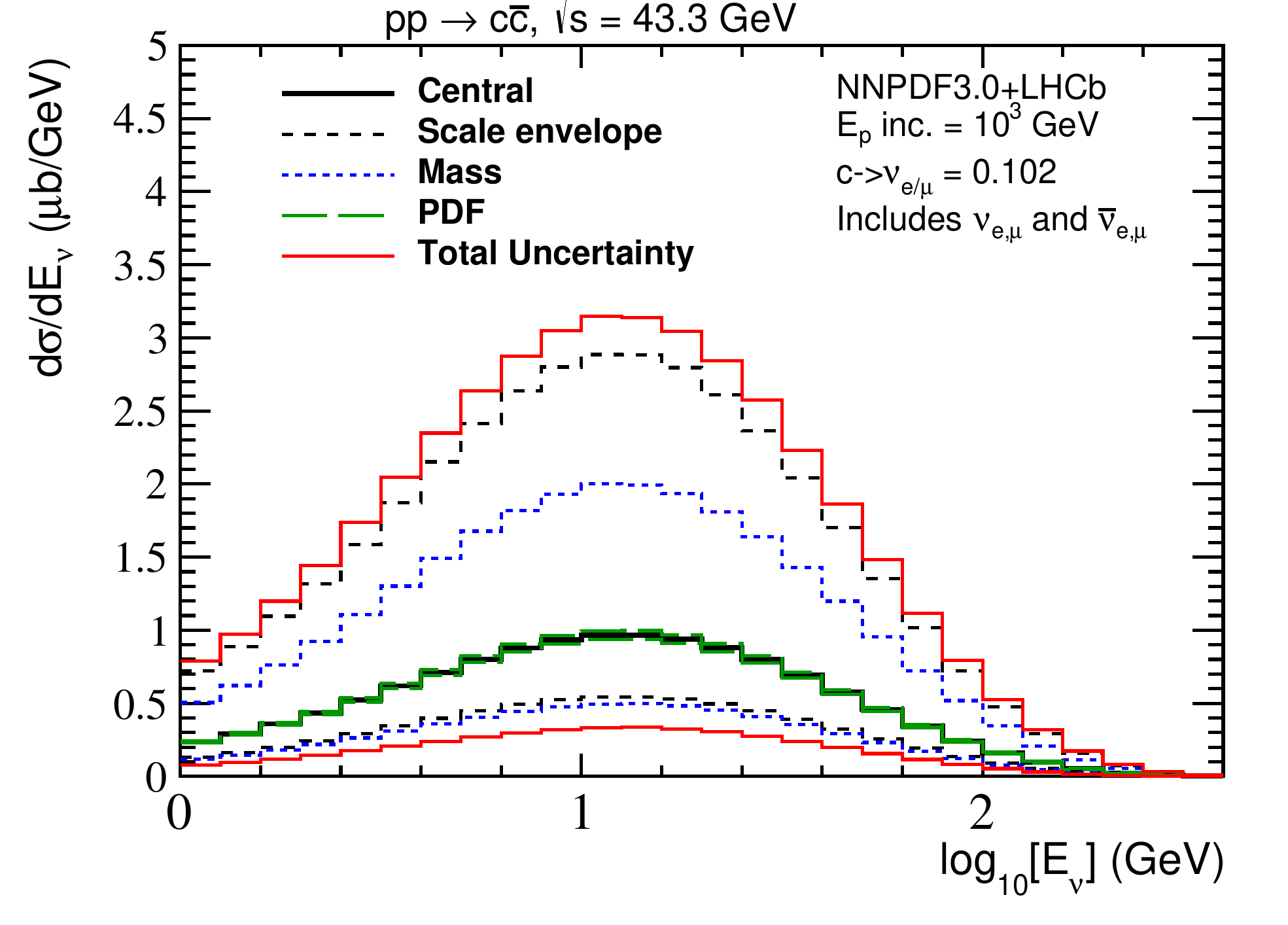}
\includegraphics[scale=.38]{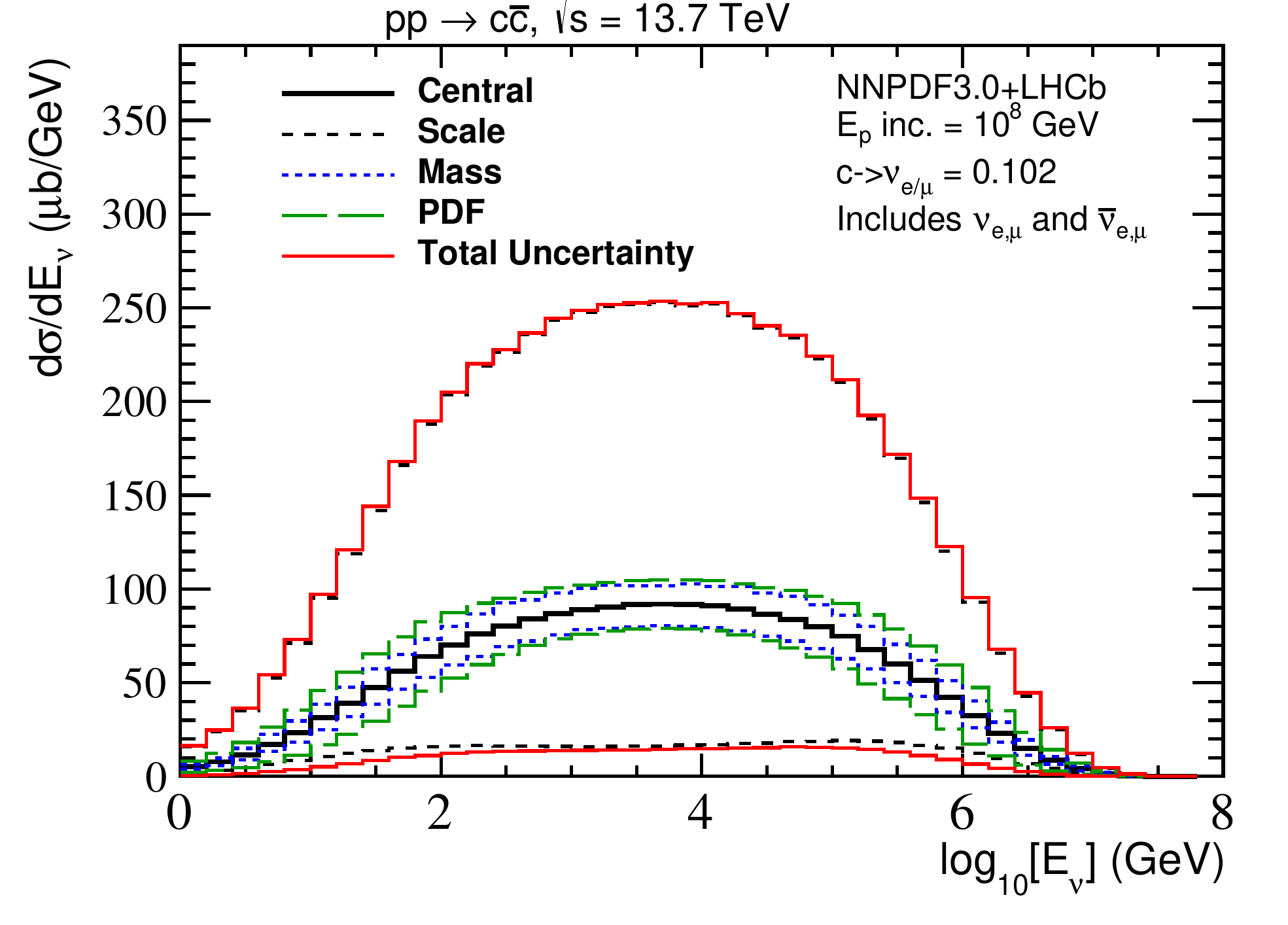}
\caption{\small The differential cross-section for the production of
  neutrinos from charm decay in $pp$ collisions,
  Eq.~\ref{eq:dsigmadEnu}, as a function of the neutrino energy,
  computed with {\sc\small POWHEG}.
  The results are provided for two values of the incoming cosmic ray
  energy, $E=10^3$ GeV (left plot) and $E=10^8$ GeV (right plot).
  The input PDF set is NNPDF3.0NLO+LHCb.
  We show the central prediction as well as the scale, PDF and $m_c$
  uncertainties, as well as the overall theoretical uncertainty band
  computed from adding in quadrature the three independent theory
  errors.  }
\label{fig:distribution}
\end{figure}

In our approach the production of charm quarks, their
hadronisation into charmed mesons and their subsequent decays into
neutrinos are completely accounted for in the matrix element
calculation matched to the parton shower.
We can therefore obtain  exact results for the various differential
distributions relevant for prompt neutrino production.
We emphasise that the the modelling of charm production and decay in
{\sc\small Pythia8} has been validated by LEP data as well as hadron 
collider data, see Refs.~\cite{Skands:2010ak,Skands:2014pea} and
references therein.

These differential cross-sections Eq.~(\ref{eq:dsigmadEnu}) have been
computed in a range of values of $E$ and $E_{\nu}$ and then suitably
interpolated.
For each point in $\lp E;E_{\nu}\rp$, we have determined the relevant
theoretical uncertainties from scales, PDFs, and $m_c$ variations.
Our calculations use the improved NNPDF3.0+LHCb which includes the 
constraints from the 7 TeV charm data.
A representative sample of our predictions are provided in
Fig.~\ref{fig:distribution}, where we show the differential cross-section
for the production of neutrinos from charm decay in $pp$ collisions,
Eq.~(\ref{eq:dsigmadEnu}), as a function of the neutrino energy,
computed with the {\sc\small POWHEG} calculation.
  Results are shown for two values of the incoming cosmic ray
  energy, $E=10^3$ GeV (left plot) and $E=10^8$ GeV (right plot).
   We show the central prediction as well as the individual contributions 
   from scale, PDF and $m_c$ uncertainties, as well as the overall theoretical uncertainty band
  computed from adding these uncertainties in quadrature.
  We see that at the highest energies, $E=10^8$ GeV, the total
  uncertainty band is dominated by scale variations, while PDF
  uncertainties are under control thanks to the constraints
  from the LHCb charm production data.
  We stress that while NLO QCD scale uncertainties
  are still large, up to a factor three, recent work towards
  the NNLO differential distributions for heavy quark
  production~\cite{Czakon:2014xsa,Czakon:2015pga}
  will provide a reduction of these higher-order
  uncertainties.

\begin{figure}[t]
  \centering
\includegraphics[scale=.60]{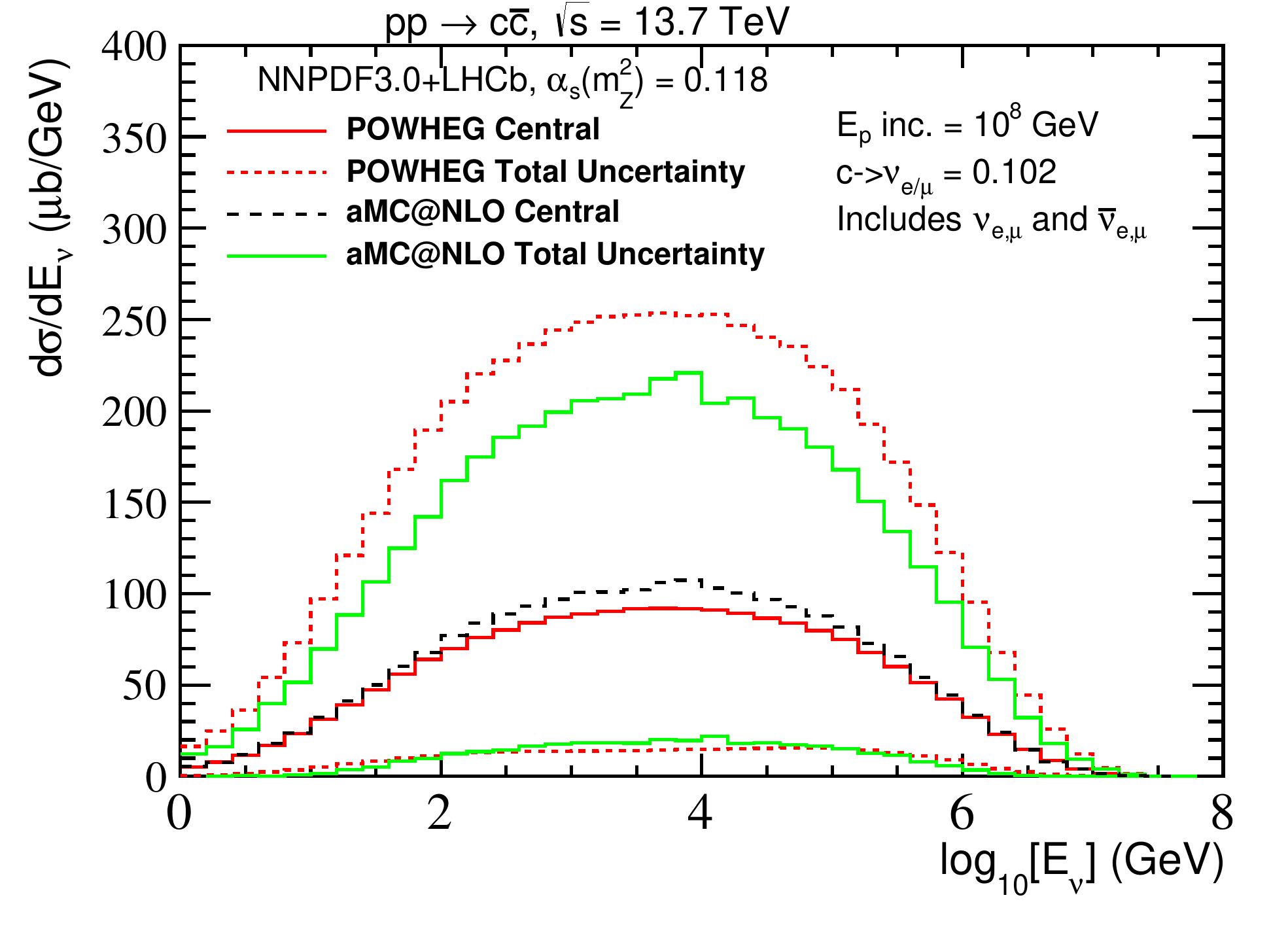}
\caption{\small Same as the right plot of Fig.~\ref{fig:distribution},
  now comparing the predictions of {\sc\small POWHEG} with those of
  {\sc\small {\rm a}MC@NLO}.
  The same theory settings are used in the two calculations.
  Only the central curve total theory uncertainty bands are shown for
  the predictions obtained with the two event generators.  }
\label{fig:distribution2}
\end{figure}

A powerful cross-check of the robustness of the predictions shown
in Fig.~\ref{fig:distribution} is provided by the fact that comparable
results are obtained using either {\sc\small POWHEG} or {\sc\small
  {\rm a}MC@NLO}, both for the central prediction and for the
upper and lower ranges of the total theory uncertainty band, as shown
in Fig.~\ref{fig:distribution2}, when the same theory settings are
used in the two calculations.
  Let us emphasise that two completely independent
  codes are used, with different underlying matrix element calculations 
  and different matching to the parton showers, so this agreement is
  an indication of the robustness of the pQCD
  predictions for the charm-induced neutrino production
  cross-sections presented here.

\begin{figure}[t]
  \centering
  \includegraphics[scale=.55]{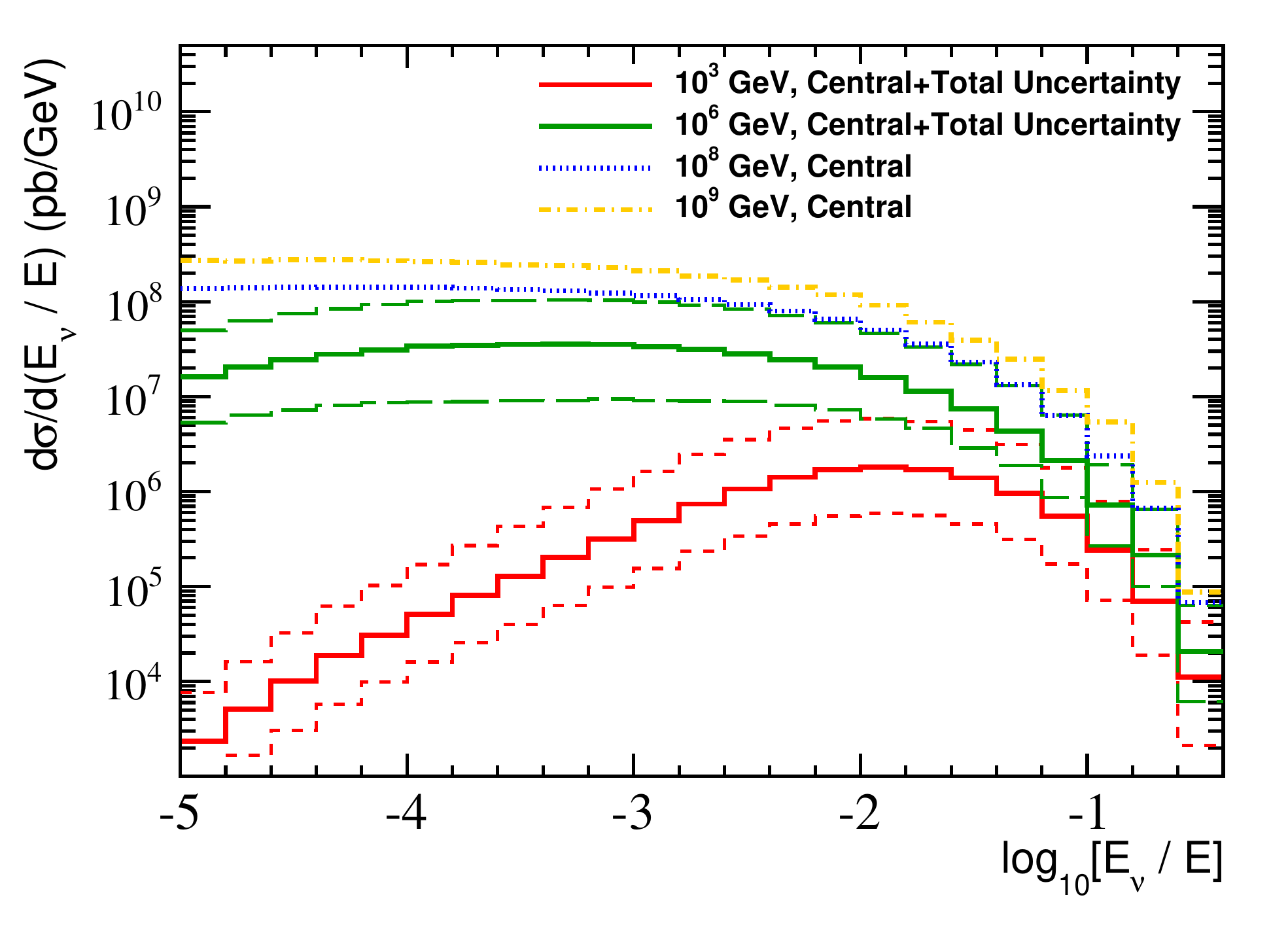}
  \caption{\small The dependence on the incoming cosmic ray energy $E$ of the
    prompt neutrino production cross-section $d\sigma/dE_{\nu}$,
    plotted as a function
    of $z\equiv E_{\nu}/E$,
Eq.~(\ref{eq:dsigmadEnu2}),
which allows to compare calculations
for different values of $E$.
    Results are shown for $E=10^3$ and $E=10^6$ GeV (both central values and total theoretical
    uncertainty) and then for  $E=10^8$ and $E=10^9$ GeV (only central values).
    The cross-sections fall steeply as one approaches the kinematical boundary $z\to 1$.
  }
\label{fig:dsigdz_unc}
\end{figure}

It is also interesting to study  the dependence of our results for the charm-induced
neutrino production  cross-sections as a function
of the incoming cosmic ray energy $E$.
In Fig.~\ref{fig:dsigdz_unc} we represent the differential cross-section for neutrino production
in charm decays, Eq.~(\ref{eq:dsigmadEnu}),
for different values of $E$, as a function of the ratio between the neutrino energy $E_{\nu}$ and the cosmic ray
energy, $z\equiv E_{\nu}/E_{p}$, that is,
\be
\label{eq:dsigmadEnu2}
\frac{d\sigma(pN \to \nu X;E;E_{\nu}=zE)}{d z} \, ,\quad z=\frac{E_{\nu}}{E} \, , 
\ee
which allows to compare the increase of the neutrino production cross-section, due to the larger value of $E$,
for the same value of $z$, the ratio of the neutrino energy over the incoming cosmic ray energy.
In  Fig.~\ref{fig:dsigdz_unc} results are shown for $E=10^3$ and $E=10^6$ GeV (both central values and total theoretical
    uncertainty) and then for  $E=10^8$ and $E=10^9$ GeV (only central values).
    Note how the cross-sections fall steeply as one approaches the kinematical boundary, $z\to 1$.
    %

Note that in pQCD, the correct expression for representing the dependence of $E$ of the prompt neutrino 
production cross-section is given by Eq.~(\ref{eq:dsigmadEnu2}), shown in Fig.~\ref{fig:dsigdz_unc}.
Previous works, for example~\cite{Bhattacharya:2015jpa}, present their calculations
of the charm production cross-section as
$d\sigma/dx_c$, where $x_c=E_c/E$, the ratio of produced charm quark energy over the incoming proton energy.
However, the charm quark energy is only well defined at leading order,
beyond which this is not true, and moreover is not accessible
experimentally.
Therefore, a robust comparison of theoretical calculations should always be presented at the level
of the physical $D$ production cross-section.
Alternatively, one might rescale by the charm branching fraction as
in Eq.~(\ref{eq:rescaling}), but this approximation is only valid for relatively inclusive observables.

Finally, let us mention that our calculations for Eq.~(\ref{eq:dsigmadEnu}),
illustrated
in Figs.~\ref{fig:distribution} and~\ref{fig:distribution2},
are available for a wide range of $E$ and $E_{\nu}$ values
in the  format of interpolated tables that
can be used as input for calculations of the prompt neutrino flux
at IceCube, and are available from
the authors upon request.

%% file: sec-delivery.tex
\section{Summary and outlook}
\label{sec:delivery}

In this work we have performed a detailed study of charm and bottom production
in the forward region, based on state-of-the-art pQCD 
with NLO calculations matched to parton showers.
Our  motivation was to provide a robust estimate of the theoretical
uncertainties associated to the prompt neutrino flux at neutrino telescopes
like IceCube, which
is the dominant background for the detection of astrophysical
neutrinos.

Our strategy was
based on the careful validation of the pQCD calculations with the LHCb charm and
bottom production data at 7 TeV, which cover the same kinematical region as
that relevant for the production of prompt neutrinos at IceCube.
We found that, with a suitable normalisation of the
differential distributions, it is possible to include the 7 TeV
$D$ meson data from LHCb in order to significantly constrain the poorly known
small-$x$
gluon.
Being able to include the LHCb charm measurements in a global
NLO PDF fit further enhances our confidence of the applicability of pQCD
to provide predictions for the prompt neutrino flux.
These improved PDFs, NNPDF3.0+LHCb, which include the information from the LHCb charm
data, are then used to construct the predictions for charm and
bottom production at LHCb for the recently started
Run II with a centre-of-mass energy of 13 TeV, as well as for the ratio of 13 over 7 TeV
cross-sections.

Our main result for the cross-sections of the
production of prompt neutrinos in charmed meson
decays originating from cosmic ray collisions in the atmosphere is summarised
in Figs.~\ref{fig:distribution},~\ref{fig:distribution2}
and~\ref{fig:dsigdz_unc}.
The main difference as compared to previous calculations of the prompt
neutrino flux is that our approach has been fully validated with the recent
LHCb differential measurements on charm production,
and that the input PDF set used in our calculations is one that already includes
the constraints from the LHCb charm data.
We would like to emphasise that our calculations, both
for the central values and for the various theoretical
uncertainties, have been carefully benchmarked using three independent codes.

The main results of our study can be summarised as follows:
\begin{itemize}
\item pQCD predictions for charm and bottom production in the forward
  region are consistent with the recent LHCb 7 TeV
measurements
  within  theoretical
  uncertainties.
  Predictions obtained with three different codes, two Monte Carlo
  parton shower programs, {\sc\small aMC@NLO} and  
  {\sc\small POWHEG}, and one semi-analytical calculation,
{\sc\small FONLL}, yield comparable results,
   both for the central value and for the total uncertainty.
 \item It is possible
   to include the LHCb charm data in the NNPDF3.0 NLO global analysis,
   achieving a substantial reduction of the  PDF uncertainties on the
   poorly known small-$x$ gluon.
   In order to reduce the large scale uncertainties of the
   NLO calculation, the LHCb data have been
   normalised to a fixed reference bin.

\item Run II of the LHC has just started, and the LHCb experiment will soon measure
  charm and bottom production in the forward region at 13~TeV,
  which will further explore the low-$x$ region of gluon PDF 
  providing unique information on the structure of the proton.
We have thus provided predictions for charm and bottom production
  at 13~TeV,
  as well as for the ratio of differential cross-sections between 13 and 7 TeV.
  These new measurements, both the 13 TeV (normalised) differential distributions
  and the 13 over 7 TeV cross-section ratio, offer new possibilities
  for PDF constraints, in particular thanks to the extended coverage at small-$x$
  as compared to the 7 TeV measurements.

\item Using the theory prediction for the ratio of inclusive fiducial cross-sections
  $R_{13/7}$ combined with the corresponding LHCb 7 TeV measurements, we are
  able to provide a prediction for the 13 TeV fiducial cross-section with
  substantially reduced uncertainties as that compared to the prediction
  from the NLO QCD calculation.

  
\item We have provided QCD predictions for the
   differential cross-sections
for the production of neutrinos from charm decay in $pA$ collisions,
Eq.~(\ref{eq:dsigmadEnu}), using two independent
NLO Monte Carlo generators, across a wide range of incoming
cosmic ray energies $E$, and accounting for all relevant
theory uncertainties.

\end{itemize}

It will be interesting to compare the upcoming 13 TeV LHCb measurements
with the predictions presented in this paper.
In particular, one should verify that the constraints on the
small-$x$ gluon obtained from
the inclusion on the PDF fit of the measurement of the ratio $R_{13/7}$
of differential distributions are consistent with those that have been obtained
from the 7 TeV normalised charm production cross-sections.
Likewise, comparing the inclusive fiducial cross-sections at 13 TeV with
our predictions based on $R_{13/7}$ will be an important test of the validity of
QCD calculations in this new kinematical region.

It is beyond the scope of this paper to explore quantitatively
the implications of our calculations for the recent IceCube
measurements of ultra high energy neutrinos.
Our results for the charm-induced neutrino 
cross-sections as a function of $E$ and $E_{\nu}$ are available
in the form of  interpolated
grids.
This information can be used as input in a full
calculation to derive robust
predictions for the rates of prompt neutrino events expected
at IceCube.

\section*{Acknowledgments}
We are grateful to Subir Sarkar for encouragement to start this
project and for collaboration during various stages
in this work.
We thank Matteo Cacciari for providing the NNPDF3.0  predictions using {\sc\small FONLL},
and for assistance with the benchmarking of the {\sc\small POWHEG} and {\sc\small aMC@NLO}
calculations, and Emanuele Re for assistance with the
charm production calculations in {\sc\small POWHEG}.
We thank Matteo Cacciari and Michelangelo Mangano for communications
about Ref.~\cite{CMN}.
We thank Anna Stasto, Mary Hall Reno and Atri Bhattacharya for providing us with the
results of the BERSS calculation.
We also acknowledge Tom Gaisser, Alexander Kappes and Teresa Montaruli for helpful discussions.
L.~R. thanks Marco Zaro for assistance with {\sc\small MadGraph5\_aMC@NLO}.
The work of J.~R. and L.~R. was supported by the ERC Starting Grant “PDF4BSM” and a STFC Rutherford Fellowship award to J.R (ST/K005227/1).
J.~T. is grateful to
Jurgen Rohrwild for useful discussions, and acknowledges financial
support from Hertford College.

%% file: sec-appendix.tex
\section{Predictions for charm and bottom production
  at 13~TeV}

\label{appendix}
\label{sec:appendix}

In this appendix we provide a tabulation of our predictions
for charm and bottom production in LHCb at TeV,
presented in Sect.~\ref{sec:appendix1}, as well as for
the ratio of cross-sections between 13 TeV and 7 TeV,
discussed in Sect.~\ref{sec:appendix2}.
For simplicity, we will restrict ourselves to the {\sc\small POWHEG} results,
since we have established from the comparison with
{\sc\small {\rm a}MC@NLO} in
Figs.~\ref{fig:LHCbcharm13} and~\ref{fig:LHCbbeauty13} that
the two calculations yield similar results.

\subsection{Predictions for differential distributions at 13 TeV}

First of all, in Table~\ref{tab:Ddiff} we provide the predictions for
the differential cross-sections for $D^0$ production at 13~TeV
in the LHCb acceptance, corresponding
to the results in Fig.~\ref{fig:LHCbcharm13}.
These results have been obtained with {\sc\small POWHEG} using NNPDF3.0+LHCb
as input PDF.
For each bin, we provide the central value and the total theoretical uncertainty.
To take into account the increased statistics that the Run II measurement will
benefit from, we have used in this tabulation an optimised binning as compared
to the 7 TeV results.
First of all, we have used a finer binning at low $p_T$ (the region which is most
sensitive to the gluon PDF) and extended our predictions
up to $p_T^D=30$ GeV (where theoretical uncertainties are smallest).
The corresponding predictions for any other choice of binning in $\lp p_T^D,y^D\rp$ as
well as for the other $D$ meson species are available from the authors upon request.
As in the case of the 7 TeV results, absolute cross-sections are affected by substantial
theoretical uncertainties, in particular due to the large scale variations of the NLO
computation.
On the other hand, PDF uncertainties are now subdominant for all values of $y^D$ and $p_{T}^D$,
thanks for the constraints from the 7 TeV LHCb charm data.

\renewcommand*{\arraystretch}{1.5}
\begin{table}[t!]
  \centering
  \footnotesize
  \begin{tabular}{@{} c|rcl|rcl|rcl|rcl|rcl@{}}
      \hline
   \multicolumn{16}{c}{$\frac{d^2\sigma(D)(y,p_T)}{dy^Ddp_T^D} \Delta y \,~ (\mu b/{\rm GeV})$}  \\ \hline\hline
      	$p_T^D$~(GeV) & \multicolumn{15}{c}{$y^D$}  \\ \hline
	 & \multicolumn{3}{c|}{$2.0-2.5$} &  \multicolumn{3}{c|}{$2.5-3.0$}
	& \multicolumn{3}{c|}{$3.0-3.5$}	 &  \multicolumn{3}{c|}{$3.5-4.0$}	& \multicolumn{3}{c}{$4.0-4.5$} \\ \hline
$0.0-0.25$ & 16.7 & \hspace{-0.5cm} & $^{+41.4}_{-15.8}$ &16.4 & \hspace{-0.5cm} & $^{+40.6}_{-15.3}$ &16.2 & \hspace{-0.5cm} & $^{+41.0}_{-14.9}$ &15.0 & \hspace{-0.5cm} & $^{+38.0}_{-13.8}$ &13.5 & \hspace{-0.5cm} & $^{+34.0}_{-12.2}$ \\ 
$0.25-0.5$ & 45.4 & \hspace{-0.5cm} & $^{+109.8}_{-43.0}$ &45.9 & \hspace{-0.5cm} & $^{+114.6}_{-42.8}$ &43.6 & \hspace{-0.5cm} & $^{+108.7}_{-40.2}$ &41.5 & \hspace{-0.5cm} & $^{+103.5}_{-37.6}$ &38.6 & \hspace{-0.5cm} & $^{+96.4}_{-34.1}$ \\
$0.5-0.75$ & 66.0 & \hspace{-0.5cm} & $^{+156.6}_{-61.4}$ &65.2 & \hspace{-0.5cm} & $^{+157.6}_{-60.0}$ &63.0 & \hspace{-0.5cm} & $^{+153.0}_{-57.1}$ &59.9 & \hspace{-0.5cm} & $^{+146.3}_{-53.8}$ &55.4 & \hspace{-0.5cm} & $^{+136.1}_{-48.8}$ \\ 
$0.75-1.0$ & 77.3 & \hspace{-0.5cm} & $^{+178.6}_{-71.7}$ &75.1 & \hspace{-0.5cm} & $^{+174.6}_{-68.5}$ &73.5 & \hspace{-0.5cm} & $^{+171.5}_{-65.7}$ &70.5 & \hspace{-0.5cm} & $^{+167.5}_{-62.8}$ &63.0 & \hspace{-0.5cm} & $^{+148.6}_{-55.2}$ \\ 
$1.0-1.25$ & 82.2 & \hspace{-0.5cm} & $^{+181.7}_{-75.0}$ &80.1 & \hspace{-0.5cm} & $^{+178.5}_{-72.2}$ &76.1 & \hspace{-0.5cm} & $^{+171.0}_{-67.5}$ &72.4 & \hspace{-0.5cm} & $^{+164.2}_{-63.4}$ &66.8 & \hspace{-0.5cm} & $^{+152.9}_{-57.9}$ \\ 
$1.25-1.5$ & 79.6 & \hspace{-0.5cm} & $^{+168.0}_{-71.8}$ &77.7 & \hspace{-0.5cm} & $^{+165.2}_{-69.2}$ &73.8 & \hspace{-0.5cm} & $^{+155.5}_{-64.8}$ &69.9 & \hspace{-0.5cm} & $^{+150.3}_{-60.2}$ &63.7 & \hspace{-0.5cm} & $^{+138.2}_{-54.6}$ \\ 
$1.5-1.75$ & 76.2 & \hspace{-0.5cm} & $^{+152.2}_{-67.7}$ &73.4 & \hspace{-0.5cm} & $^{+148.8}_{-64.1}$ &69.9 & \hspace{-0.5cm} & $^{+143.0}_{-60.0}$ &65.8 & \hspace{-0.5cm} & $^{+137.1}_{-56.0}$ &59.3 & \hspace{-0.5cm} & $^{+124.0}_{-49.8}$ \\ 
$1.75-2.0$ & 69.7 & \hspace{-0.5cm} & $^{+132.7}_{-60.9}$ &67.7 & \hspace{-0.5cm} & $^{+129.5}_{-58.5}$ &64.8 & \hspace{-0.5cm} & $^{+127.6}_{-54.8}$ &59.7 & \hspace{-0.5cm} & $^{+117.2}_{-49.2}$ &53.1 & \hspace{-0.5cm} & $^{+106.5}_{-44.0}$ \\ 
$2.0-2.25$ & 62.7 & \hspace{-0.5cm} & $^{+112.3}_{-53.8}$ &59.8 & \hspace{-0.5cm} & $^{+108.0}_{-50.1}$ &57.5 & \hspace{-0.5cm} & $^{+107.0}_{-47.4}$ &52.0 & \hspace{-0.5cm} & $^{+97.1}_{-42.6}$ &46.9 & \hspace{-0.5cm} & $^{+89.1}_{-37.6}$ \\ 
$2.25-2.5$ & 55.3 & \hspace{-0.5cm} & $^{+95.1}_{-46.3}$ &53.1 & \hspace{-0.5cm} & $^{+91.5}_{-43.8}$ &50.2 & \hspace{-0.5cm} & $^{+88.4}_{-40.8}$ &46.3 & \hspace{-0.5cm} & $^{+82.2}_{-36.8}$ &40.2 & \hspace{-0.5cm} & $^{+72.6}_{-31.8}$ \\ 
$2.5-2.75$ & 49.7 & \hspace{-0.5cm} & $^{+82.8}_{-41.1}$ &46.5 & \hspace{-0.5cm} & $^{+77.1}_{-37.7}$ &43.7 & \hspace{-0.5cm} & $^{+73.2}_{-34.7}$ &39.3 & \hspace{-0.5cm} & $^{+67.2}_{-30.8}$ &34.7 & \hspace{-0.5cm} & $^{+58.8}_{-26.7}$ \\ 
$2.75-3.0$ & 43.4 & \hspace{-0.5cm} & $^{+68.7}_{-35.0}$ &41.2 & \hspace{-0.5cm} & $^{+66.2}_{-32.5}$ &37.9 & \hspace{-0.5cm} & $^{+60.6}_{-29.2}$ &34.0 & \hspace{-0.5cm} & $^{+56.2}_{-25.8}$ &29.3 & \hspace{-0.5cm} & $^{+49.2}_{-21.8}$ \\ 
$3.0-3.5$ & 35.7 & \hspace{-0.5cm} & $^{+53.6}_{-28.1}$ &33.5 & \hspace{-0.5cm} & $^{+50.6}_{-25.7}$ &30.7 & \hspace{-0.5cm} & $^{+46.8}_{-23.1}$ &27.6 & \hspace{-0.5cm} & $^{+42.6}_{-20.4}$ &22.9 & \hspace{-0.5cm} & $^{+35.9}_{-16.8}$ \\ 
$3.5-4.0$ & 26.9 & \hspace{-0.5cm} & $^{+37.5}_{-20.2}$ &25.0 & \hspace{-0.5cm} & $^{+34.8}_{-18.3}$ &22.9 & \hspace{-0.5cm} & $^{+32.4}_{-16.5}$ &20.0 & \hspace{-0.5cm} & $^{+28.8}_{-14.3}$ &16.7 & \hspace{-0.5cm} & $^{+24.5}_{-11.6}$ \\ 
$4.0-4.5$ & 19.9 & \hspace{-0.5cm} & $^{+26.0}_{-14.3}$ &18.9 & \hspace{-0.5cm} & $^{+25.0}_{-13.2}$ &16.6 & \hspace{-0.5cm} & $^{+21.6}_{-11.5}$ &14.2 & \hspace{-0.5cm} & $^{+18.7}_{-9.6}$ &11.8 & \hspace{-0.5cm} & $^{+15.9}_{-7.9}$ \\ 
$4.5-5.0$ & 15.2 & \hspace{-0.5cm} & $^{+19.0}_{-10.5}$ &13.9 & \hspace{-0.5cm} & $^{+17.3}_{-9.4}$ &12.1 & \hspace{-0.5cm} & $^{+15.3}_{-8.0}$ &10.3 & \hspace{-0.5cm} & $^{+13.0}_{-6.7}$ &8.3 & \hspace{-0.5cm} & $^{+10.9}_{-5.3}$ \\ 
$5.0-6.0$ & 10.2 & \hspace{-0.5cm} & $^{+11.7}_{-6.7}$ &9.0 & \hspace{-0.5cm} & $^{+10.3}_{-5.8}$ &8.1 & \hspace{-0.5cm} & $^{+9.3}_{-5.0}$ &6.7 & \hspace{-0.5cm} & $^{+8.0}_{-4.1}$ &5.0 & \hspace{-0.5cm} & $^{+6.0}_{-3.0}$ \\ 
$6.0-7.0$ & 5.85 & \hspace{-0.5cm} & $^{+6.02}_{-3.53}$ &5.11 & \hspace{-0.5cm} & $^{+5.41}_{-3.0}$ &4.43 & \hspace{-0.5cm} & $^{+4.62}_{-2.55}$ &3.55 & \hspace{-0.5cm} & $^{+3.91}_{-1.99}$ &2.67 & \hspace{-0.5cm} & $^{+2.88}_{-1.48}$ \\ 
$7.0-8.0$ & 3.63 & \hspace{-0.5cm} & $^{+3.41}_{-2.0}$ &3.11 & \hspace{-0.5cm} & $^{+3.02}_{-1.7}$ &2.65 & \hspace{-0.5cm} & $^{+2.61}_{-1.38}$ &2.06 & \hspace{-0.5cm} & $^{+2.0}_{-1.08}$ &1.52 & \hspace{-0.5cm} & $^{+1.55}_{-0.78}$ \\ 
$8.0-9.0$ & 2.21 & \hspace{-0.5cm} & $^{+1.98}_{-1.14}$ &1.89 & \hspace{-0.5cm} & $^{+1.68}_{-0.97}$ &1.57 & \hspace{-0.5cm} & $^{+1.42}_{-0.77}$ &1.25 & \hspace{-0.5cm} & $^{+1.2}_{-0.6}$ &0.78 & \hspace{-0.5cm} & $^{+0.74}_{-0.37}$ \\ 
$9.0-10.0$ & 1.45 & \hspace{-0.5cm} & $^{+1.23}_{-0.71}$ &1.21 & \hspace{-0.5cm} & $^{+1.05}_{-0.57}$ &1.0 & \hspace{-0.5cm} & $^{+0.87}_{-0.47}$ &0.75 & \hspace{-0.5cm} & $^{+0.67}_{-0.34}$ &0.46 & \hspace{-0.5cm} & $^{+0.41}_{-0.2}$ \\ 
$10.0-15.0$ & 0.54 & \hspace{-0.5cm} & $^{+0.42}_{-0.24}$ &0.45 & \hspace{-0.5cm} & $^{+0.35}_{-0.19}$ &0.34 & \hspace{-0.5cm} & $^{+0.26}_{-0.14}$ &0.25 & \hspace{-0.5cm} & $^{+0.2}_{-0.1}$ &0.16 & \hspace{-0.5cm} & $^{+0.12}_{-0.06}$ \\ 
$15.0-20.0$ & 0.12 & \hspace{-0.5cm} & $^{+0.08}_{-0.04}$ &0.1 & \hspace{-0.5cm} & $^{+0.07}_{-0.04}$ &0.07 & \hspace{-0.5cm} & $^{+0.05}_{-0.03}$ &0.04 & \hspace{-0.5cm} & $^{+0.03}_{-0.01}$ &0.02 & \hspace{-0.5cm} & $^{+0.02}_{-0.01}$ \\ 
$20.0-30.0$ & 0.029 & \hspace{-0.5cm} & $^{+0.018}_{-0.008}$ &0.025 & \hspace{-0.5cm} & $^{+0.014}_{-0.008}$ &0.016 & \hspace{-0.5cm} & $^{+0.009}_{-0.005}$ &0.009 & \hspace{-0.5cm} & $^{+0.006}_{-0.003}$ &0.003 & \hspace{-0.5cm} & $^{+0.002}_{-0.001}$ \\
\hline
     \end{tabular}
  \caption{\small Predictions for the differential cross-sections for
    $D^0$ meson production at LHCb at 13~TeV, computed
    using {\sc\small POWHEG} and NNPDF3.0+LHCb.
    For each bin we indicate the central value and the total theoretical uncertainty.
    Predictions for different binnings and for other $D$ meson species are available upon request.
    }
  \label{tab:Ddiff}
\end{table}

The corresponding predictions
for the differential
cross-sections of the production of $B^0$ mesons at LHCb Run II are shown in Table~\ref{tab:Bdiff}, which
 is the analog of Table~\ref{tab:Ddiff} for charm production.
 These predictions were represented graphically (and compared to the {\sc\small {\rm a}MC@NLO} calculation)
 in Fig.~\ref{fig:LHCbbeauty13}.
 In this case
we have assumed the same binning as in the 7 TeV measurement.
As compared to the 13 TeV charm predictions, the higher scales and the larger values of
Bjorken-$x$ probed in the case
of bottom production result in reduced  theory uncertainties.
The total uncertainty is around 
$50\%$, with differences depending on the specific bin, and is
again dominated by scale uncertainties.

\renewcommand*{\arraystretch}{1.5}
\begin{table}[t!]
  \centering
  \footnotesize
  \begin{tabular}{@{} c|rcl|rcl|rcl|rcl|rcl@{}}
   \hline
   \multicolumn{16}{c}{$\frac{d^2\sigma(B)(y,p_T)}{dy^Bdp_T^B} (\mu b/{\rm GeV})$}  \\ \hline\hline
      	$p_T^B$~(GeV) & \multicolumn{15}{c}{$y^B$}  \\ \hline
	 & \multicolumn{3}{c|}{$2.0-2.5$} &  \multicolumn{3}{c|}{$2.5-3.0$}
	& \multicolumn{3}{c|}{$3.0-3.5$}	 &  \multicolumn{3}{c|}{$3.5-4.0$}	& \multicolumn{3}{c}{$4.0-4.5$} \\ \hline
$0.0-0.5$ & 0.6 & \hspace{-0.5cm} & $^{+0.37}_{-0.28}$ &0.53 & \hspace{-0.5cm} & $^{+0.34}_{-0.25}$ &0.45 & \hspace{-0.5cm} & $^{+0.29}_{-0.21}$ &0.4 & \hspace{-0.5cm} & $^{+0.25}_{-0.18}$ &0.32 & \hspace{-0.5cm} & $^{+0.2}_{-0.13}$ \\ 
$0.5-1.0$ & 1.67 & \hspace{-0.5cm} & $^{+1.0}_{-0.8}$ &1.53 & \hspace{-0.5cm} & $^{+0.94}_{-0.73}$ &1.36 & \hspace{-0.5cm} & $^{+0.84}_{-0.62}$ &1.14 & \hspace{-0.5cm} & $^{+0.71}_{-0.51}$ &0.91 & \hspace{-0.5cm} & $^{+0.56}_{-0.4}$ \\ 
$1.0-1.5$ & 2.56 & \hspace{-0.5cm} & $^{+1.58}_{-1.21}$ &2.32 & \hspace{-0.5cm} & $^{+1.42}_{-1.07}$ &2.05 & \hspace{-0.5cm} & $^{+1.27}_{-0.93}$ &1.75 & \hspace{-0.5cm} & $^{+1.07}_{-0.77}$ &1.39 & \hspace{-0.5cm} & $^{+0.83}_{-0.59}$ \\ 
$1.5-2.0$ & 3.27 & \hspace{-0.5cm} & $^{+1.98}_{-1.54}$ &2.96 & \hspace{-0.5cm} & $^{+1.77}_{-1.36}$ &2.58 & \hspace{-0.5cm} & $^{+1.57}_{-1.15}$ &2.21 & \hspace{-0.5cm} & $^{+1.33}_{-0.97}$ &1.77 & \hspace{-0.5cm} & $^{+1.07}_{-0.76}$ \\ 
$2.0-2.5$ & 3.68 & \hspace{-0.5cm} & $^{+2.19}_{-1.68}$ &3.34 & \hspace{-0.5cm} & $^{+1.99}_{-1.49}$ &2.93 & \hspace{-0.5cm} & $^{+1.76}_{-1.29}$ &2.45 & \hspace{-0.5cm} & $^{+1.46}_{-1.04}$ &1.98 & \hspace{-0.5cm} & $^{+1.18}_{-0.83}$ \\ 
$2.5-3.0$ & 3.91 & \hspace{-0.5cm} & $^{+2.3}_{-1.76}$ &3.55 & \hspace{-0.5cm} & $^{+2.09}_{-1.57}$ &3.1 & \hspace{-0.5cm} & $^{+1.82}_{-1.33}$ &2.58 & \hspace{-0.5cm} & $^{+1.52}_{-1.07}$ &2.05 & \hspace{-0.5cm} & $^{+1.2}_{-0.84}$ \\ 
$3.0-3.5$ & 3.99 & \hspace{-0.5cm} & $^{+2.25}_{-1.76}$ &3.58 & \hspace{-0.5cm} & $^{+2.06}_{-1.53}$ &3.16 & \hspace{-0.5cm} & $^{+1.82}_{-1.32}$ &2.64 & \hspace{-0.5cm} & $^{+1.51}_{-1.09}$ &2.04 & \hspace{-0.5cm} & $^{+1.16}_{-0.82}$ \\ 
$3.5-4.0$ & 3.91 & \hspace{-0.5cm} & $^{+2.17}_{-1.68}$ &3.49 & \hspace{-0.5cm} & $^{+1.96}_{-1.46}$ &3.06 & \hspace{-0.5cm} & $^{+1.7}_{-1.25}$ &2.54 & \hspace{-0.5cm} & $^{+1.44}_{-1.01}$ &1.97 & \hspace{-0.5cm} & $^{+1.11}_{-0.76}$ \\ 
$4.0-4.5$ & 3.71 & \hspace{-0.5cm} & $^{+2.03}_{-1.54}$ &3.36 & \hspace{-0.5cm} & $^{+1.83}_{-1.37}$ &2.9 & \hspace{-0.5cm} & $^{+1.6}_{-1.17}$ &2.4 & \hspace{-0.5cm} & $^{+1.3}_{-0.93}$ &1.82 & \hspace{-0.5cm} & $^{+0.98}_{-0.67}$ \\ 
$4.5-5.0$ & 3.5 & \hspace{-0.5cm} & $^{+1.87}_{-1.43}$ &3.16 & \hspace{-0.5cm} & $^{+1.68}_{-1.27}$ &2.67 & \hspace{-0.5cm} & $^{+1.42}_{-1.04}$ &2.21 & \hspace{-0.5cm} & $^{+1.18}_{-0.84}$ &1.69 & \hspace{-0.5cm} & $^{+0.91}_{-0.62}$ \\ 
$5.0-5.5$ & 3.17 & \hspace{-0.5cm} & $^{+1.69}_{-1.25}$ &2.87 & \hspace{-0.5cm} & $^{+1.49}_{-1.12}$ &2.43 & \hspace{-0.5cm} & $^{+1.27}_{-0.91}$ &1.98 & \hspace{-0.5cm} & $^{+1.05}_{-0.72}$ &1.5 & \hspace{-0.5cm} & $^{+0.77}_{-0.53}$ \\ 
$5.5-6.0$ & 2.89 & \hspace{-0.5cm} & $^{+1.48}_{-1.12}$ &2.6 & \hspace{-0.5cm} & $^{+1.31}_{-0.98}$ &2.22 & \hspace{-0.5cm} & $^{+1.13}_{-0.82}$ &1.77 & \hspace{-0.5cm} & $^{+0.89}_{-0.62}$ &1.34 & \hspace{-0.5cm} & $^{+0.68}_{-0.46}$ \\ 
$6.0-6.5$ & 2.63 & \hspace{-0.5cm} & $^{+1.29}_{-1.0}$ &2.33 & \hspace{-0.5cm} & $^{+1.15}_{-0.87}$ &1.97 & \hspace{-0.5cm} & $^{+0.97}_{-0.69}$ &1.58 & \hspace{-0.5cm} & $^{+0.76}_{-0.55}$ &1.15 & \hspace{-0.5cm} & $^{+0.57}_{-0.37}$ \\ 
$6.5-7.0$ & 2.36 & \hspace{-0.5cm} & $^{+1.14}_{-0.87}$ &2.05 & \hspace{-0.5cm} & $^{+1.0}_{-0.73}$ &1.74 & \hspace{-0.5cm} & $^{+0.85}_{-0.61}$ &1.38 & \hspace{-0.5cm} & $^{+0.67}_{-0.45}$ &0.98 & \hspace{-0.5cm} & $^{+0.48}_{-0.31}$ \\ 
$7.0-7.5$ & 2.07 & \hspace{-0.5cm} & $^{+0.99}_{-0.74}$ &1.84 & \hspace{-0.5cm} & $^{+0.88}_{-0.65}$ &1.52 & \hspace{-0.5cm} & $^{+0.73}_{-0.5}$ &1.2 & \hspace{-0.5cm} & $^{+0.56}_{-0.38}$ &0.86 & \hspace{-0.5cm} & $^{+0.41}_{-0.27}$ \\ 
$7.5-8.0$ & 1.82 & \hspace{-0.5cm} & $^{+0.87}_{-0.63}$ &1.61 & \hspace{-0.5cm} & $^{+0.76}_{-0.54}$ &1.33 & \hspace{-0.5cm} & $^{+0.62}_{-0.43}$ &1.04 & \hspace{-0.5cm} & $^{+0.47}_{-0.32}$ &0.72 & \hspace{-0.5cm} & $^{+0.34}_{-0.22}$ \\ 
$8.0-8.5$ & 1.62 & \hspace{-0.5cm} & $^{+0.74}_{-0.55}$ &1.41 & \hspace{-0.5cm} & $^{+0.65}_{-0.47}$ &1.15 & \hspace{-0.5cm} & $^{+0.51}_{-0.36}$ &0.88 & \hspace{-0.5cm} & $^{+0.39}_{-0.26}$ &0.62 & \hspace{-0.5cm} & $^{+0.28}_{-0.18}$ \\ 
$8.5-9.0$ & 1.41 & \hspace{-0.5cm} & $^{+0.63}_{-0.46}$ &1.23 & \hspace{-0.5cm} & $^{+0.55}_{-0.39}$ &1.01 & \hspace{-0.5cm} & $^{+0.45}_{-0.31}$ &0.77 & \hspace{-0.5cm} & $^{+0.35}_{-0.23}$ &0.52 & \hspace{-0.5cm} & $^{+0.24}_{-0.14}$ \\ 
$9.0-9.5$ & 1.22 & \hspace{-0.5cm} & $^{+0.55}_{-0.39}$ &1.07 & \hspace{-0.5cm} & $^{+0.48}_{-0.33}$ &0.87 & \hspace{-0.5cm} & $^{+0.38}_{-0.26}$ &0.66 & \hspace{-0.5cm} & $^{+0.29}_{-0.19}$ &0.45 & \hspace{-0.5cm} & $^{+0.2}_{-0.12}$ \\ 
$9.5-10.0$ & 1.11 & \hspace{-0.5cm} & $^{+0.48}_{-0.35}$ &0.93 & \hspace{-0.5cm} & $^{+0.4}_{-0.28}$ &0.77 & \hspace{-0.5cm} & $^{+0.33}_{-0.23}$ &0.58 & \hspace{-0.5cm} & $^{+0.26}_{-0.16}$ &0.39 & \hspace{-0.5cm} & $^{+0.16}_{-0.11}$ \\ 
$10.0-10.5$ & 0.95 & \hspace{-0.5cm} & $^{+0.41}_{-0.29}$ &0.82 & \hspace{-0.5cm} & $^{+0.35}_{-0.24}$ &0.66 & \hspace{-0.5cm} & $^{+0.28}_{-0.19}$ &0.48 & \hspace{-0.5cm} & $^{+0.21}_{-0.12}$ &0.33 & \hspace{-0.5cm} & $^{+0.14}_{-0.09}$ \\ 
$10.5-11.5$ & 0.79 & \hspace{-0.5cm} & $^{+0.33}_{-0.23}$ &0.67 & \hspace{-0.5cm} & $^{+0.28}_{-0.19}$ &0.53 & \hspace{-0.5cm} & $^{+0.22}_{-0.14}$ &0.4 & \hspace{-0.5cm} & $^{+0.16}_{-0.1}$ &0.26 & \hspace{-0.5cm} & $^{+0.11}_{-0.07}$ \\ 
$11.5-12.5$ & 0.61 & \hspace{-0.5cm} & $^{+0.25}_{-0.17}$ &0.51 & \hspace{-0.5cm} & $^{+0.21}_{-0.14}$ &0.4 & \hspace{-0.5cm} & $^{+0.17}_{-0.1}$ &0.3 & \hspace{-0.5cm} & $^{+0.12}_{-0.08}$ &0.18 & \hspace{-0.5cm} & $^{+0.07}_{-0.05}$ \\ 
$12.5-14.0$ & 0.45 & \hspace{-0.5cm} & $^{+0.18}_{-0.12}$ &0.38 & \hspace{-0.5cm} & $^{+0.15}_{-0.1}$ &0.29 & \hspace{-0.5cm} & $^{+0.12}_{-0.08}$ &0.21 & \hspace{-0.5cm} & $^{+0.08}_{-0.05}$ &0.13 & \hspace{-0.5cm} & $^{+0.05}_{-0.03}$ \\ 
$14.0-16.5$ & 0.28 & \hspace{-0.5cm} & $^{+0.11}_{-0.07}$ &0.23 & \hspace{-0.5cm} & $^{+0.09}_{-0.06}$ &0.18 & \hspace{-0.5cm} & $^{+0.07}_{-0.04}$ &0.12 & \hspace{-0.5cm} & $^{+0.05}_{-0.03}$ &0.07 & \hspace{-0.5cm} & $^{+0.03}_{-0.02}$ \\ 
$16.5-23.5$ & 0.11 & \hspace{-0.5cm} & $^{+0.04}_{-0.03}$ &0.09 & \hspace{-0.5cm} & $^{+0.03}_{-0.02}$ &0.06 & \hspace{-0.5cm} & $^{+0.02}_{-0.01}$ &0.04 & \hspace{-0.5cm} & $^{+0.01}_{-0.01}$ &0.02 & \hspace{-0.5cm} & $^{+0.01}_{-0.0}$ \\ 
$23.5-40.0$ & 0.019 & \hspace{-0.5cm} & $^{+0.006}_{-0.004}$ &0.014 & \hspace{-0.5cm} & $^{+0.005}_{-0.003}$ &0.01 & \hspace{-0.5cm} & $^{+0.003}_{-0.002}$ &0.005 & \hspace{-0.5cm} & $^{+0.002}_{-0.001}$ &0.002 & \hspace{-0.5cm} & $^{+0.001}_{-0.001}$ \\
\hline
     \end{tabular}
  \caption{\small
    Same as Table~\ref{tab:Ddiff}, now for the production of $B^0$ mesons at 13 TeV.
    Predictions for different binnings and for other $B$ mesons are available upon request.
  }
  \label{tab:Bdiff}
\end{table}

\subsection{Predictions for the ratio $R_{13/7}$}

Next  we turn to the predictions for the ratio $R_{13/7}$ of the differential
distributions for heavy quark production at LHCb
between 13 TeV and 7 TeV, Eqns.~(\ref{eq:ratioD0}) and~(\ref{eq:ratioB0}),
discussed in Sect.~\ref{sec:appendix2}.
In Table~\ref{tab:Dratio} we show
the ratio $R_{13/7}$ for  $D^0$ mesons using {\sc\small POWHEG} and
NNPDF3.0+LHCb.
These predictions were represented graphically in
Fig.~\ref{fig:LHCbcharmRatio}.
We provide the  central value of $R_{13/7}$  and the total
theoretical uncertainty, which as can be seen from Fig.~\ref{fig:LHCbcharmRatio}
arises predominantly due to scale variations.
When evaluating Eq.~(\ref{eq:ratioD0}), scale variations, charm mass variations
and PDF variations are considered to be fully correlated between 13 and
7~TeV.

\renewcommand*{\arraystretch}{1.5}
\begin{table}[t!]
\centering
\begin{tabular}{@{} c|rcl|rcl|rcl|rcl|rcl@{}}
  \hline
   \multicolumn{16}{c}{$R_{13/7}^D=\frac{d^2\sigma(D)(y,p_T,13~{\rm TeV})}{dy^Ddp_T^D}\Big/
     \frac{d^2\sigma(D)(y,p_T,7~{\rm TeV})}{dy^Ddp_T^D}$}  \\
   \hline\hline
 $p_T^D$~(GeV)    	 & \multicolumn{15}{c}{$y^D$}  \\ \hline
 & \multicolumn{3}{c|}{$2.0-2.5$} &  \multicolumn{3}{c|}{$2.5-3.0$}
	& \multicolumn{3}{c|}{$3.0-3.5$}	 &  \multicolumn{3}{c|}{$3.5-4.0$}	& \multicolumn{3}{c}{$4.0-4.5$} \\ \hline
	$0.0-1.0$ & 1.23 & \hspace{-0.5cm} & $^{+0.13}_{-0.35}$ &1.25 & \hspace{-0.5cm} & $^{+0.13}_{-0.35}$ &1.28 & \hspace{-0.5cm} & $^{+0.15}_{-0.32}$ &1.33 & \hspace{-0.5cm} & $^{+0.15}_{-0.32}$ &1.38 & \hspace{-0.5cm} & $^{+0.17}_{-0.27}$ \\ 
$1.0-2.0$ & 1.26 & \hspace{-0.5cm} & $^{+0.12}_{-0.34}$ &1.29 & \hspace{-0.5cm} & $^{+0.12}_{-0.37}$ &1.32 & \hspace{-0.5cm} & $^{+0.13}_{-0.35}$ &1.38 & \hspace{-0.5cm} & $^{+0.13}_{-0.31}$ &1.46 & \hspace{-0.5cm} & $^{+0.14}_{-0.31}$ \\ 
$2.0-3.0$ & 1.31 & \hspace{-0.5cm} & $^{+0.11}_{-0.31}$ &1.33 & \hspace{-0.5cm} & $^{+0.12}_{-0.29}$ &1.39 & \hspace{-0.5cm} & $^{+0.12}_{-0.31}$ &1.45 & \hspace{-0.5cm} & $^{+0.13}_{-0.33}$ &1.57 & \hspace{-0.5cm} & $^{+0.15}_{-0.3}$ \\ 
$3.0-4.0$ & 1.39 & \hspace{-0.5cm} & $^{+0.1}_{-0.26}$ &1.43 & \hspace{-0.5cm} & $^{+0.11}_{-0.26}$ &1.5 & \hspace{-0.5cm} & $^{+0.11}_{-0.27}$ &1.61 & \hspace{-0.5cm} & $^{+0.13}_{-0.33}$ &1.72 & \hspace{-0.5cm} & $^{+0.16}_{-0.32}$ \\ 
$4.0-5.0$ & 1.49 & \hspace{-0.5cm} & $^{+0.1}_{-0.22}$ &1.53 & \hspace{-0.5cm} & $^{+0.13}_{-0.24}$ &1.61 & \hspace{-0.5cm} & $^{+0.12}_{-0.26}$ &1.72 & \hspace{-0.5cm} & $^{+0.14}_{-0.29}$ &2.01 & \hspace{-0.5cm} & $^{+0.16}_{-0.39}$ \\ 
$5.0-6.0$ & 1.57 & \hspace{-0.5cm} & $^{+0.11}_{-0.24}$ &1.65 & \hspace{-0.5cm} & $^{+0.12}_{-0.25}$ &1.79 & \hspace{-0.5cm} & $^{+0.11}_{-0.26}$ &1.99 & \hspace{-0.5cm} & $^{+0.11}_{-0.25}$ &2.18 & \hspace{-0.5cm} & $^{+0.19}_{-0.35}$ \\ 
$6.0-7.0$ & 1.67 & \hspace{-0.5cm} & $^{+0.13}_{-0.24}$ &1.75 & \hspace{-0.5cm} & $^{+0.12}_{-0.23}$ &1.85 & \hspace{-0.5cm} & $^{+0.14}_{-0.28}$ &1.98 & \hspace{-0.5cm} & $^{+0.21}_{-0.27}$ & & \hspace{-0.5cm} &  \\ 
$7.0-8.0$ & 1.78 & \hspace{-0.5cm} & $^{+0.1}_{-0.18}$ &1.84 & \hspace{-0.5cm} & $^{+0.09}_{-0.18}$ &2.02 & \hspace{-0.5cm} & $^{+0.12}_{-0.15}$ &2.17 & \hspace{-0.5cm} & $^{+0.16}_{-0.31}$ & & \hspace{-0.5cm} &  \\
\hline
     \end{tabular}
\caption{\small Predictions for the ratio $R_{13/7}^D$ of double
  differential cross-sections for $D^0$ meson production between 13
  and 7 TeV at LHCb, Eq.~(\ref{eq:ratioD0})
  Results have obtained using {\sc\small POWHEG}
  with the  NNPDF3.0+LHCb NLO PDF set.
  The same binning as in the 7~TeV measurement is assumed at
  13 TeV.
  In each bin, we provide the central prediction and the
  total theoretical uncertainty, obtained from the sum in quadrature
  of scales, PDFs and charm mass variations.
  See Fig.~\ref{fig:LHCbcharmRatio} for the graphical representation
  of these predictions.  \label{tab:Dratio}
}
\end{table}

In order to evaluate the impact of the reduction of PDF
uncertainties on the observable $R^D_{13/7}$, that has been achieved by including in NNPDF3.0
the LHCb 7 TeV charm production data, it is useful
to compare with the corresponding predictions with the
original NNPDF3.0 set.
With this motivation, in Table~\ref{tab:DratioOld}
we show ratio $R_{13/7}({\rm orig})$ computed with the original NNPDF3.0 PDF set,
which should be compared with the  predictions obtained with
    the NNPDF3.0+LHCb set in Table~\ref{tab:Dratio}.
The data is ordered in increasing rapidity bins, and within each of these in increasing $p_T$ bins.
For each bin, we show the central prediction, the PDF uncertainty
and the the total theory uncertainty for  $R^D_{13/7}({\rm orig})$,
as well as the ratio between the predictions
for the ratio itself computed with NNPDF3.0+LHCb and with the original
NNPDF3.0, $R^D_{13/7}({\rm new})/R^D_{13/7}({\rm orig})$.

From the comparison between Tables~\ref{tab:DratioOld}
and~\ref{tab:Dratio} we see first of all that the predictions
for the central value of $R^D_{13/7}$ are reasonably stable:
differences for the central value
computed between the original and new PDFs are typically a few percent,
rather smaller than the total theory uncertainties.
This nicely illustrates the compatibility of the NNPDF3.0
small-$x$ gluon with the 7 TeV LHCb charm production
data.
The real difference comes from the reduction in PDF uncertainties:
since scale and charm mass uncertainties are essentially the
same in $R^D_{13/7}({\rm new})$ and $R^D_{13/7}({\rm old})$,
the differences between the total theory errors stem
from the reduction of PDF uncertainties in $R_{13/7}({\rm new})$.
For instance, in the lowest $p_T$ and most forward region (data bin 33),
the relative total theory uncertainty of $R^D_{13/7}({\rm orig})$ is $~^{+30\%}_{-38\%}$,
while for $R^D_{13/7}({\rm new})$ the corresponding uncertainty is substantially reduced
reduced down to $~^{+17\%}_{-27\%}$.
Similar comparisons can be performed for other bins.

\renewcommand*{\arraystretch}{1.15}
\begin{table}[t!]
  \centering
  \footnotesize
  \begin{tabular}{@{} c|c|c|c|c@{}}
    Data Index & $R_{13/7}({\rm new})$/$R_{13/7}({\rm orig})$ &
    $R_{13/7}({\rm orig})$ cv & $R_{13/7}({\rm orig})$ PDF &
    $R_{13/7}({\rm orig})$ Tot \\ \hline
1 & 0.9 & 1.36 & $\pm$0.19 & $^{+0.23}_{-0.43}$ \\ 
2 & 0.93 & 1.35 & $\pm$0.18 & $^{+0.22}_{-0.4}$ \\ 
3 & 0.94 & 1.39 & $\pm$0.19 & $^{+0.21}_{-0.37}$ \\ 
4 & 0.95 & 1.47 & $\pm$0.18 & $^{+0.2}_{-0.32}$ \\ 
5 & 0.97 & 1.55 & $\pm$0.15 & $^{+0.18}_{-0.27}$ \\ 
6 & 0.99 & 1.59 & $\pm$0.13 & $^{+0.17}_{-0.27}$ \\ 
7 & 1.0 & 1.67 & $\pm$0.12 & $^{+0.18}_{-0.27}$ \\ 
8 & 1.01 & 1.75 & $\pm$0.09 & $^{+0.13}_{-0.2}$ \\ 
9 & 0.88 & 1.42 & $\pm$0.19 & $^{+0.23}_{-0.43}$ \\ 
10 & 0.92 & 1.41 & $\pm$0.17 & $^{+0.21}_{-0.44}$ \\ 
11 & 0.93 & 1.42 & $\pm$0.17 & $^{+0.2}_{-0.35}$ \\ 
12 & 0.96 & 1.5 & $\pm$0.16 & $^{+0.19}_{-0.31}$ \\ 
13 & 0.96 & 1.6 & $\pm$0.15 & $^{+0.2}_{-0.29}$ \\ 
14 & 0.95 & 1.73 & $\pm$0.18 & $^{+0.22}_{-0.31}$ \\ 
15 & 0.96 & 1.81 & $\pm$0.16 & $^{+0.2}_{-0.29}$ \\ 
16 & 0.99 & 1.85 & $\pm$0.14 & $^{+0.16}_{-0.22}$ \\ 
17 & 0.88 & 1.46 & $\pm$0.2 & $^{+0.25}_{-0.4}$ \\ 
18 & 0.91 & 1.44 & $\pm$0.18 & $^{+0.22}_{-0.42}$ \\ 
19 & 0.93 & 1.48 & $\pm$0.16 & $^{+0.2}_{-0.36}$ \\ 
20 & 0.95 & 1.58 & $\pm$0.14 & $^{+0.18}_{-0.32}$ \\ 
21 & 0.95 & 1.69 & $\pm$0.15 & $^{+0.19}_{-0.3}$ \\ 
22 & 1.02 & 1.75 & $\pm$0.16 & $^{+0.19}_{-0.3}$ \\ 
23 & 1.0 & 1.85 & $\pm$0.16 & $^{+0.21}_{-0.32}$ \\ 
24 & 1.02 & 1.97 & $\pm$0.17 & $^{+0.19}_{-0.21}$ \\ 
25 & 0.88 & 1.51 & $\pm$0.25 & $^{+0.29}_{-0.43}$ \\ 
26 & 0.91 & 1.51 & $\pm$0.19 & $^{+0.23}_{-0.38}$ \\ 
27 & 0.92 & 1.57 & $\pm$0.16 & $^{+0.21}_{-0.39}$ \\ 
28 & 0.96 & 1.68 & $\pm$0.15 & $^{+0.2}_{-0.37}$ \\ 
29 & 0.97 & 1.78 & $\pm$0.15 & $^{+0.2}_{-0.33}$ \\ 
30 & 1.01 & 1.98 & $\pm$0.16 & $^{+0.18}_{-0.29}$ \\ 
31 & 0.95 & 2.09 & $\pm$0.17 & $^{+0.27}_{-0.32}$ \\ 
32 & 1.04 & 2.09 & $\pm$0.16 & $^{+0.21}_{-0.33}$ \\ 
33 & 0.86 & 1.61 & $\pm$0.24 & $^{+0.3}_{-0.38}$ \\ 
34 & 0.9 & 1.61 & $\pm$0.21 & $^{+0.25}_{-0.39}$ \\ 
35 & 0.92 & 1.7 & $\pm$0.18 & $^{+0.23}_{-0.36}$ \\ 
36 & 0.93 & 1.85 & $\pm$0.17 & $^{+0.23}_{-0.37}$ \\ 
37 & 0.98 & 2.04 & $\pm$0.16 & $^{+0.22}_{-0.42}$ \\ 
38 & 0.99 & 2.21 & $\pm$0.2 & $^{+0.27}_{-0.4}$ \\	
\hline
     \end{tabular}
  \caption{\small The ratio $R_{13/7}$ computed with the original NNPDF3.0 PDF set, in order to compare with the predictions obtained with
    the NNPDF3.0+LHCb set in Table~\ref{tab:Dratio}.
The data is ordered in increasing rapidity bins, an within each of these, in increasing $p_T$ bins.
For each bin, we show the central prediction, the PDF uncertainty
and the the total theory uncertainty for  $R_{13/7}({\rm orig})$,
as well as the ratio between the new and orig predictions
for the ratio itself, $R_{13/7}({\rm new})/R_{13/7}({\rm orig})$.
 \label{tab:DratioOld}
  }
\end{table}

We should mention that, once a measurement of $R_{13/7}$ becomes available, it should be possible
to include this data in a global PDF fit in a similar way as we have done with the 7 TeV charm
normalised cross-sections.
One expects similar improvements in the low-$x$ gluon, though perhaps the increased
lever arm in $x$ of the 13 TeV data will increase the constraining power
towards smaller values of $x$.
As discussed in Sect.~\ref{sec:appendix2}, the main advantage of the ratio
measurement is the cancellation of theory systematics, in particular
from scale variations.

We have also computed the value of ratio of inclusive fiducial cross-sections,
as explained in Sect.~\ref{sec:predictionsforratio}, but this time
for original NNPDF3.0 set,
which turns out to be
\beq
\label{eq:r1372}
R_{D}({\rm orig}) = 1.52 \, ^{+0.19 \,(12.6\%)}_{-0.34 \,(22.6\%)} \,,
\eeq
where we provide the total theoretical uncertainty of the prediction.
This should be compared with the result obtained with the NNPDF3.0+LHCb
set, Eq.~(\ref{eq:r137}).
The reduction of the total theory uncertainty in  Eq.~(\ref{eq:r137})
as compared to Eq.~(\ref{eq:r1372}) is a consequence of the constraints
from the 7 TeV LHCb charm measurements.

We now provide the differential predictions for $B$ meson production at LHC Run II in LHCb.
First of all, in Table~\ref{tab:Bratio} we provide the predictions for
the ratio of double differential cross-sections for the production of $B^0$ mesons
between 13 TeV and 7 TeV, see Eq.~(\ref{eq:ratioB0}).
These results were represented graphically in Fig.~\ref{fig:LHCbbeauty13}.
This is the analog table as that for charm production in
Table~\ref{tab:Dratio}.
As in the case of charm, we have assumed the same binning in $\lp p_T^B,y^B\rp$ than
the corresponding 7 TeV measurement.
The magnitude of $R_{13/7}^B$ increases rapidly with increasing $ p_T^B$ and $y^B$,
where the 7 TeV cross-sections are close to their kinematical boundaries.

\renewcommand*{\arraystretch}{1.5}
\begin{table}[t!]
  \centering
    \footnotesize
\begin{tabular}{@{} c|rcl|rcl|rcl|rcl|rcl@{}}
\hline
   \multicolumn{16}{c}{$R_{13/7}^B=\frac{d^2\sigma(B)(y,p_T,13~{\rm TeV})}{dy^Bdp_T^B}\Big/
     \frac{d^2\sigma(B)(y,p_T,7~{\rm TeV})}{dy^Bdp_T^B}$}  \\
   \hline\hline
 $p_T^B$~(GeV)    	 & \multicolumn{15}{c}{$y^B$}  \\ \hline
 & \multicolumn{3}{c|}{$2.0-2.5$} &  \multicolumn{3}{c|}{$2.5-3.0$}
	& \multicolumn{3}{c|}{$3.0-3.5$}	 &  \multicolumn{3}{c|}{$3.5-4.0$}	& \multicolumn{3}{c}{$4.0-4.5$} \\ \hline
	$0.0-0.5$ & 1.56 & \hspace{-0.5cm} & $^{+0.07}_{-0.11}$ &1.6 & \hspace{-0.5cm} & $^{+0.08}_{-0.13}$ &1.59 & \hspace{-0.5cm} & $^{+0.12}_{-0.14}$ &1.77 & \hspace{-0.5cm} & $^{+0.08}_{-0.17}$ &1.85 & \hspace{-0.5cm} & $^{+0.25}_{-0.17}$ \\ 
$0.5-1.0$ & 1.54 & \hspace{-0.5cm} & $^{+0.08}_{-0.14}$ &1.6 & \hspace{-0.5cm} & $^{+0.07}_{-0.15}$ &1.66 & \hspace{-0.5cm} & $^{+0.07}_{-0.15}$ &1.74 & \hspace{-0.5cm} & $^{+0.11}_{-0.15}$ &1.93 & \hspace{-0.5cm} & $^{+0.11}_{-0.16}$ \\ 
$1.0-1.5$ & 1.53 & \hspace{-0.5cm} & $^{+0.07}_{-0.13}$ &1.59 & \hspace{-0.5cm} & $^{+0.07}_{-0.15}$ &1.65 & \hspace{-0.5cm} & $^{+0.09}_{-0.15}$ &1.75 & \hspace{-0.5cm} & $^{+0.1}_{-0.15}$ &1.95 & \hspace{-0.5cm} & $^{+0.1}_{-0.14}$ \\ 
$1.5-2.0$ & 1.55 & \hspace{-0.5cm} & $^{+0.06}_{-0.13}$ &1.59 & \hspace{-0.5cm} & $^{+0.08}_{-0.13}$ &1.66 & \hspace{-0.5cm} & $^{+0.08}_{-0.14}$ &1.8 & \hspace{-0.5cm} & $^{+0.08}_{-0.14}$ &1.99 & \hspace{-0.5cm} & $^{+0.09}_{-0.2}$ \\ 
$2.0-2.5$ & 1.53 & \hspace{-0.5cm} & $^{+0.07}_{-0.13}$ &1.58 & \hspace{-0.5cm} & $^{+0.08}_{-0.14}$ &1.66 & \hspace{-0.5cm} & $^{+0.07}_{-0.13}$ &1.78 & \hspace{-0.5cm} & $^{+0.1}_{-0.16}$ &2.01 & \hspace{-0.5cm} & $^{+0.09}_{-0.18}$ \\ 
$2.5-3.0$ & 1.55 & \hspace{-0.5cm} & $^{+0.08}_{-0.13}$ &1.62 & \hspace{-0.5cm} & $^{+0.07}_{-0.13}$ &1.69 & \hspace{-0.5cm} & $^{+0.08}_{-0.14}$ &1.79 & \hspace{-0.5cm} & $^{+0.11}_{-0.15}$ &1.98 & \hspace{-0.5cm} & $^{+0.13}_{-0.17}$ \\ 
$3.0-3.5$ & 1.58 & \hspace{-0.5cm} & $^{+0.07}_{-0.13}$ &1.62 & \hspace{-0.5cm} & $^{+0.08}_{-0.14}$ &1.71 & \hspace{-0.5cm} & $^{+0.07}_{-0.14}$ &1.87 & \hspace{-0.5cm} & $^{+0.08}_{-0.16}$ &2.03 & \hspace{-0.5cm} & $^{+0.11}_{-0.17}$ \\ 
$3.5-4.0$ & 1.59 & \hspace{-0.5cm} & $^{+0.07}_{-0.13}$ &1.65 & \hspace{-0.5cm} & $^{+0.07}_{-0.14}$ &1.73 & \hspace{-0.5cm} & $^{+0.09}_{-0.15}$ &1.89 & \hspace{-0.5cm} & $^{+0.08}_{-0.13}$ &2.09 & \hspace{-0.5cm} & $^{+0.11}_{-0.18}$ \\ 
$4.0-4.5$ & 1.58 & \hspace{-0.5cm} & $^{+0.08}_{-0.13}$ &1.67 & \hspace{-0.5cm} & $^{+0.07}_{-0.14}$ &1.76 & \hspace{-0.5cm} & $^{+0.09}_{-0.15}$ &1.95 & \hspace{-0.5cm} & $^{+0.08}_{-0.17}$ &2.15 & \hspace{-0.5cm} & $^{+0.12}_{-0.17}$ \\ 
$4.5-5.0$ & 1.63 & \hspace{-0.5cm} & $^{+0.07}_{-0.13}$ &1.71 & \hspace{-0.5cm} & $^{+0.07}_{-0.15}$ &1.77 & \hspace{-0.5cm} & $^{+0.09}_{-0.15}$ &1.96 & \hspace{-0.5cm} & $^{+0.09}_{-0.17}$ &2.26 & \hspace{-0.5cm} & $^{+0.1}_{-0.19}$ \\ 
$5.0-5.5$ & 1.62 & \hspace{-0.5cm} & $^{+0.07}_{-0.13}$ &1.71 & \hspace{-0.5cm} & $^{+0.08}_{-0.14}$ &1.81 & \hspace{-0.5cm} & $^{+0.08}_{-0.14}$ &2.01 & \hspace{-0.5cm} & $^{+0.09}_{-0.16}$ &2.27 & \hspace{-0.5cm} & $^{+0.11}_{-0.16}$ \\ 
$5.5-6.0$ & 1.65 & \hspace{-0.5cm} & $^{+0.08}_{-0.14}$ &1.73 & \hspace{-0.5cm} & $^{+0.07}_{-0.13}$ &1.88 & \hspace{-0.5cm} & $^{+0.07}_{-0.15}$ &2.06 & \hspace{-0.5cm} & $^{+0.12}_{-0.2}$ &2.39 & \hspace{-0.5cm} & $^{+0.1}_{-0.2}$ \\ 
$6.0-6.5$ & 1.69 & \hspace{-0.5cm} & $^{+0.08}_{-0.15}$ &1.78 & \hspace{-0.5cm} & $^{+0.07}_{-0.15}$ &1.89 & \hspace{-0.5cm} & $^{+0.08}_{-0.15}$ &2.11 & \hspace{-0.5cm} & $^{+0.1}_{-0.2}$ &2.42 & \hspace{-0.5cm} & $^{+0.11}_{-0.17}$ \\ 
$6.5-7.0$ & 1.73 & \hspace{-0.5cm} & $^{+0.08}_{-0.15}$ &1.76 & \hspace{-0.5cm} & $^{+0.08}_{-0.14}$ &1.92 & \hspace{-0.5cm} & $^{+0.09}_{-0.15}$ &2.15 & \hspace{-0.5cm} & $^{+0.1}_{-0.18}$ &2.44 & \hspace{-0.5cm} & $^{+0.15}_{-0.19}$ \\ 
$7.0-7.5$ & 1.74 & \hspace{-0.5cm} & $^{+0.07}_{-0.14}$ &1.87 & \hspace{-0.5cm} & $^{+0.08}_{-0.16}$ &1.95 & \hspace{-0.5cm} & $^{+0.1}_{-0.15}$ &2.23 & \hspace{-0.5cm} & $^{+0.08}_{-0.17}$ &2.54 & \hspace{-0.5cm} & $^{+0.22}_{-0.21}$ \\ 
$7.5-8.0$ & 1.76 & \hspace{-0.5cm} & $^{+0.07}_{-0.13}$ &1.85 & \hspace{-0.5cm} & $^{+0.08}_{-0.15}$ &1.97 & \hspace{-0.5cm} & $^{+0.1}_{-0.14}$ &2.24 & \hspace{-0.5cm} & $^{+0.1}_{-0.17}$ &2.53 & \hspace{-0.5cm} & $^{+0.19}_{-0.2}$ \\ 
$8.0-8.5$ & 1.81 & \hspace{-0.5cm} & $^{+0.08}_{-0.13}$ &1.93 & \hspace{-0.5cm} & $^{+0.08}_{-0.16}$ &2.03 & \hspace{-0.5cm} & $^{+0.09}_{-0.15}$ &2.26 & \hspace{-0.5cm} & $^{+0.14}_{-0.18}$ &2.75 & \hspace{-0.5cm} & $^{+0.14}_{-0.19}$ \\ 
$8.5-9.0$ & 1.8 & \hspace{-0.5cm} & $^{+0.1}_{-0.13}$ &1.9 & \hspace{-0.5cm} & $^{+0.07}_{-0.13}$ &2.1 & \hspace{-0.5cm} & $^{+0.1}_{-0.16}$ &2.33 & \hspace{-0.5cm} & $^{+0.11}_{-0.17}$ &2.78 & \hspace{-0.5cm} & $^{+0.16}_{-0.19}$ \\ 
$9.0-9.5$ & 1.83 & \hspace{-0.5cm} & $^{+0.09}_{-0.12}$ &1.98 & \hspace{-0.5cm} & $^{+0.07}_{-0.16}$ &2.1 & \hspace{-0.5cm} & $^{+0.1}_{-0.17}$ &2.42 & \hspace{-0.5cm} & $^{+0.1}_{-0.19}$ &2.82 & \hspace{-0.5cm} & $^{+0.16}_{-0.18}$ \\ 
$9.5-10.0$ & 1.89 & \hspace{-0.5cm} & $^{+0.07}_{-0.14}$ &1.96 & \hspace{-0.5cm} & $^{+0.09}_{-0.15}$ &2.11 & \hspace{-0.5cm} & $^{+0.11}_{-0.16}$ &2.49 & \hspace{-0.5cm} & $^{+0.15}_{-0.21}$ &3.01 & \hspace{-0.5cm} & $^{+0.15}_{-0.24}$ \\ 
$10.0-10.5$ & 1.88 & \hspace{-0.5cm} & $^{+0.07}_{-0.13}$ &2.02 & \hspace{-0.5cm} & $^{+0.07}_{-0.16}$ &2.2 & \hspace{-0.5cm} & $^{+0.09}_{-0.17}$ &2.41 & \hspace{-0.5cm} & $^{+0.17}_{-0.15}$ &3.09 & \hspace{-0.5cm} & $^{+0.17}_{-0.21}$ \\ 
$10.5-11.5$ & 1.94 & \hspace{-0.5cm} & $^{+0.07}_{-0.14}$ &2.05 & \hspace{-0.5cm} & $^{+0.08}_{-0.13}$ &2.19 & \hspace{-0.5cm} & $^{+0.15}_{-0.16}$ &2.59 & \hspace{-0.5cm} & $^{+0.12}_{-0.14}$ &3.31 & \hspace{-0.5cm} & $^{+0.14}_{-0.32}$ \\ 
$11.5-12.5$ & 1.94 & \hspace{-0.5cm} & $^{+0.08}_{-0.13}$ &2.07 & \hspace{-0.5cm} & $^{+0.08}_{-0.14}$ &2.3 & \hspace{-0.5cm} & $^{+0.1}_{-0.17}$ &2.78 & \hspace{-0.5cm} & $^{+0.11}_{-0.22}$ &3.32 & \hspace{-0.5cm} & $^{+0.15}_{-0.21}$ \\ 
$12.5-14.0$ & 2.03 & \hspace{-0.5cm} & $^{+0.08}_{-0.14}$ &2.15 & \hspace{-0.5cm} & $^{+0.08}_{-0.14}$ &2.48 & \hspace{-0.5cm} & $^{+0.08}_{-0.19}$ &2.88 & \hspace{-0.5cm} & $^{+0.1}_{-0.23}$ &3.72 & \hspace{-0.5cm} & $^{+0.18}_{-0.28}$ \\ 
$14.0-16.5$ & 2.11 & \hspace{-0.5cm} & $^{+0.09}_{-0.14}$ &2.27 & \hspace{-0.5cm} & $^{+0.09}_{-0.15}$ &2.6 & \hspace{-0.5cm} & $^{+0.1}_{-0.2}$ &3.06 & \hspace{-0.5cm} & $^{+0.11}_{-0.2}$ &4.12 & \hspace{-0.5cm} & $^{+0.21}_{-0.22}$ \\ 
$16.5-23.5$ & 2.24 & \hspace{-0.5cm} & $^{+0.08}_{-0.13}$ &2.42 & \hspace{-0.5cm} & $^{+0.11}_{-0.14}$ &2.86 & \hspace{-0.5cm} & $^{+0.13}_{-0.17}$ &3.47 & \hspace{-0.5cm} & $^{+0.16}_{-0.22}$ &4.77 & \hspace{-0.5cm} & $^{+0.27}_{-0.44}$ \\ 
$23.5-40.0$ & 2.6 & \hspace{-0.5cm} & $^{+0.09}_{-0.14}$ &2.88 & \hspace{-0.5cm} & $^{+0.11}_{-0.13}$ &3.6 & \hspace{-0.5cm} & $^{+0.15}_{-0.23}$ &4.81 & \hspace{-0.5cm} & $^{+0.21}_{-0.39}$ &7.45 & \hspace{-0.5cm} & $^{+1.21}_{-0.59}$ \\
\hline
     \end{tabular}
\caption{\small Same as Table~\ref{tab:Dratio} for the ratio $R_{13/7}^B$ of double
  differential cross-sections for $B^0$ mesons between 13 TeV and 7 TeV.
}
  \label{tab:Bratio}
\end{table}

%% file: GRRT.bbl
\providecommand{\href}[2]{#2}\begingroup\raggedright\endgroup

%% file: GRRT.bbl
\begin{thebibliography}{10}

\bibitem{Aartsen:2013bka}
{\bf IceCube} Collaboration, M.~Aartsen et~al., {\it {First observation of
  PeV-energy neutrinos with IceCube}},  {\em Phys.Rev.Lett.} {\bf 111} (2013)
  021103, [\href{http://arxiv.org/abs/1304.5356}{{\tt arXiv:1304.5356}}].

\bibitem{Aartsen:2013jdh}
{\bf IceCube} Collaboration, M.~Aartsen et~al., {\it {Evidence for High-Energy
  Extraterrestrial Neutrinos at the IceCube Detector}},  {\em Science} {\bf
  342} (2013) 1242856, [\href{http://arxiv.org/abs/1311.5238}{{\tt
  arXiv:1311.5238}}].

\bibitem{Aartsen:2014gkd}
{\bf IceCube} Collaboration, M.~Aartsen et~al., {\it {Observation of
  High-Energy Astrophysical Neutrinos in Three Years of IceCube Data}},  {\em
  Phys.Rev.Lett.} {\bf 113} (2014) 101101,
  [\href{http://arxiv.org/abs/1405.5303}{{\tt arXiv:1405.5303}}].

\bibitem{Barr:2004br}
G.~Barr, T.~Gaisser, P.~Lipari, S.~Robbins, and T.~Stanev, {\it {A Three -
  dimensional calculation of atmospheric neutrinos}},  {\em Phys.Rev.} {\bf
  D70} (2004) 023006, [\href{http://arxiv.org/abs/astro-ph/0403630}{{\tt
  astro-ph/0403630}}].

\bibitem{GonzalezGarcia:2006ay}
M.~C. Gonzalez-Garcia, M.~Maltoni, and J.~Rojo, {\it {Determination of the
  atmospheric neutrino fluxes from atmospheric neutrino data}},  {\em JHEP}
  {\bf 10} (2006) 075, [\href{http://arxiv.org/abs/hep-ph/0607324}{{\tt
  hep-ph/0607324}}].

\bibitem{Honda:2006qj}
M.~Honda, T.~Kajita, K.~Kasahara, S.~Midorikawa, and T.~Sanuki, {\it
  {Calculation of atmospheric neutrino flux using the interaction model
  calibrated with atmospheric muon data}},  {\em Phys.Rev.} {\bf D75} (2007)
  043006, [\href{http://arxiv.org/abs/astro-ph/0611418}{{\tt
  astro-ph/0611418}}].

\bibitem{Lipari:1993hd}
P.~Lipari, {\it {Lepton spectra in the earth's atmosphere}},  {\em
  Astropart.Phys.} {\bf 1} (1993) 195--227.

\bibitem{Pasquali:1998ji}
L.~Pasquali, M.~Reno, and I.~Sarcevic, {\it {Lepton fluxes from atmospheric
  charm}},  {\em Phys.Rev.} {\bf D59} (1999) 034020,
  [\href{http://arxiv.org/abs/hep-ph/9806428}{{\tt hep-ph/9806428}}].

\bibitem{Enberg:2008te}
R.~Enberg, M.~H. Reno, and I.~Sarcevic, {\it {Prompt neutrino fluxes from
  atmospheric charm}},  {\em Phys.Rev.} {\bf D78} (2008) 043005,
  [\href{http://arxiv.org/abs/0806.0418}{{\tt arXiv:0806.0418}}].

\bibitem{Gondolo:1995fq}
P.~Gondolo, G.~Ingelman, and M.~Thunman, {\it {Charm production and high-energy
  atmospheric muon and neutrino fluxes}},  {\em Astropart.Phys.} {\bf 5} (1996)
  309--332, [\href{http://arxiv.org/abs/hep-ph/9505417}{{\tt hep-ph/9505417}}].

\bibitem{Martin:2003us}
A.~Martin, M.~Ryskin, and A.~Stasto, {\it {Prompt neutrinos from atmospheric $c
  \bar{c}$ and $b \bar{b}$ production and the gluon at very small x}},  {\em
  Acta Phys.Polon.} {\bf B34} (2003) 3273--3304,
  [\href{http://arxiv.org/abs/hep-ph/0302140}{{\tt hep-ph/0302140}}].

\bibitem{Gelmini:1999ve}
G.~Gelmini, P.~Gondolo, and G.~Varieschi, {\it {Prompt atmospheric neutrinos
  and muons: NLO versus LO QCD predictions}},  {\em Phys.Rev.} {\bf D61} (2000)
  036005, [\href{http://arxiv.org/abs/hep-ph/9904457}{{\tt hep-ph/9904457}}].

\bibitem{Bhattacharya:2015jpa}
A.~Bhattacharya, R.~Enberg, M.~H. Reno, I.~Sarcevic, and A.~Stasto, {\it
  {Perturbative charm production and the prompt atmospheric neutrino flux in
  light of RHIC and LHC}},  \href{http://arxiv.org/abs/1502.01076}{{\tt
    arXiv:1502.01076}}.

\bibitem{Garzelli:2015psa}
  M.~V.~Garzelli, S.~Moch and G.~Sigl,
  {\it Lepton fluxes from atmospheric charm revisited},
  \href{http://arxiv.org/abs/1507.01570}{{\tt
    arXiv:1507.01570}}.

\bibitem{Engel:2015dxa}
F.~Riehn, R.~Engel, A.~Fedynitch, T.~K. Gaisser, and T.~Stanev, {\it {Charm
  production in SIBYLL}},  \href{http://arxiv.org/abs/1502.06353}{{\tt
  arXiv:1502.06353}}.

\bibitem{Arguelles:2015wba}
C.~A. Arguelles, F.~Halzen, L.~Will, M.~Kroll, and M.~H. Reno, {\it {The
  High-Energy Behavior of Photon, Neutrino and Proton Cross Sections}},
  \href{http://arxiv.org/abs/1504.06639}{{\tt arXiv:1504.06639}}.

\bibitem{Watt:2011kp}
G.~Watt, {\it {Parton distribution function dependence of benchmark Standard
  Model total cross sections at the 7 TeV LHC}},  {\em JHEP} {\bf 1109} (2011)
069, [\href{http://arxiv.org/abs/1106.5788}{{\tt arXiv:1106.5788}}].

\bibitem{Rojo:2015acz} 
  J.~Rojo {\it et al.},
  {\it The PDF4LHC report on PDFs and LHC data: Results from Run I and preparation for Run II,}
  \href{http://arxiv.org/abs/1507.00556}{{\tt arXiv:1507.00556}}.
  
\bibitem{Ball:2012wy}
R.~D. Ball, S.~Carrazza, L.~Del~Debbio, S.~Forte, J.~Gao, et~al., {\it {Parton
  Distribution Benchmarking with LHC Data}},  {\em JHEP} {\bf 1304} (2013) 125,
  [\href{http://arxiv.org/abs/1211.5142}{{\tt arXiv:1211.5142}}].

\bibitem{Gao:2013xoa}
J.~Gao, M.~Guzzi, J.~Huston, H.-L. Lai, Z.~Li, et~al., {\it {The CT10 NNLO
  Global Analysis of QCD}},  \href{http://arxiv.org/abs/1302.6246}{{\tt
  arXiv:1302.6246}}.

\bibitem{Ball:2010de}
{\bf {The NNPDF }} Collaboration, R.~D. Ball et~al., {\it {A first unbiased
  global NLO determination of parton distributions and their uncertainties}},
  {\em Nucl. Phys.} {\bf B838} (2010) 136--206,
  [\href{http://arxiv.org/abs/1002.4407}{{\tt arXiv:1002.4407}}].

\bibitem{Harland-Lang:2014zoa}
L.~Harland-Lang, A.~Martin, P.~Motylinski, and R.~Thorne, {\it {Parton
  distributions in the LHC era: MMHT 2014 PDFs}},  {\em Eur.Phys.J.} {\bf C75}
  (2015), no.~5 204, [\href{http://arxiv.org/abs/1412.3989}{{\tt
  arXiv:1412.3989}}].

\bibitem{Ball:2001pq}
R.~Ball and R.~K. Ellis, {\it {Heavy quark production at high-energy}},  {\em
  JHEP} {\bf 0105} (2001) 053, [\href{http://arxiv.org/abs/hep-ph/0101199}{{\tt
  hep-ph/0101199}}].

\bibitem{Altarelli:2008aj}
G.~Altarelli, R.~D. Ball, and S.~Forte, {\it {Small x Resummation with Quarks:
  Deep-Inelastic Scattering}},  {\em Nucl. Phys.} {\bf B799} (2008) 199--240,
  [\href{http://arxiv.org/abs/0802.0032}{{\tt arXiv:0802.0032}}].

\bibitem{Ciafaloni:2007gf}
M.~Ciafaloni, D.~Colferai, G.~Salam, and A.~Stasto, {\it {A Matrix formulation
  for small-x singlet evolution}},  {\em JHEP} {\bf 0708} (2007) 046,
  [\href{http://arxiv.org/abs/0707.1453}{{\tt arXiv:0707.1453}}].

\bibitem{Aaron:2009aa}
{\bf H1 and ZEUS} Collaboration, F.~Aaron et~al., {\it {Combined Measurement
  and QCD Analysis of the Inclusive $e^{\pm}p$ Scattering Cross Sections at
  HERA}},  {\em JHEP} {\bf 1001} (2010) 109,
  [\href{http://arxiv.org/abs/0911.0884}{{\tt arXiv:0911.0884}}].

\bibitem{Caola:2009iy}
F.~Caola, S.~Forte, and J.~Rojo, {\it {Deviations from NLO QCD evolution in
  inclusive HERA data}},  {\em Phys. Lett.} {\bf B686} (2010) 127--135,
  [\href{http://arxiv.org/abs/0910.3143}{{\tt arXiv:0910.3143}}].

\bibitem{H1:2015mha}
{\bf ZEUS and H1} Collaboration, S.~Schmidtt, {\it {Combination of Measurements
  of Inclusive Deep Inelastic $e^{\pm}p$ Scattering Cross Sections and QCD
  Analysis of HERA Data}},  \href{http://arxiv.org/abs/1506.06042}{{\tt
  arXiv:1506.06042}}.

\bibitem{Cacciari:1993mq}
M.~Cacciari and M.~Greco, {\it {Large $p_{T}$ hadroproduction of heavy
  quarks}},  {\em Nucl.Phys.} {\bf B421} (1994) 530--544,
  [\href{http://arxiv.org/abs/hep-ph/9311260}{{\tt hep-ph/9311260}}].

\bibitem{Alwall:2014hca}
J.~Alwall, R.~Frederix, S.~Frixione, V.~Hirschi, F.~Maltoni, et~al., {\it {The
  automated computation of tree-level and next-to-leading order differential
  cross sections, and their matching to parton shower simulations}},  {\em
  JHEP} {\bf 1407} (2014) 079, [\href{http://arxiv.org/abs/1405.0301}{{\tt
  arXiv:1405.0301}}].

\bibitem{Sjostrand:2007gs}
T.~Sjostrand, S.~Mrenna, and P.~Z. Skands, {\it {A Brief Introduction to PYTHIA
  8.1}},  {\em Comput. Phys. Commun.} {\bf 178} (2008) 852--867,
  [\href{http://arxiv.org/abs/0710.3820}{{\tt arXiv:0710.3820}}].

\bibitem{Sjostrand:2014zea}
T.~Sjöstrand, S.~Ask, J.~R. Christiansen, R.~Corke, N.~Desai, et~al., {\it {An
  Introduction to PYTHIA 8.2}},  {\em Comput.Phys.Commun.} {\bf 191} (2015)
  159--177, [\href{http://arxiv.org/abs/1410.3012}{{\tt arXiv:1410.3012}}].

\bibitem{Ball:2014uwa}
{\bf NNPDF} Collaboration, R.~D. Ball et~al., {\it {Parton distributions for
  the LHC Run II}},  {\em JHEP} {\bf 1504} (2015) 040,
  [\href{http://arxiv.org/abs/1410.8849}{{\tt arXiv:1410.8849}}].

\bibitem{Lai:2010vv}
H.-L. Lai et~al., {\it {New parton distributions for collider physics}},  {\em
  Phys. Rev.} {\bf D82} (2010) 074024,
  [\href{http://arxiv.org/abs/1007.2241}{{\tt arXiv:1007.2241}}].

\bibitem{Aaij:2013noa}
{\bf LHCb} Collaboration, R.~Aaij et~al., {\it {Measurement of B meson
  production cross-sections in proton-proton collisions at $\sqrt{s}$ = 7
  TeV}},  {\em JHEP} {\bf 1308} (2013) 117,
  [\href{http://arxiv.org/abs/1306.3663}{{\tt arXiv:1306.3663}}].

\bibitem{Aaij:2013mga}
{\bf LHCb} Collaboration, R.~Aaij et~al., {\it {Prompt charm production in pp
  collisions at sqrt(s)=7 TeV}},  {\em Nucl.Phys.} {\bf B871} (2013) 1--20,
  [\href{http://arxiv.org/abs/1302.2864}{{\tt arXiv:1302.2864}}].

\bibitem{Zenaiev:2015rfa}
O.~Zenaiev, A.~Geiser, K.~Lipka, J.~Blümlein, A.~Cooper-Sarkar, et~al., {\it
  {Impact of heavy-flavour production cross sections measured by the LHCb
  experiment on parton distribution functions at low x}},
  \href{http://arxiv.org/abs/1503.04581}{{\tt arXiv:1503.04581}}.

\bibitem{PROSAurl} \url{https://indico.desy.de/getFile.py/access?contribId=60&sessionId=13&resId=0&materialId=slides&confId=9319}.
  
\bibitem{Ball:2010gb}
{\bf The NNPDF} Collaboration, R.~D. Ball et~al., {\it {Reweighting NNPDFs: the
  W lepton asymmetry}},  {\em Nucl. Phys.} {\bf B849} (2011) 112--143,
  [\href{http://arxiv.org/abs/1012.0836}{{\tt arXiv:1012.0836}}].

\bibitem{Ball:2011gg}
R.~D. Ball, V.~Bertone, F.~Cerutti, L.~Del~Debbio, S.~Forte, et~al., {\it
  {Reweighting and Unweighting of Parton Distributions and the LHC W lepton
  asymmetry data}},  {\em Nucl.Phys.} {\bf B855} (2012) 608--638,
  [\href{http://arxiv.org/abs/1108.1758}{{\tt arXiv:1108.1758}}].

\bibitem{Mangano:2012mh}
M.~L. Mangano and J.~Rojo, {\it {Cross Section Ratios between different CM
  energies at the LHC: opportunities for precision measurements and BSM
  sensitivity}},  {\em JHEP} {\bf 1208} (2012) 010,
  [\href{http://arxiv.org/abs/1206.3557}{{\tt arXiv:1206.3557}}].

\bibitem{Nason:1987xz}
P.~Nason, S.~Dawson, and R.~K. Ellis, {\it {The Total Cross-Section for the
  Production of Heavy Quarks in Hadronic Collisions}},  {\em Nucl.Phys.} {\bf
  B303} (1988) 607.

\bibitem{Nason:1989zy}
P.~Nason, S.~Dawson, and R.~K. Ellis, {\it {The One Particle Inclusive
  Differential Cross-Section for Heavy Quark Production in Hadronic
  Collisions}},  {\em Nucl.Phys.} {\bf B327} (1989) 49--92.

\bibitem{Beenakker:1990maa} 
  W.~Beenakker, W.~L.~van Neerven, R.~Meng, G.~A.~Schuler and J.~Smith,
  {\it QCD corrections to heavy quark production in hadron hadron collisions},
  Nucl.\ Phys.\ B {\bf 351}, 507 (1991).

  \bibitem{Beenakker:1988bq} 
  W.~Beenakker, H.~Kuijf, W.~L.~van Neerven and J.~Smith,
  {\it QCD Corrections to Heavy Quark Production in p anti-p Collisions},
  Phys.\ Rev.\ D {\bf 40}, 54 (1989).
 
\bibitem{Mangano:1991jk}
M.~L. Mangano, P.~Nason, and G.~Ridolfi, {\it {Heavy quark correlations in
  hadron collisions at next-to-leading order}},  {\em Nucl.Phys.} {\bf B373}
  (1992) 295--345.

\bibitem{Bonciani:1998vc}
R.~Bonciani, S.~Catani, M.~L. Mangano, and P.~Nason, {\it {NLL resummation of
  the heavy quark hadroproduction cross-section}},  {\em Nucl.Phys.} {\bf B529}
  (1998) 424--450, [\href{http://arxiv.org/abs/hep-ph/9801375}{{\tt
      hep-ph/9801375}}].

\bibitem{Kidonakis:1997gm} 
  N.~Kidonakis and G.~F.~Sterman,
  {\it Resummation for QCD hard scattering},
  Nucl.\ Phys.\ B {\bf 505}, 321 (1997),
  [\href{http://arxiv.org/abs/hep-ph/9705234}{{\tt
      hep-ph/9705234}}].
 
  \bibitem{Ahrens:2010zv} 
  V.~Ahrens, A.~Ferroglia, M.~Neubert, B.~D.~Pecjak and L.~L.~Yang,
  {\it Renormalization-Group Improved Predictions for Top-Quark Pair Production at Hadron Colliders},
  JHEP {\bf 1009}, 097 (2010),
  [\href{http://arxiv.org/abs/hep-ph/1003.5827}{{\tt hep-ph/1003.5827}}].
      
\bibitem{Czakon:2009zw}
M.~Czakon, A.~Mitov, and G.~F. Sterman, {\it {Threshold Resummation for
  Top-Pair Hadroproduction to Next-to-Next-to-Leading Log}},  {\em Phys.Rev.}
  {\bf D80} (2009) 074017, [\href{http://arxiv.org/abs/0907.1790}{{\tt
        arXiv:0907.1790}}].

\bibitem{Cacciari:1998it}
M.~Cacciari, M.~Greco, and P.~Nason, {\it {The P(T) spectrum in heavy flavor
  hadroproduction}},  {\em JHEP} {\bf 9805} (1998) 007,
  [\href{http://arxiv.org/abs/hep-ph/9803400}{{\tt hep-ph/9803400}}].

\bibitem{Kniehl:2008zza}
B.~A. Kniehl, G.~Kramer, I.~Schienbein, and H.~Spiesberger, {\it {Finite-mass
  effects on inclusive $B$ meson hadroproduction}},  {\em Phys.Rev.} {\bf D77}
  (2008) 014011, [\href{http://arxiv.org/abs/0705.4392}{{\tt
  arXiv:0705.4392}}].

\bibitem{Czakon:2012pz}
M.~Czakon and A.~Mitov, {\it {NNLO corrections to top pair production at hadron
  colliders: the quark-gluon reaction}},  {\em JHEP} {\bf 1301} (2013) 080,
  [\href{http://arxiv.org/abs/1210.6832}{{\tt arXiv:1210.6832}}].

\bibitem{Czakon:2013goa}
M.~Czakon, P.~Fiedler, and A.~Mitov, {\it {The total top quark pair production
  cross-section at hadron colliders through O($\alpha_S^4$)}},  {\em
  Phys.Rev.Lett.} {\bf 110} (2013) 252004,
  [\href{http://arxiv.org/abs/1303.6254}{{\tt arXiv:1303.6254}}].

\bibitem{Baernreuther:2012ws}
P.~Bärnreuther, M.~Czakon, and A.~Mitov, {\it {Percent Level Precision Physics
  at the Tevatron: First Genuine NNLO QCD Corrections to $q \bar{q} \to t
  \bar{t} + X$}},  {\em Phys.Rev.Lett.} {\bf 109} (2012) 132001,
  [\href{http://arxiv.org/abs/1204.5201}{{\tt arXiv:1204.5201}}].

\bibitem{Czakon:2014xsa}
M.~Czakon, P.~Fiedler, and A.~Mitov, {\it {Resolving the Tevatron top quark
  forward-backward asymmetry puzzle}},
  \href{http://arxiv.org/abs/1411.3007}{{\tt arXiv:1411.3007}}.

\bibitem{Czakon:2015pga}
M.~Czakon, A.~Mitov, and J.~Rojo, {\it {Summary of the Topical Workshop on Top
  Quark Differential Distributions 2014}},
  \href{http://arxiv.org/abs/1501.01112}{{\tt arXiv:1501.01112}}.

\bibitem{Cacciari:2012ny}
M.~Cacciari, S.~Frixione, N.~Houdeau, M.~L. Mangano, P.~Nason, et~al., {\it
  {Theoretical predictions for charm and bottom production at the LHC}},  {\em
  JHEP} {\bf 1210} (2012) 137, [\href{http://arxiv.org/abs/1205.6344}{{\tt
  arXiv:1205.6344}}].

\bibitem{Cacciari:2001td}
M.~Cacciari, S.~Frixione, and P.~Nason, {\it {The p(T) spectrum in heavy flavor
  photoproduction}},  {\em JHEP} {\bf 0103} (2001) 006,
  [\href{http://arxiv.org/abs/hep-ph/0102134}{{\tt hep-ph/0102134}}].

\bibitem{Forte:2010ta}
S.~Forte, E.~Laenen, P.~Nason, and J.~Rojo, {\it {Heavy quarks in
  deep-inelastic scattering}},  {\em Nucl. Phys.} {\bf B834} (2010) 116--162,
  [\href{http://arxiv.org/abs/1001.2312}{{\tt arXiv:1001.2312}}].

\bibitem{Cacciari:2005uk}
M.~Cacciari, P.~Nason, and C.~Oleari, {\it {A Study of heavy flavored meson
  fragmentation functions in e+ e- annihilation}},  {\em JHEP} {\bf 0604}
  (2006) 006, [\href{http://arxiv.org/abs/hep-ph/0510032}{{\tt
  hep-ph/0510032}}].

\bibitem{Nason:2004rx}
P.~Nason, {\it {A New method for combining NLO QCD with shower Monte Carlo
  algorithms}},  {\em JHEP} {\bf 0411} (2004) 040,
  [\href{http://arxiv.org/abs/hep-ph/0409146}{{\tt hep-ph/0409146}}].

\bibitem{Frixione:2007vw}
S.~Frixione, P.~Nason, and C.~Oleari, {\it {Matching NLO QCD computations with
  Parton Shower simulations: the POWHEG method}},  {\em JHEP} {\bf 0711} (2007)
  070, [\href{http://arxiv.org/abs/0709.2092}{{\tt arXiv:0709.2092}}].

\bibitem{Alioli:2010xd}
S.~Alioli, P.~Nason, C.~Oleari, and E.~Re, {\it {A general framework for
  implementing NLO calculations in shower Monte Carlo programs: the POWHEG
  BOX}},  {\em JHEP} {\bf 1006} (2010) 043,
  [\href{http://arxiv.org/abs/1002.2581}{{\tt arXiv:1002.2581}}].

\bibitem{Frixione:2007nw}
S.~Frixione, P.~Nason, and G.~Ridolfi, {\it {A Positive-weight
  next-to-leading-order Monte Carlo for heavy flavour hadroproduction}},  {\em
  JHEP} {\bf 0709} (2007) 126, [\href{http://arxiv.org/abs/0707.3088}{{\tt
  arXiv:0707.3088}}].

\bibitem{Skands:2014pea}
P.~Skands, S.~Carrazza, and J.~Rojo, {\it {Tuning PYTHIA 8.1: the Monash 2013
  Tune}},  {\em European Physical Journal} {\bf 74} (2014) 3024,
  [\href{http://arxiv.org/abs/1404.5630}{{\tt arXiv:1404.5630}}].

\bibitem{Frixione:2002ik}
S.~Frixione and B.~R. Webber, {\it {Matching NLO QCD computations and parton
  shower simulations}},  {\em JHEP} {\bf 0206} (2002) 029,
  [\href{http://arxiv.org/abs/hep-ph/0204244}{{\tt hep-ph/0204244}}].

\bibitem{Ball:2011mu}
{\bf {NNPDF }} Collaboration, R.~D. Ball et~al., {\it {Impact of Heavy Quark
  Masses on Parton Distributions and LHC Phenomenology}},  {\em Nucl. Phys.}
  {\bf B849} (2011) 296--363, [\href{http://arxiv.org/abs/1101.1300}{{\tt
  arXiv:1101.1300}}].

\bibitem{Beringer:1900zz}
{\bf Particle Data Group} Collaboration, J.~Beringer et~al., {\it {Review of
  Particle Physics (RPP)}},  {\em Phys.Rev.} {\bf D86} (2012) 010001.

\bibitem{Dowling:2013baa} 
  M.~Dowling and S.~O.~Moch,
  {\it Differential distributions for top-quark hadro-production with a running mass,}
  Eur.\ Phys.\ J.\ C {\bf 74}, no. 11, 3167 (2014),
  [\href{http://arxiv.org/abs/1305.6422}{{\tt
        arXiv:1305.6422}}].
  
\bibitem{Gauld:2015lxa} 
  R.~Gauld,
  {\it Forward $D$ predictions for $p\rm Pb$ collisions, and sensitivity to cold nuclear matter effects,}
    \href{http://arxiv.org/abs/1508.07629}{{\tt arXiv:1508.07629}}



\bibitem{Aaij:2011jp}
{\bf LHCb} Collaboration, R.~Aaij et~al., {\it {Measurement of $b$-hadron
  production fractions in $7~\rm{TeV} pp$ collisions}},  {\em Phys.Rev.} {\bf
  D85} (2012) 032008, [\href{http://arxiv.org/abs/1111.2357}{{\tt
  arXiv:1111.2357}}].

\bibitem{Bertone:2013vaa}
V.~Bertone, S.~Carrazza, and J.~Rojo, {\it {APFEL: A PDF Evolution Library with
  QED corrections}},  {\em Comput.Phys.Commun.} {\bf 185} (2014) 1647--1668,
  [\href{http://arxiv.org/abs/1310.1394}{{\tt arXiv:1310.1394}}].

\bibitem{Carrazza:2014gfa}
S.~Carrazza, A.~Ferrara, D.~Palazzo, and J.~Rojo, {\it {APFEL Web: a web-based
  application for the graphical visualization of parton distribution
  functions}},  {\em J.Phys.} {\bf G42} (2015), no.~5 057001,
  [\href{http://arxiv.org/abs/1410.5456}{{\tt arXiv:1410.5456}}].

\bibitem{Alekhin:2014irh}
S.~Alekhin, O.~Behnke, P.~Belov, S.~Borroni, M.~Botje, et~al., {\it
  {HERAFitter, Open Source QCD Fit Project}},
  \href{http://arxiv.org/abs/1410.4412}{{\tt arXiv:1410.4412}}.

\bibitem{Buckley:2014ana}
A.~Buckley, J.~Ferrando, S.~Lloyd, K.~Nordström, B.~Page, et~al., {\it
  {LHAPDF6: parton density access in the LHC precision era}},  {\em
  Eur.Phys.J.} {\bf C75} (2015), no.~3 132,
  [\href{http://arxiv.org/abs/1412.7420}{{\tt arXiv:1412.7420}}].

\bibitem{d'Enterria:2012yj}
D.~d'Enterria and J.~Rojo, {\it {Quantitative constraints on the gluon
  distribution function in the proton from collider isolated-photon data}},
  {\em Nucl.Phys.} {\bf B860} (2012) 311--338,
  [\href{http://arxiv.org/abs/1202.1762}{{\tt arXiv:1202.1762}}].

\bibitem{Carminati:2012mm}
L.~Carminati, G.~Costa, D.~D'Enterria, I.~Koletsou, G.~Marchiori, et~al., {\it
  {Sensitivity of the LHC isolated-gamma+jet data to the parton distribution
  functions of the proton}},  {\em Europhys.Lett.} {\bf 101} (2013) 61002,
  [\href{http://arxiv.org/abs/1212.5511}{{\tt arXiv:1212.5511}}].

\bibitem{Czakon:2013tha}
M.~Czakon, M.~L. Mangano, A.~Mitov, and J.~Rojo, {\it {Constraints on the gluon
  PDF from top quark pair production at hadron colliders}},  {\em JHEP} {\bf
  1307} (2013) 167, [\href{http://arxiv.org/abs/1303.7215}{{\tt
  arXiv:1303.7215}}].

\bibitem{Nocera:2014gqa}
{\bf NNPDF} Collaboration, E.~R. Nocera, R.~D. Ball, S.~Forte, G.~Ridolfi, and
  J.~Rojo, {\it {A first unbiased global determination of polarized PDFs and
  their uncertainties}},  {\em Nucl.Phys.} {\bf B887} (2014) 276--308,
  [\href{http://arxiv.org/abs/1406.5539}{{\tt arXiv:1406.5539}}].

\bibitem{amcfast}
V.~Bertone, R.~Frederix, S.~Frixione, J.~Rojo, and M.~Sutton, {\it {aMCfast:
  automation of fast NLO computations for PDF fits}},  {\em JHEP} {\bf 1408}
  (2014) 166, [\href{http://arxiv.org/abs/1406.7693}{{\tt arXiv:1406.7693}}].

\bibitem{Carli:2010rw}
T.~Carli, D.~Clements, A.~Cooper-Sarkar, C.~Gwenlan, G.~P. Salam, et~al., {\it
  {A posteriori inclusion of parton density functions in NLO QCD final-state
  calculations at hadron colliders: The APPLGRID Project}},  {\em Eur.Phys.J.}
  {\bf C66} (2010) 503--524, [\href{http://arxiv.org/abs/0911.2985}{{\tt
  arXiv:0911.2985}}].

\bibitem{Aad:2013lpa}
{\bf ATLAS} Collaboration, G.~Aad et~al., {\it {Measurement of the inclusive
  jet cross section in pp collisions at $\sqrt{s}$=2.76 TeV and comparison to
  the inclusive jet cross section at $\sqrt{s}$=7 TeV using the ATLAS
  detector}},  {\em Eur.Phys.J.} {\bf C73} (2013) 2509,
[\href{http://arxiv.org/abs/1304.4739}{{\tt arXiv:1304.4739}}].

\bibitem{Alekhin:2013nda} 
  S.~Alekhin, J.~Blumlein and S.~Moch,
  {\it The ABM parton distributions tuned to LHC data}
  Phys.\ Rev.\ D {\bf 89}, no. 5, 054028 (2014)
  [\href{http://arxiv.org/abs/1310.3059}{{\tt arXiv:1310.3059}}].

  \bibitem{Dulat:2015mca} 
  S.~Dulat {\it et al.},
  {\it The CT14 Global Analysis of Quantum Chromodynamics},
  \href{http://arxiv.org/abs/1506.07443}{{\tt arXiv:1506.07443}}.

  \bibitem{Abramowicz:2015mha} 
  H.~Abramowicz {\it et al.} [H1 and ZEUS Collaborations],
  {\it Combination of Measurements of Inclusive Deep Inelastic $e^{\pm}p$ Scattering Cross Sections and QCD Analysis of HERA Data},
  \href{http://arxiv.org/abs/1506.06042}{{\tt arXiv:1506.06042}}.  


\bibitem{Aaij:2015bpa} 
  {\bf LHCb} Collaboration,  R.~Aaij {\it et al.}, {\it
    Measurements of prompt charm production cross-sections in $pp$ collisions at $\sqrt{s} = 13\,\mathrm{TeV}$},
  [\href{http://arxiv.org/abs/1510.01707}{{\tt arXiv:1510.01707}}].
    
\bibitem{CMS:2014jea}
{\bf CMS} Collaboration, V.~Khachatryan et~al., {\it {Measurements of
  differential and double-differential Drell-Yan cross sections in
  proton-proton collisions at 8 TeV}},  {\em Eur.Phys.J.} {\bf C75} (2015),
  no.~4 147, [\href{http://arxiv.org/abs/1412.1115}{{\tt arXiv:1412.1115}}].

\bibitem{CMN}
  M.~Cacciari, M.~L.~Mangano and P.~Nason,
  {\it Gluon PDF constraints from the ratio of forward heavy quark production at the LHC at $\sqrt{s}=7$ and 13 TeV},
\href{http://arxiv.org/abs/1507.06197}{{\tt arXiv:1507.06197}}


\bibitem{CMS-PAS-HIN-14-004}
{\bf CMS} Collaboration, S.~Chatrchyan et~al., {\it {Measurements of the
  ${B}^{\rm +}$, ${B}^{\rm 0}$ and ${B}_{\rm s}^{\rm 0}$ production cross
  sections in pPb collisions at $\sqrt{s_{_{\text{NN}}}}$ = 5.02 TeV}},
  \href{http://arxiv.org/abs/CMS-PAS-HIN-14-004}{{\tt CMS-PAS-HIN-14-004}}.

\bibitem{Eskola:2009uj}
K.~Eskola, H.~Paukkunen, and C.~Salgado, {\it {EPS09: A New Generation of NLO
  and LO Nuclear Parton Distribution Functions}},  {\em JHEP} {\bf 0904} (2009)
  065, [\href{http://arxiv.org/abs/0902.4154}{{\tt arXiv:0902.4154}}].

\bibitem{deFlorian:2003qf}
D.~de~Florian and R.~Sassot, {\it {Nuclear parton distributions at next to
  leading order}},  {\em Phys. Rev.} {\bf D69} (2004) 074028,
  [\href{http://arxiv.org/abs/hep-ph/0311227}{{\tt hep-ph/0311227}}].

\bibitem{deFlorian:2011fp}
D.~de~Florian, R.~Sassot, P.~Zurita, and M.~Stratmann, {\it {Global Analysis of
  Nuclear Parton Distributions}},  {\em Phys.Rev.} {\bf D85} (2012) 074028,
  [\href{http://arxiv.org/abs/1112.6324}{{\tt arXiv:1112.6324}}].

\bibitem{Skands:2010ak}
P.~Z. Skands, {\it {Tuning Monte Carlo Generators: The Perugia Tunes}},  {\em
  Phys.Rev.} {\bf D82} (2010) 074018,
  [\href{http://arxiv.org/abs/1005.3457}{{\tt arXiv:1005.3457}}].

\end{thebibliography}
